\documentclass{article}

\usepackage{arxiv}

\usepackage[utf8]{inputenc} 
\usepackage[T1]{fontenc}    
\usepackage{hyperref}       
\usepackage{url}            
\usepackage{booktabs}       
\usepackage{amsfonts}       
\usepackage{nicefrac}       
\usepackage{microtype}      
\usepackage{lipsum}		
\usepackage{graphicx}
\usepackage{natbib}
\usepackage{doi}
\usepackage[dvipsnames]{xcolor}
\usepackage[pagewise]{lineno}
\usepackage{tikz}
\usepackage{subfigure} 
\usepackage[all]{xy}

\usepackage{amsfonts,amssymb,amsmath,amsgen,amsopn,amsbsy,theorem,graphicx,epsfig}

\newtheorem{Obs}{\quad Remarks}

\title{A Markov Chain Model for COVID19 in Mexico City}

\author{ \href{https://orcid.org/0000-0003-3848-9369}{\includegraphics[scale=0.06]{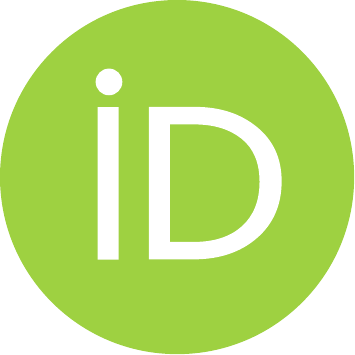}\hspace{1mm}Carlos ~Martinez-Rodriguez} \\
	Academia de Matem\'aticas\\
	Universidad Aut\'onoma de la Ciudad de M\'exico\\
	Iztapalapa,  Ermita Iztapalapa 4163\\
	\texttt{carlos.martinez@uacm.edu.mx} \\
}

\renewcommand{\shorttitle}{Covid19 in Mexico City}

\hypersetup{
pdftitle={A Markov Chain Model for COVID19 in Mexico City},
pdfsubject={q-bio.NC, q-bio.QM},
pdfauthor={Carlos~Martinez-Rodriguez},
pdfkeywords={Probability, Stochastic Processes, Markov Chain, COVID19},
}

\begin{document}
\maketitle

\begin{abstract}
This paper presents a model for COVID19 in Mexico City. The data analyzed were considered from the appearance of the first case in Mexico until July 2021.  In this first approximation the states considered were Susceptible, Infected, Hospitalized, Intensive Care Unit, Intubated, and Dead.  As a consequence of the lack of coronavirus testing, the number of infected and dead people is underestimated, although the results obtained give a good approximation to the evolution of the pandemic in Mexico City. The model is based on a discrete-time Markov chain considering data provided by the Mexican government, the main objective  is to estimate the transient probabilities from one state to another for the Mexico City case.

\end{abstract}


\section{Introduction}
\label{Sec:1}

Coronavirus is a disease caused by several viruses, those who gets this disease are likely to develop respiratory infections or complications that can even lead to death.  However, there are people infected by the virus who do not show any symptoms, that is, they are carriers of the virus and do not show some of the following symptoms: fever, headache, runny nose, respiratory problems, dry cough, even loss of smell and taste.

According to  Chinese health authorities,  the first time it was detected in the seafood market of the city of Wuhan, on November 17, 2019.  At the end of December,  the Chinese government informed the World Health Organization about these new cases of unknown origin. On January 1, 2020,  the World Health Organization (WHO) asked the local authorities for more information on these detected cases to establish the level of risk of both contagion and spread of this new disease.   On January 3,  after 55 cases of atypical pneumonia were reported,  the Chinese government identified this new family of viruses,  Coronavirus,  as the cause of this new and dangerous disease.  Finally,  on February 11,  the WHO formally named the disease COVID-19. On March 11,  2020,  the WHO because of the spread of the virus in several countries,  concluded that COVID-19 should be considered a pandemic,  and expressed concern about the existence of a significant number of cases in such a short time in most parts of the world,  and urged governments to implement response measures to this emerging health emergency.   More information about the pandemic spread can be found in  \citeauthor{17,18,20,29}.  


COVID-19 is a respiratory-type virus that is transferred from person to person either by contact with drops of saliva from an infected person, which can be expelled when coughing or sneezing and even when talking. The general symptoms are fever, fatigue, respiratory problems, dry cough, even loss of smell and taste, with eventual diarrhea. Some patients have mild symptoms, but some more have serious health problems and may even lose their lives.  

The main recommended health contingency measures were the following: keep at least one meter of distance between people, use masks when having relationships with people in any type of situation; avoid closed,  congested spaces or in which it is impossible to keep a distance of at least one meter away, keep windows and doors open to promote ventilation in closed spaces. The sanitary measures to prevent the spread of the virus that causes COVID19 disease are: wash your hands regularly and carefully with soap and water; in the absence of soap and water use hydro alcoholic gel, avoid touching the eyes, nose, and mouth with dirty hands, when coughing or sneezing, cover the mouth and nose with the elbow flexed or with a handkerchief, as well as cleaning and disinfecting frequently surfaces, particularly those that are touched regularly. After a year of pandemic,  it is known that the population sector most affected or more likely to lose their lives as a result of having been infected by the virus,  is that people of advanced age or with comorbidities such diabetes,  hypertension, obesity,  diseases lung problems,  kidney problems,  immunity problems,  asthma,  or heart problems.
On February 28, 2020, jointly, the Pan American Health Organization and the WHO, through the document epidemiological updates, indicated the existence of the first confirmed cases of COVID19. Both cases coincided with the antecedent of having made a trip to Italy before the presentation of symptoms of the disease. After that the government of Mexico City made known to its population the means through which a person infected by the SARS-CoV-2 virus could infect others, mainly through drops of saliva that expelled a person infected by coughing or sneezing, or by contact with a surface contaminated by the virus and subsequently putting contaminated hands to the face, touching the mouth, nose or eyes.

Unfortunately, Mexico is one of the countries most affected by the spread of the disease, at the moment this is written, it ranks fourth in the number of deaths. As part of the response to the health emergency caused by this new SARS-CoV-2 virus, scientists, academics, and experts from different areas have been collaborating and designing strategies to combat the COVID19 pandemic. Specifically, the National Council of Science and Technology works together with the government of Mexico, who, under the coordination of the Ministry of Health, began collaboration in various projects: design of data science tools that make data available to the population data and information in real-time.

\section{The epidemic SIR model}

Scientists use mathematical models to predict the progression of diseases and understand how the interventions of the authorities affect the spread of the disease. More complex models serve to help in decision-making and make more efficient use of resources as well as the consequences of health contingency measures implemented by health authorities. One of the simplest, to estimate the spread of the disease divides the population that is susceptible to infection into three categories according to their stage concerning the disease. Those who have not yet become ill but who can contract it are called susceptible. Those people who have already been infected go from being susceptible to being infected. The third group is of recovered people, that is, those who after suffering the disease have recovered and become immune or unfortunately have died, these people can no longer be infected or spread the disease, this model is known as the SIR model \citeauthor{14,16, 19, 34}.This model can provide an estimate of the total number of infections with and without the intervention of the health authorities. One of the parameters of interest to estimate is the reproduction number, the average number of infections that an infected person can cause in a given time, it is known that if this reproduction number takes values lower than the unity, then the disease will disappear \citeauthor{32,36}.  The proportion of infected people who die as a result of the disease is called the fatality rate. The COVID19 case fatality rate depends significantly on the age and presence of comorbidities of people with the disease.

Among the health contingency measures, the most important are quarantine and isolation, which contribute to reducing the speed of transmission, and therefore the number of reproduction. Isolating infected people reduces the rate of spread while quarantining healthy people reduces the susceptible population and therefore the virus is less likely to spread and prevail among the population \citeauthor{3}. One of the most effective measures to control and eradicate the disease is vaccination, with the application of vaccines, people move from the susceptible sector to the removed sector without becoming infected, reducing the size of the susceptible population, so the disease will finally be controlled.

\subsection*{Description of the SIR model}

The SIR model describes the behavior of the disease, this model was developed by Kermack \citeauthor{23},  this model assumes that each of the members of the population belongs to one of the following three states: Susceptible, Infected and Recovered. Those who are susceptible are those who do not yet have the disease, but who are not immune to it, therefore they can be infected if they have contact with someone who is sick. Infected people are those people who currently have the disease and can transmit it to other people. Recovered people are those who have already suffered from the disease and who can no longer transmit it to other people, nor can they be reinfected.

This model can incorporate more specifications regarding the spread of the virus, however, in most cases, they are a simplification and approximation of reality. Its level of precision will depend on the assumptions made and how much they reflect reality. Some assumptions do not necessarily apply to reality, such as that each individual has the same probability of coming into contact with any other member of the population, that is, it does not consider the fact that contacts are more likely between people geographically and socially closer. Another assumption is that there are people more susceptible to being infected than others, in addition, there are people who have a greater number of contacts than others, also it is assumed is that once a person has been infected, the recovery time depends on the time that has elapsed since the infection, among others. 

Recent work about SIR model with applications and generalizations can be found in \citep{5,9,11,12,22,25}, also \citep{7,13,15,21,37}.

\section{Fundamentals}
General stochastic processes are widely used in epidemic analysis, bayesian models also have been used in the study of COVID19, for example \citep{1,4,6,8,10,24,27,31,35,38}.

\subsection*{Markov Chains preliminars}

Markov chains have proven to be useful tools for modeling important problems that appear in medicine, in particular, they have been successfully applied to model the dynamics of epidemics, that is, the rapid spread (transmission) of an infectious disease to large numbers of individuals in a given population. Factors whose nature is uncertain and whose current state depends on the previous levels of the disease, the application of probabilistic models, such as Markov chains, is a natural and powerful approach.

 A Markov Chain is a sequence of events for which the probabilities of outcomes or states depend on what happened previously. The data are obtained in time intervals (for instance hours, days, weeks, years) that is, there is a succession of values of a certain variable that are successively obtained in time.

A stochastic process is defined as a collection of random variables, $\left\{X_{t},t\in T\right\} $,  $T$ index set, ordered by the sub\'index $t$ which usually denotes time. The possible value that the random variable can take is called a state space,  $S$, so this state space can be discrete or continuous. A discrete-time Markov chain is a stochastic process with a discrete state space,  $S$, such that for\\
 $\left\{X_{0}=x_{0},X_{1}=x_{1},\ldots,X_{n}=x_ {n}\right\} $, the property (\ref{Eq.Prop.Markov}) is satisfied

\begin{eqnarray}\label{Eq.Prop.Markov}
P\left(X_{n+1}=x_{n+1}|X_{n-1}=x_{n-1},\ldots, X_{0}=x_{0}\right)=P\left(X_{n+1}=x_{n}|X_{n}=x_{n}\right).
\end{eqnarray}

A Markov chain is said to be homogeneous if for any value of $n\geq0 $ and any elements of the state space $i, j\in S$ holds that $P\left(X_{n+1}=j|X_{n}=i\right)=P\left(X_{1}=j|X_{0}=i\right)$. This probabilities are denoted by $p_{ij}$ and is called the transition matrix $P=\left(p_{ij}\right)$ for $i, j\in S$.

If for each element of the state space $ S $ we have that $P\left(X_{0}=i\right)=\mu_{i}$, $i\in S $, where $\mu=\left(\mu_{i}\right)_ {i\in S}$ meets the condition $\sum_{i\in S}\mu_{i}=1$, $ \mu$ is called the initial distribution. The inputs of the $P$ matrix verify that $\sum_{j\in S} p_{ij}=1$ is for all $i\in S $, so it is called a stochastic matrix. Also, a Markov chain is determined by its state space, its initial distribution, and its transition matrix, the stochastic matrix.

As regards the SIR model, this model assumes that time is discrete,  the variable $t$ takes values in $\left\{0,1,2, \ldots, \right\} $,  therefore the random variables that define the various states of the Markov chain are discrete. In the SIR model, the elements of the population are classified into three states: Susceptible ($S$), Infected ($I$), and Recovered ($R$). The basic assumptions considered are: the population remains constant over time, those who make up the population go from susceptible to infected; people become infected regardless of age, sex, social status, etc. The population leaves the infected state recovering from the disease, and those who manage to do so acquire immunity and therefore can no longer be infected again. Several applications of Markov chains to COVID19 problem has been developed since the pandemic started, some of them  can be found in \citep{2,26,28,30}  

\section{Markov Chain Model}
The model was motivated by the work developed by \citep{2,7,30,33}.  In this work, the model considered is based on the information provided daily by the health authorities of the Mexican government. From the variables involved in the database, those considered to be analyzed are: infected, hospitalized, intensive care unit, intubated and those who unfortunately have passed away as a consequence of being infected by the SARS-Cov2 virus. With these variables, a data subset was constructed for Mexico City, including its delegations. 

\begin{figure}[!ht]
\centering
\subfigure[Daily cases]{\includegraphics[width=6cm, height=5cm]{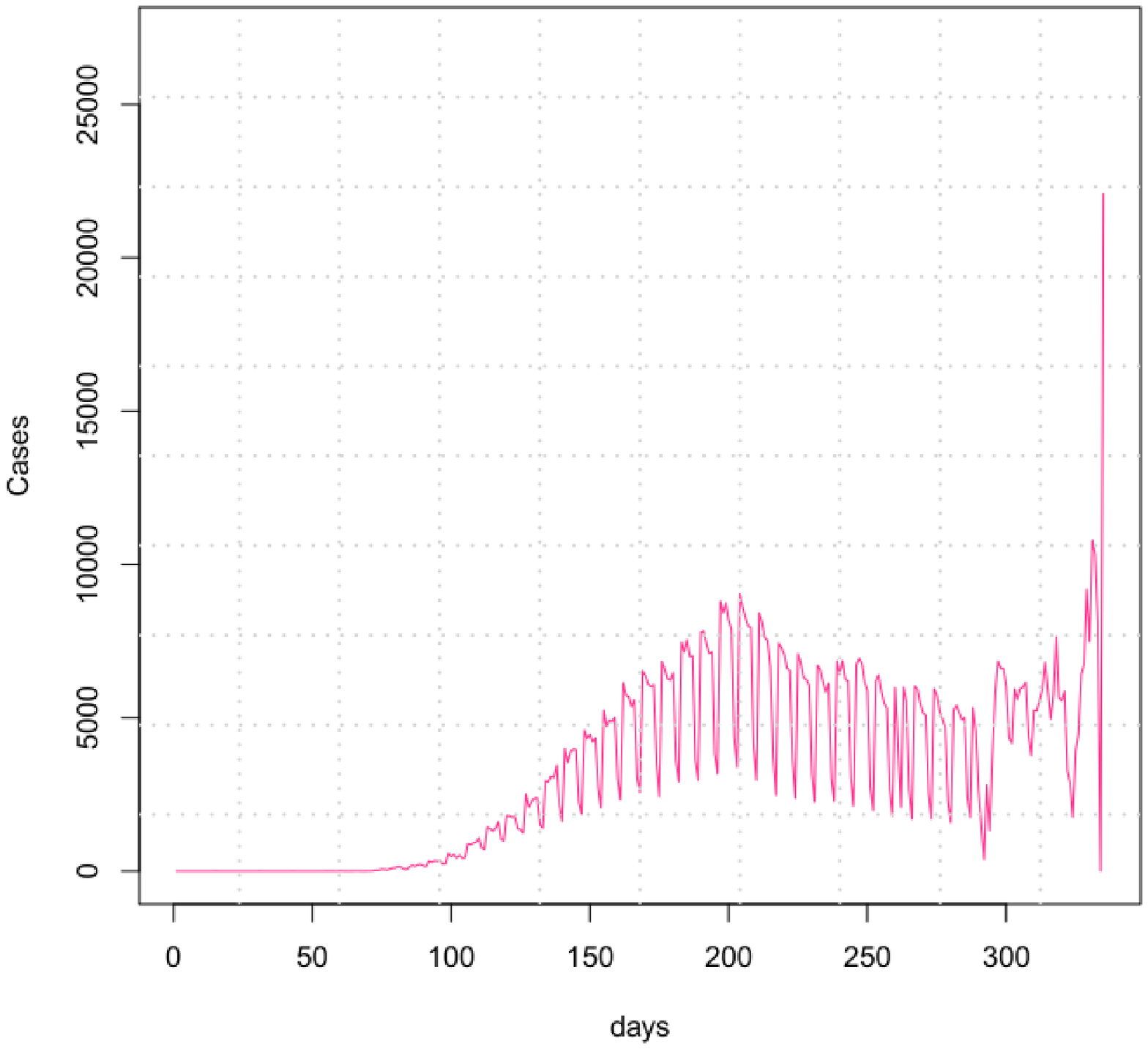}}\hspace{3mm}
\subfigure[Daily Deaths]{\includegraphics[width=6cm, height=5cm]{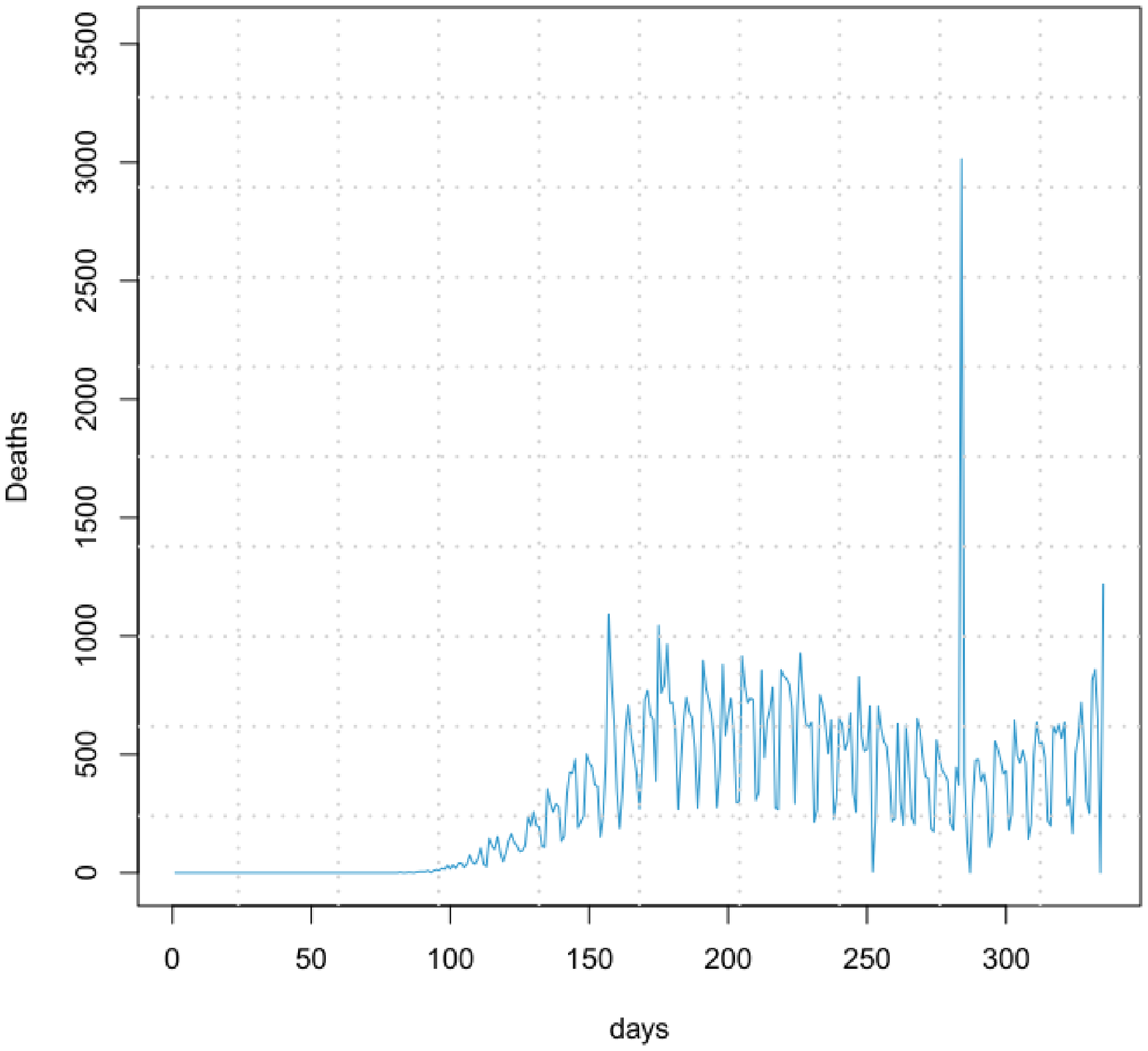}}\hspace{3mm}
\subfigure[Total Cases]{\includegraphics[width=6cm, height=5cm]{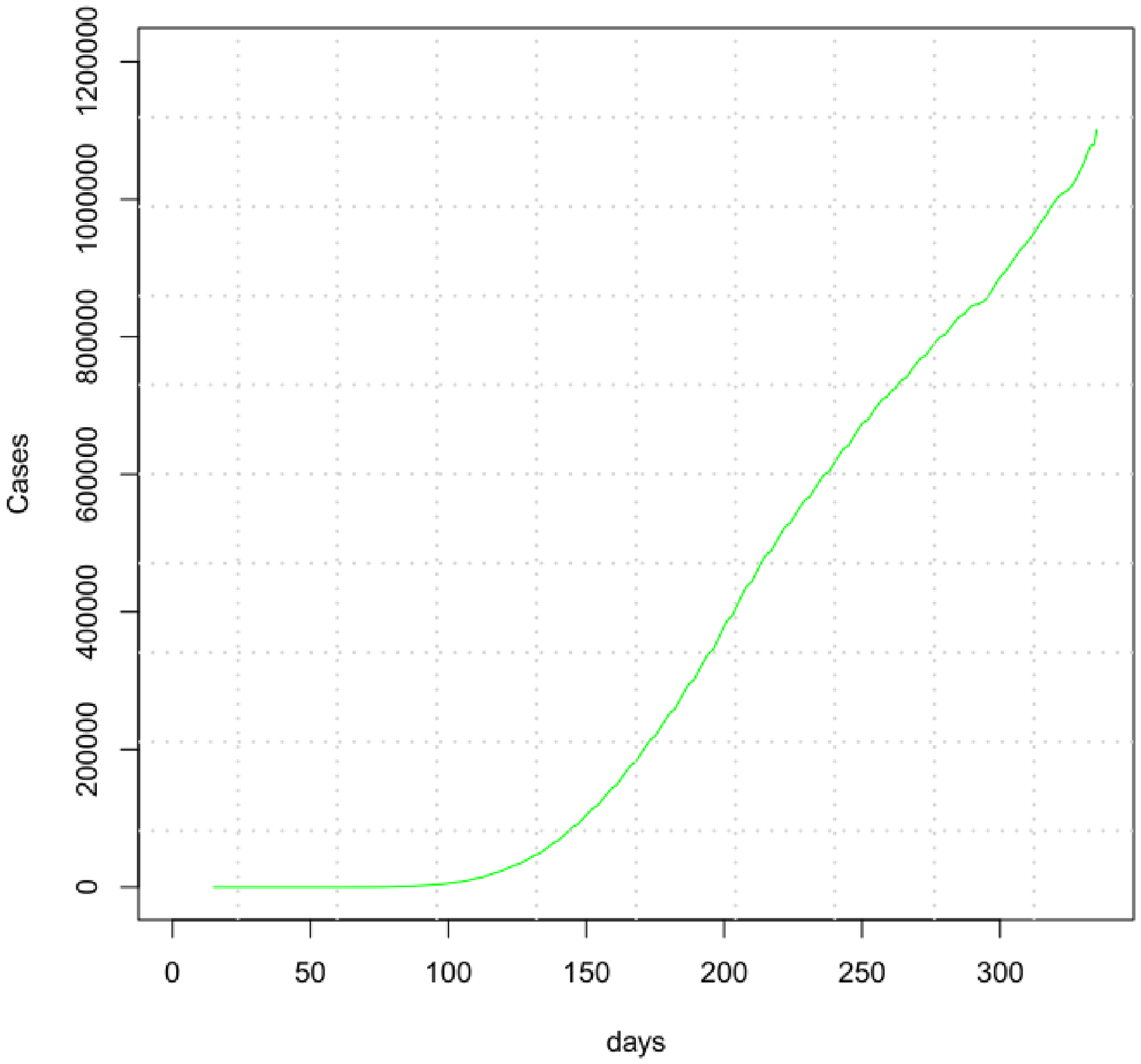}}\vspace{3mm}
\subfigure[Total Deaths ]{\includegraphics[width=6cm, height=5cm]{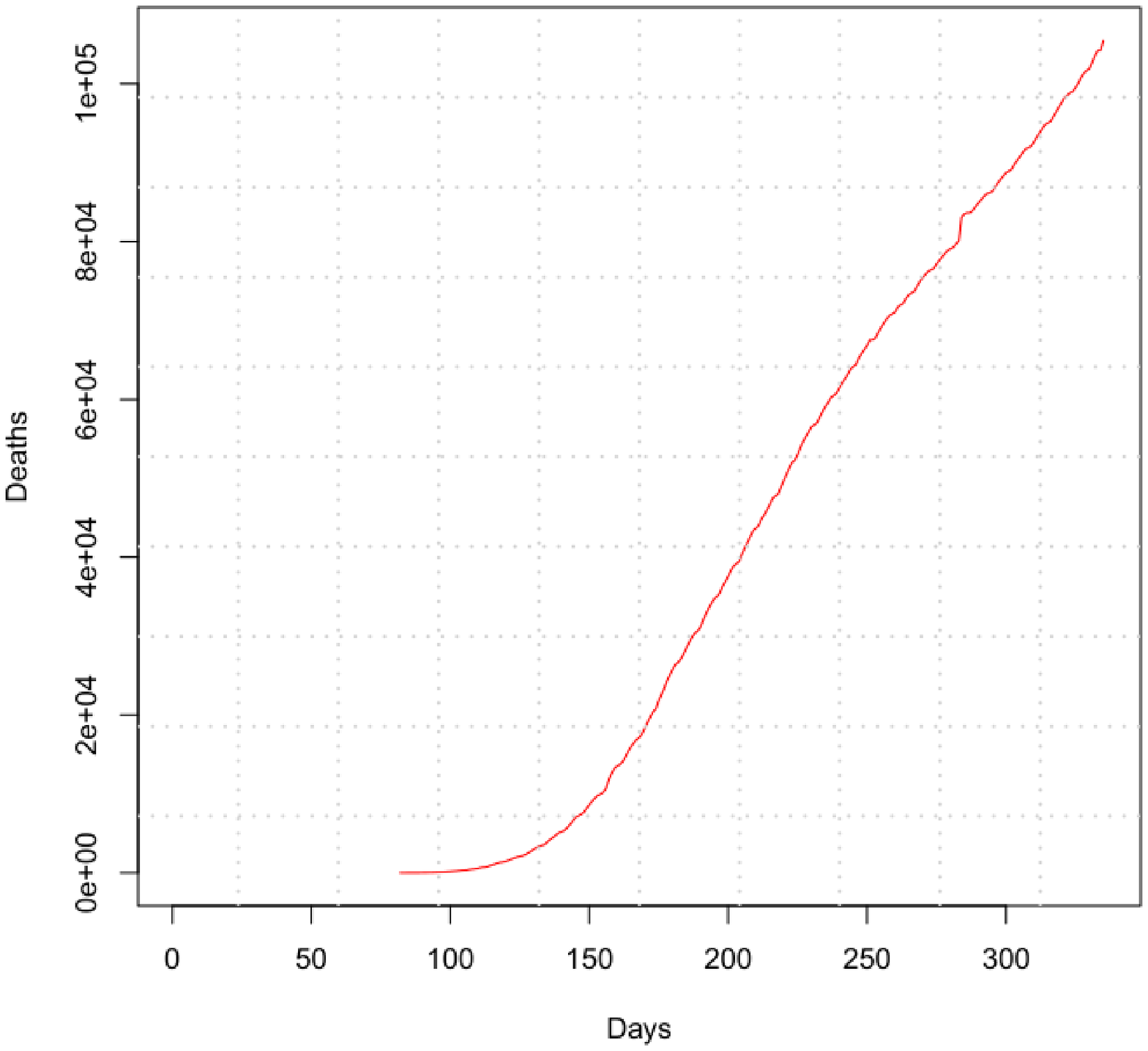}}\hspace{3mm}
\caption{Covid Graphs for M\'exico} \label{fig:Acumulados Mx}
\end{figure}

\begin{figure}[!ht]
\centering
\subfigure[Daily Cases]{\includegraphics[width=6cm, height=5cm]{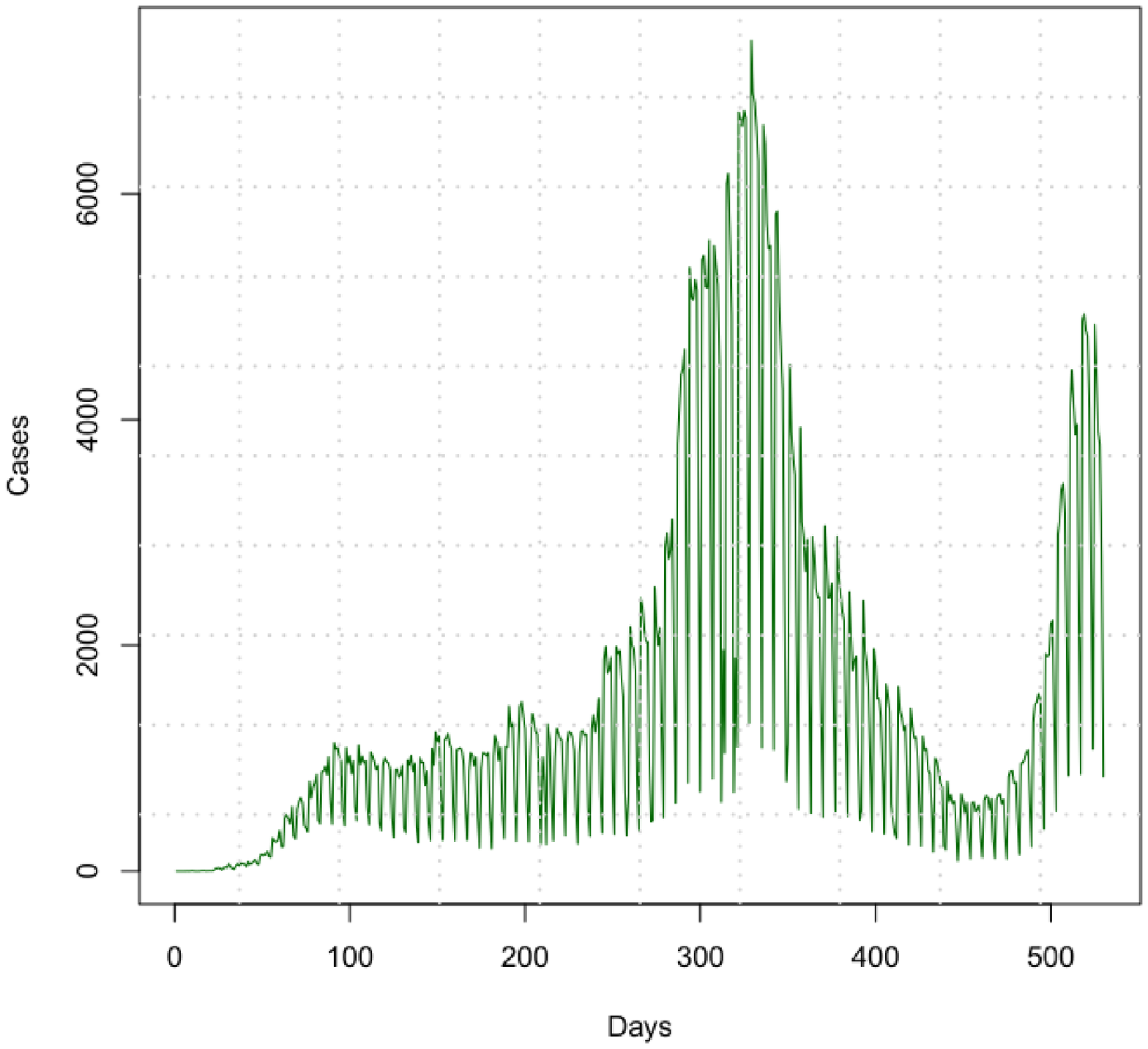}}\hspace{3mm}
\subfigure[Daily Deaths]{\includegraphics[width=6cm, height=5cm]{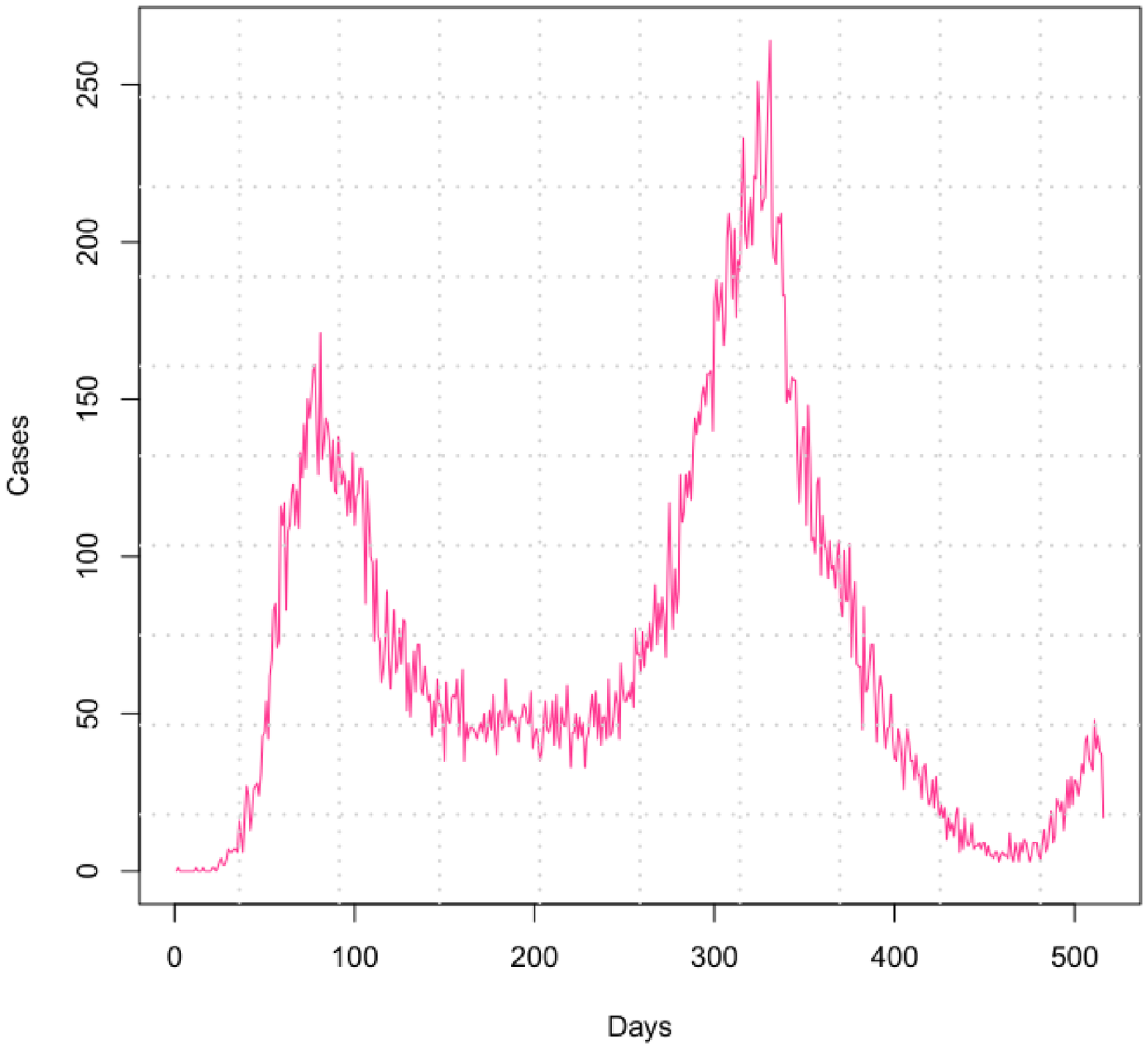}}\hspace{3mm}
\subfigure[Total Cases]{\includegraphics[width=6cm, height=5cm]{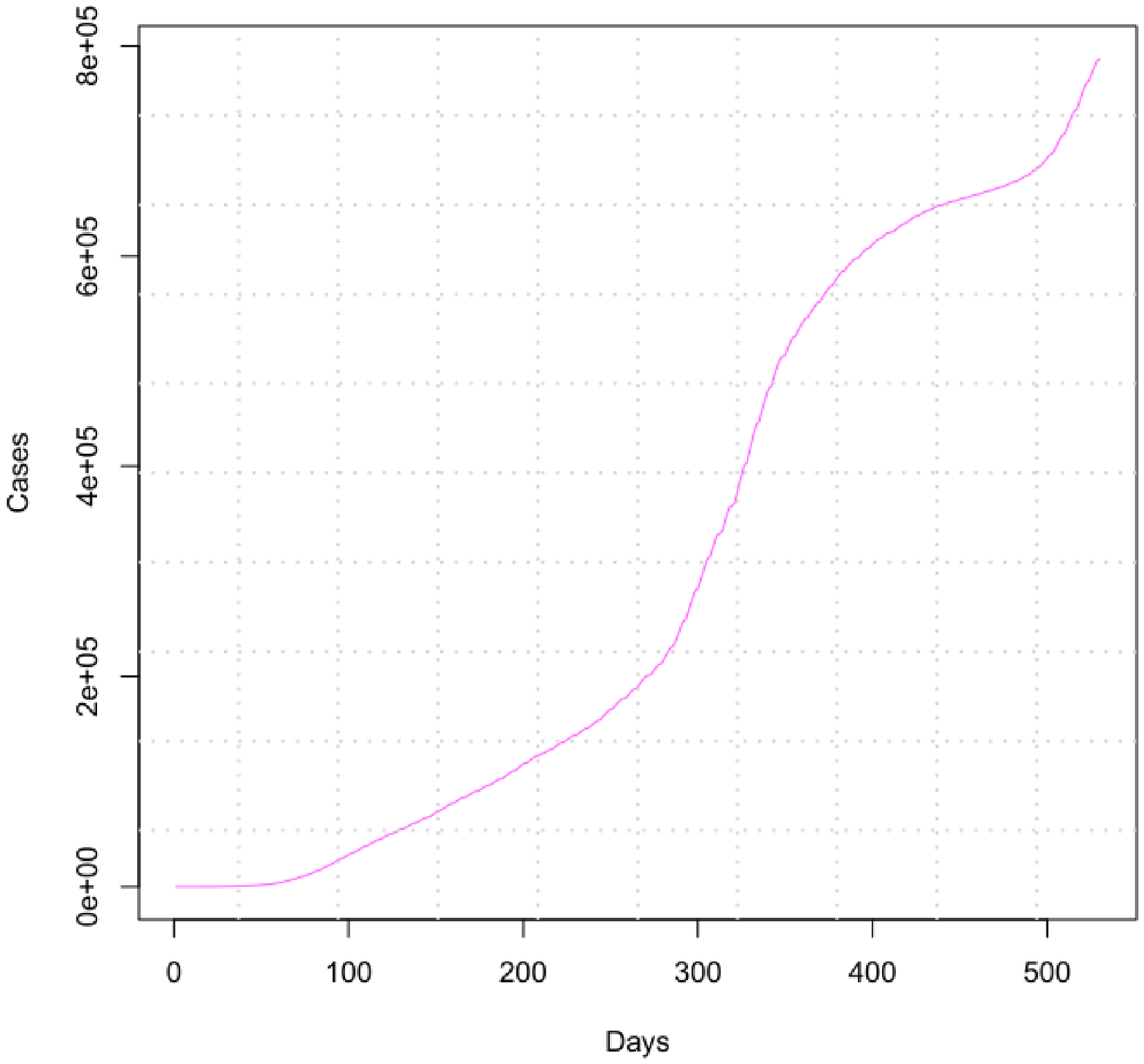}}\vspace{3mm}
\subfigure[Total Deaths]{\includegraphics[width=6cm, height=5cm]{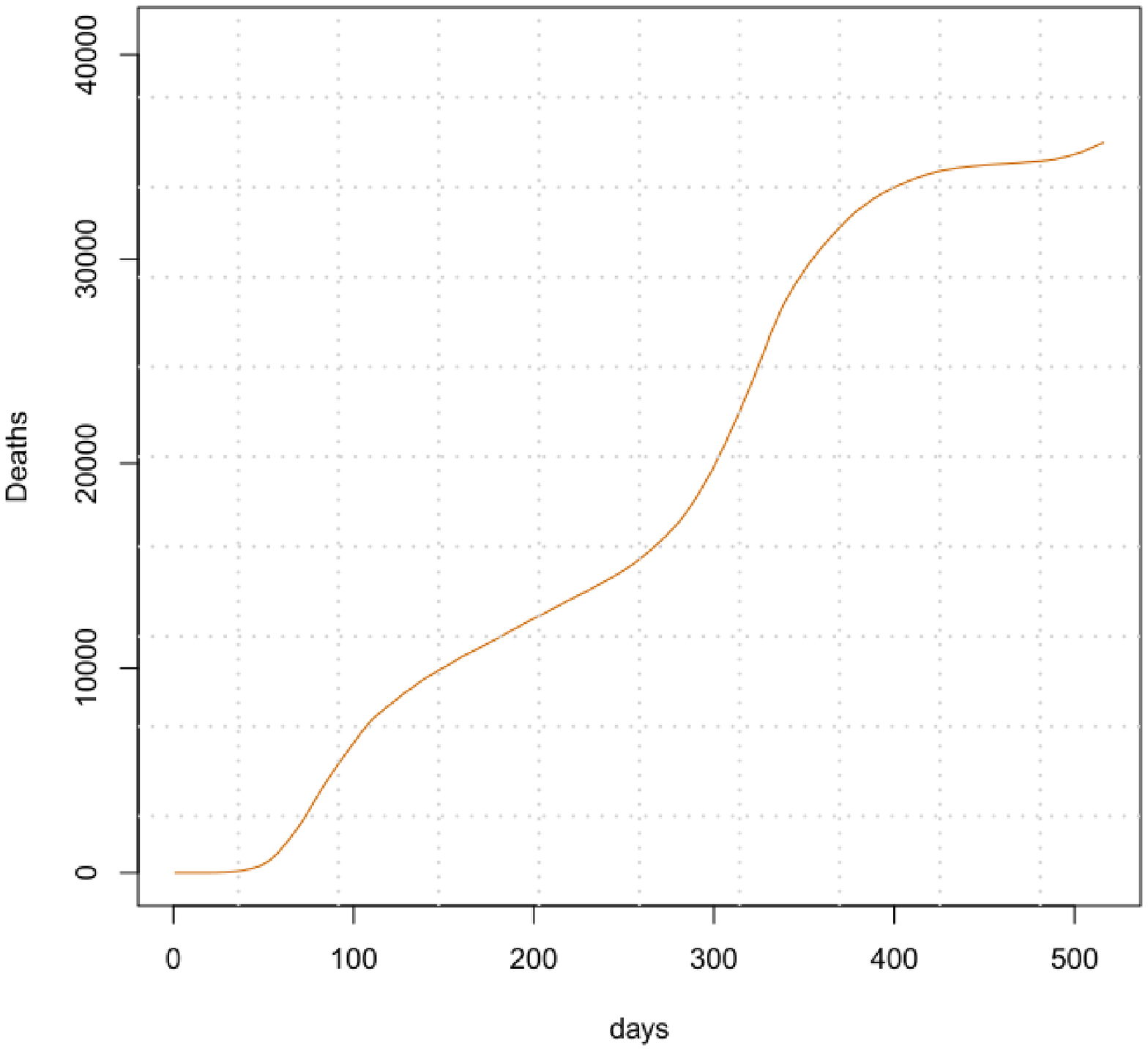}}\hspace{3mm}
\caption{Covid graphs for Mexico City} \label{fig:NuevosCasosCDMx}
\end{figure}

\begin{figure}[!ht]
\centering
\includegraphics[width=8cm, height=8cm]{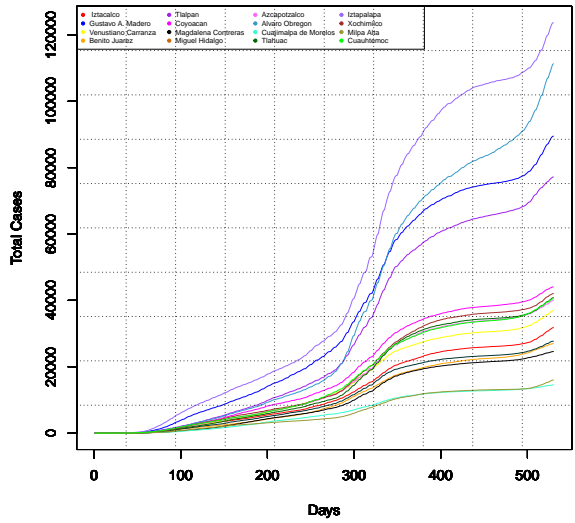}
\caption{Cases for the Mexico City delegations} \label{fig:Delegaciones.Casos}
\end{figure}

From a first review of the data on infections and accumulated cases, daily deaths, and totals, we obtained the graphs shown in the Figures \ref{fig:Acumulados Mx} and \ref{fig:NuevosCasosCDMx},  the case of the delegations in Mexico City is presented in Figure \ref{fig:Delegaciones.Casos}.

\begin{center}
\begin{table}[!ht]
\scalebox{0.65}{
\begin{tabular}{|l||r||r|r||r|r||r|r||r|}\hline
Locality&Population  &\textbf{Cases}& \textbf{Deaths} & Hospitalized& Non-Hospitalized &UCI &Recovered&Intubated \\\hline\hline
CDMX       & 9018645 & 2867582   &    44829  &    119240 &   2748384 &   7694 &   2822795  &  16793\\\hline
Azcapotzalco  &  408441 & 120044   &    3036   &    7610  &   112434 & 323   & 117008  &  1068\\\hline
Milpa Alta   &  139371  & 62580   &    506 &      1388 &    61192 & 90 &    62074 &    225\\\hline\hline
\textbf{Iztapalapa}  &   1815551  & \textbf{517663}    &   \textbf{8810}    &   21947   &  495716&  1364   & 508853   &  3528\\\hline\hline
\textbf{Tlalpan }   &  682234  & \textbf{226018} &      \textbf{2457}    &   8073  &   217945 &  667    & 223561    &1249\\\hline\hline
Xochimilco   &   418060  &161640  &     1662  &     4934   &  156706 &  405   &  159978  &   772\\\hline
MagContreras  &   245147  & 110230  &     894    &   2416  &   107814 & 176  &   109336  &   323\\\hline\hline
\textbf{Gustavo A. Madero}      &   1176967  & \textbf{345557}    &   \textbf{6759}  &    16448   &  329109 &  905   &  338798   &  2246\\\hline\hline
Miguel Hidalgo   &379624  & 101038   &    1770   &    5006   &  96032 &377    & 99268  &   656\\\hline
Iztacalco   &   393821  & 120927    &   2829     &  6992    & 113935 & 355    &118098   &  868\\\hline
Tlahuac    &   366586  &147349    &   1343   &    3668  &   143681& 255  &  146006   &  599\\\hline
Coyoacan   &    621952  &176323   &    3119    &   8778   &  167545 &   620  &  173204  &  1200\\\hline
VCarranza   &   433231  & 145418   &    2496  &     6345   &  139073 & 422  &   142922  &   922\\\hline
Cuajimalpa   &   199809  & 55683   &    640    &   1930   &   53753& 138 &    55043  &   238\\\hline
Cuauhtemoc   &  548606  &173372   &    3003   &    7901   &  165471& 565  &  170369 &    971\\\hline
BJuarez     &  433708  &121962   &    1693   &    5176   &  116786 &  416  &  120269   & 652\\\hline\hline
\textbf{Alvaro O.}   & 755537  & \textbf{281778}   &   \textbf{ 3809}  &    10621  &   271157 & 615    & 277969  &  1276\\\hline\hline
\end{tabular}}
\caption{Official data for Mexico City and it's delegations}\label{Table.Datos.CDMX}
\end{table}
\end{center}

\begin{figure}[!ht]
\centering
\includegraphics[width=8cm, height=8cm]{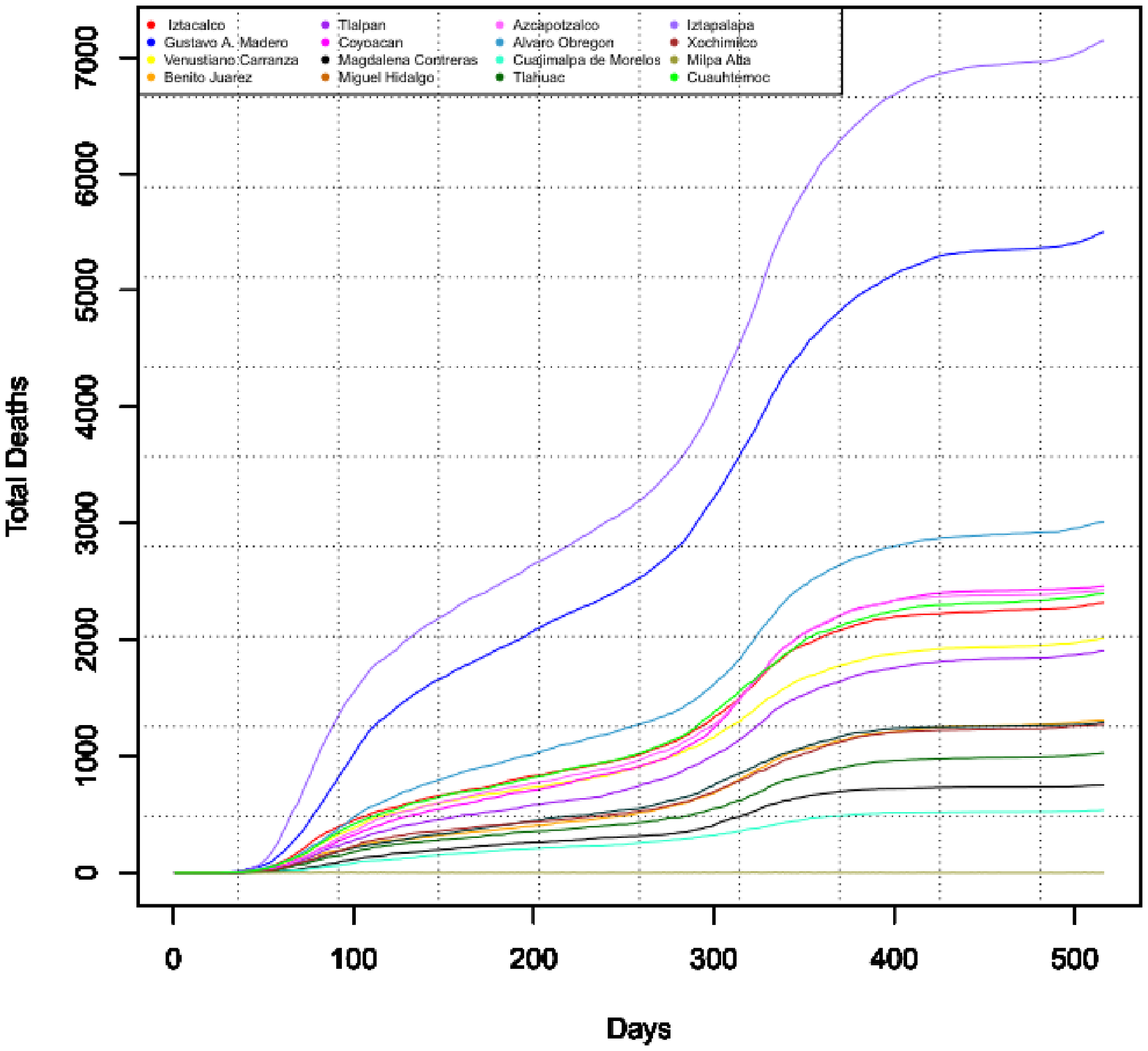}
\caption{Deaths for the Mexico City delegations} \label{fig:Delegaciones.Fallecimientos}
\end{figure}

\begin{figure}[!ht]
\centering
\includegraphics[width=8cm, height=8cm]{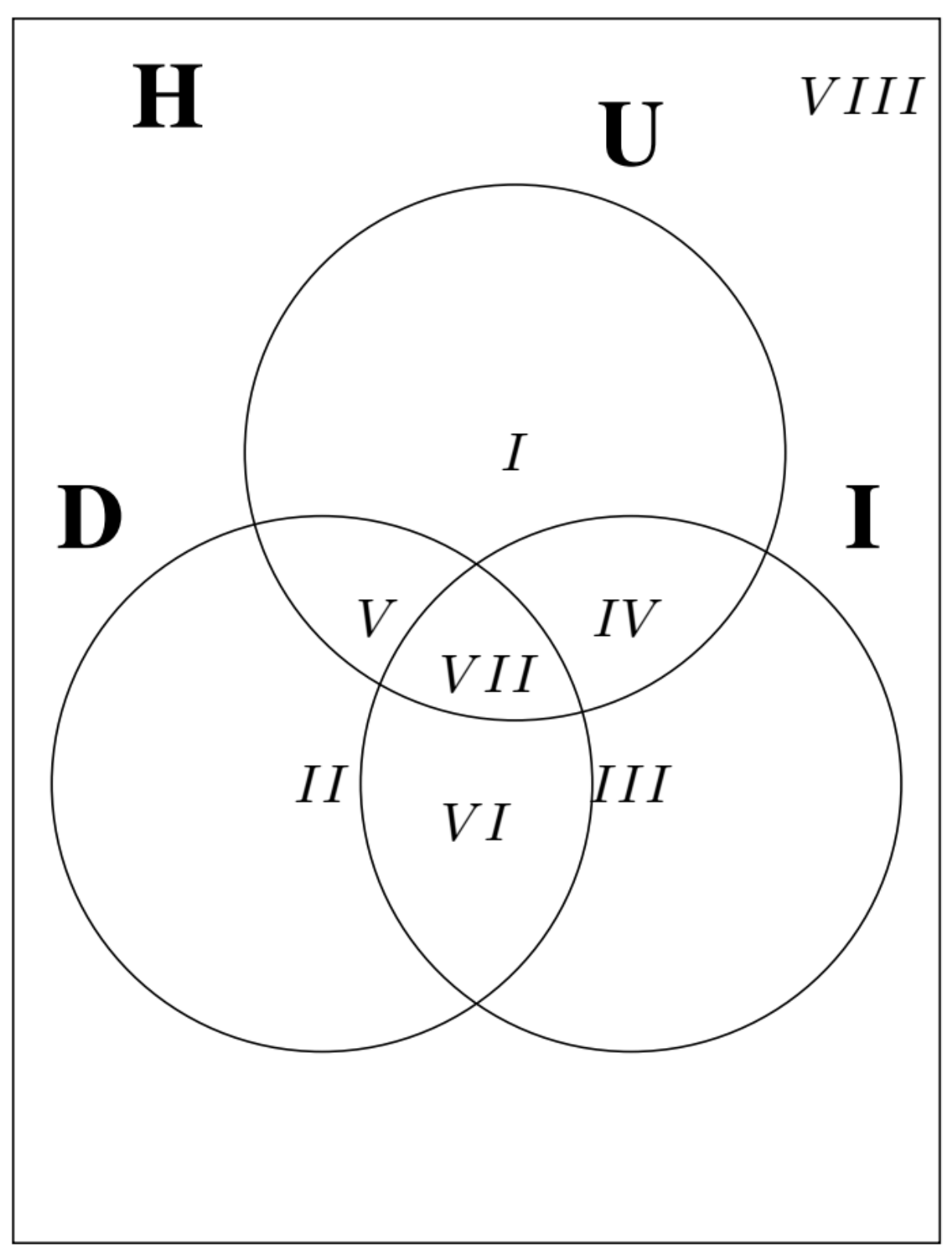}
\caption{Data for Hospitalized patients: U (UCI), I (Intubated), D (Dead)} \label{fig:Regiones}
\end{figure}

Based on the data analyzed, it is possible to determine the exact number of hospitalized patients ($H$),  in one of the following conditions: Intensive Care Unit ($U$), Intubated ($I$) and Death ($D$), this can be illustrated in Figure \ref{fig:Regiones}, the values for each of the regions are given in Table \ref{Table.Regiones. Delegaciones.CDMX}.

\begin{center}
\begin{table}[!ht]
\scalebox{0.85}{
\begin{tabular}{|l||r|r|r|r|r|r|r|r||}\hline\hline
Delegation & I & II & III & IV & V & VI & VII & VIII\\\hline\hline
\textbf{CDMX}      &  1830  & 27900  & 2193  & 1971  & \textbf{348}  & \textbf{9086}  & \textbf{3545}  & \textbf{119240}\\\hline\hline
Azcapotzalco   &  99  & 1924  & 98  & 64 &  18 &  764  & 142  &  7610\\\hline
Coyoacan    &  125 &  2056  & 197  & 183 & 20  & 528  & 292   & 8778\\\hline
Cuajimalpa    &  34  &  399  & 20  & 45  & 6  & 120  & 53  &  1930\\\hline\hline
\textbf{Gustavo A. }       &  235  & 4457  & 237  & 203  & \textbf{52}  & \textbf{1391}  & \textbf{415}  & \textbf{16448}\\\hline\hline
Iztacalco    &  74  & 1945  & 109  &  89  & 16  & 494  & 176  & 6992\\\hline\hline
\textbf{Iztapalapa}    &  285  & 5225  & 445  & 339  & \textbf{54}  & \textbf{2058}  & \textbf{686}  & \textbf{21947}\\\hline\hline
Magdalena C    &  58  &  425  &  64  & 43  &  7  & 148 &  68 &  2416\\\hline
Milpa A      &  28  & 238  & 56  & 21  &  5  & 112  &  36  &  1388\\\hline\hline
\textbf{Alvaro O}     &  164  & 2384  & 148  & 154  & \textbf{29}  & \textbf{706} & \textbf{268}  & \textbf{10621}\\\hline\hline
Tlahuac     &  54  & 756  &  87  &  64 &  7  & 318  & 130  & 3668\\\hline\hline
\textbf{Tlalpan}     &  145  & 1363  & 218  & 188 &  \textbf{24}  & \textbf{533}  & \textbf{310}  & \textbf{8073}\\\hline\hline
Xochimilco     & 81  & 989  & 127  & 112  & 14  & 335  & 198  &  4934\\\hline
Benito J     & 115  & 1108  &  68  & 132   & 9  & 292 &  160  &  5176\\\hline
Cuauhtemoc    & 153  & 1949  & 117  & 135  & 46  & 488  & 231   & 7901\\\hline
Miguel H      & 81  & 1066  & 59  & 78  & 20  & 321 &  198  &  5006\\\hline
Venustiano C    & 99  & 1615  & 143 &  121  & 21  & 477  & 181   & 6345\\\hline\hline
\end{tabular}}
\caption{Hospitalization data for the City of Mexico and its delegations}\label{Table.Regiones. Delegaciones.CDMX}
\end{table}
\end{center}

\section{Description of the model and numerical results}
\begin{figure}[!ht]
\centering
\centering
\includegraphics[width=8cm, height=8cm]{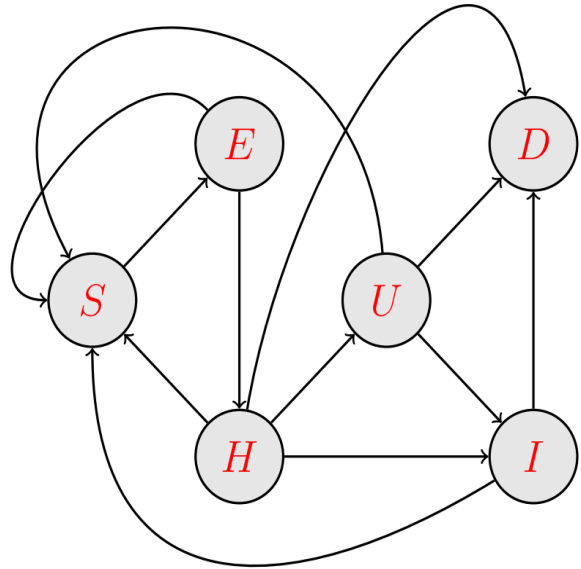}
\caption{Transition between states:$S,E,H,U,F,I$ } \label{fig:RedCadenaMarkov}
\end{figure}

According to the information that can be extracted from the database, the state space for the Markov chain consists of six elements. Lets define the state space $\mathcal{C}$
\begin{eqnarray}
\mathcal{C}&=&\left\{S,E,H,U,I,D\right\}
\end{eqnarray}
where $S$ corresponds to the susceptible population, $E$ for those infected, $H$ for those hospitalized, $U$ for those in the Intensive Care Unit, $I$ for intubated, and $D$ for deceased. The representation is given in Figure \ref{fig:RedCadenaMarkov}.

The values used to determine the probability transition matrix are shown in Table \ref{Table.Regiones.CDMX}, this values are obtained from Table \ref{Table.Regiones.CDMX} and Figure \ref{fig:Regiones}.
\begin{center}
\begin{table}[!ht]
\begin{tabular}{|c||cccccc|}
\hline 
   & S &    E&    H&   U &  I &   F\\ \hline \hline 
S & 9018645  &  - &   -  & -  &  - &  -\\ \hline 
E   &  -& 2867624 &   - &  - &  -  & -\\ \hline 
H   & - &   - & 119240 &  -  &  -  &  -\\ \hline 
U   & -  &   - &   - &7694 & 5516&  3893\\ \hline 
I   & - &   - &   -  & - & 16795& 12631\\ \hline 
F   & - &    - &   -  & -  &  - & 44829\\ \hline 
\end{tabular} 
\caption{Crossed values for Mexico City}\label{Table.Regiones.CDMX}
\end{table}
\end{center}
As shown in Figure \ref{fig:RedCadenaMarkov}, the transition between the states that conform to the state space is assumed to take place per unit of time, this unit of time being one day. When a person enters into the dead state, cannot move to another state, while, a healthy person can continue to be healthy. The people who have recovered from the disease were intubated, or in the intensive care unit, are considered susceptible to getting the disease again, since having overcome the disease does not guarantee immunity and therefore they can get the disease again. The matrix $P$ with the transistion probabilities is given by:

\begin{eqnarray*}
P=\left(
\begin{array}{cccccc}
0.68 &0.32 &0 &0 & 0 &0\\
0.31 & 0.65 &0.04 &0 & 0 &0\\
0.66 & 0 &0.08 &0.01  &0.02 & 0.23\\
0.49 & 0 &0 & 0.20  &0.26 & 0.05\\
0.25 & 0 &0 &0 &0 &0.75\\
0 & 0 &0 & 0  &0 &1
\end{array}
\right).
\end{eqnarray*}
Once the matrix with the transition probabilities has been defined, we are interested in determining the transitions from one state to another after the following days $7,15,30,45,60$, $90,120,180,240$, and $365$, intending to study the evolution of the pandemic over time.\\

\begin{figure}[!ht]
\centering
\includegraphics[scale=0.75]{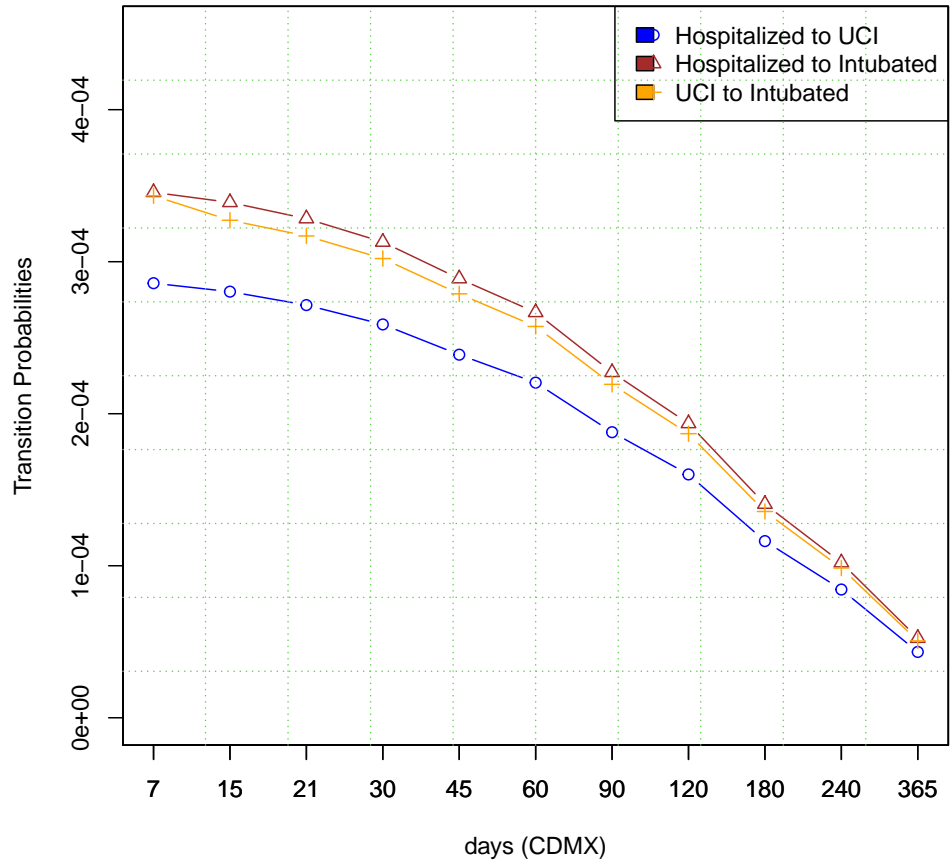}
\caption{Transition probabilities: Hospitalized to UCI, Hospitalized to Intubated, and UCI to Intubated state}
\label{Fig.ToRecoveredState}
\end{figure}

\begin{figure}[!ht]
\centering
\includegraphics[scale=0.75]{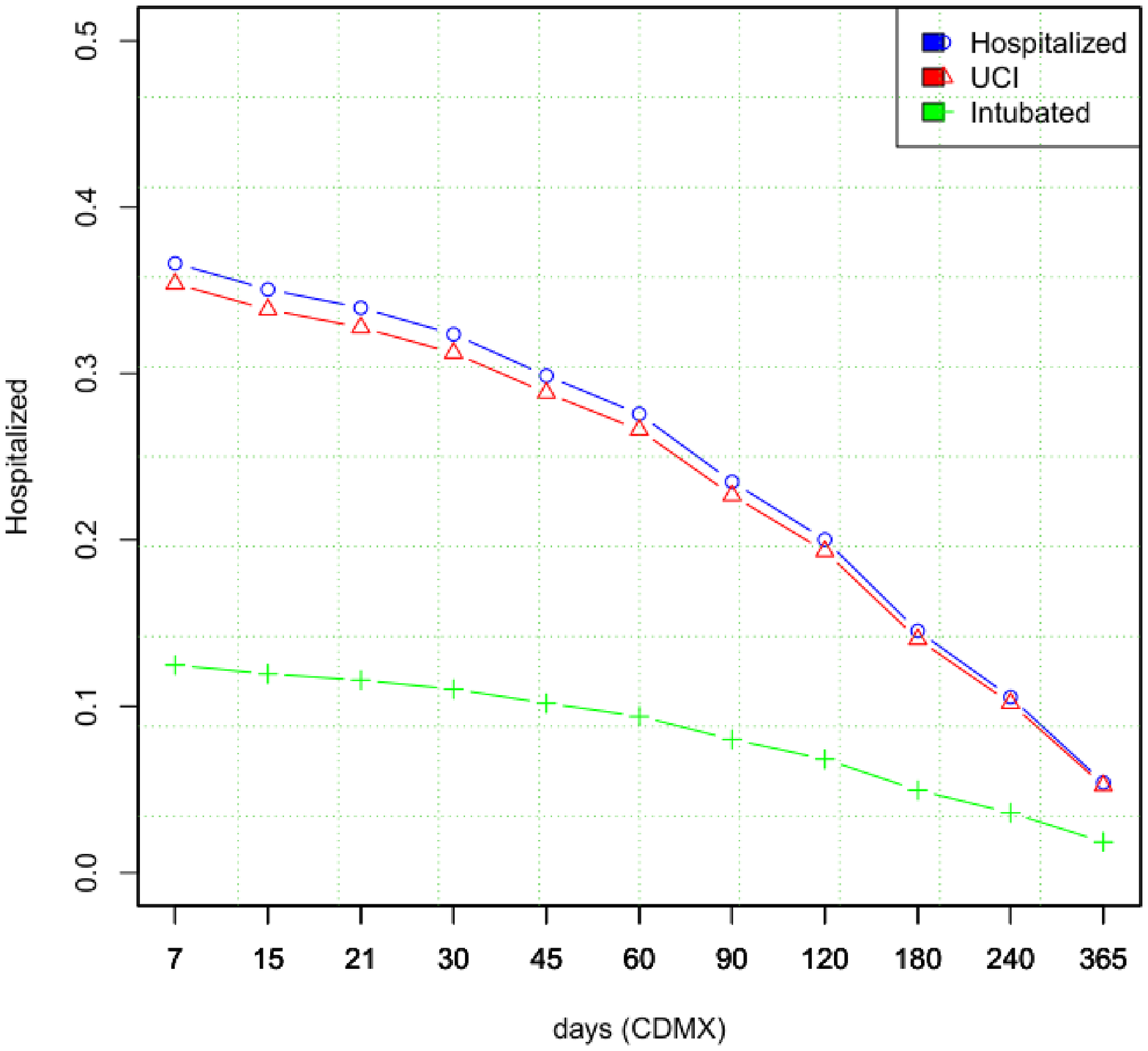}
\caption{Transition probabilities to recovered state}\label{Fig.ToSeveralStates}
\end{figure}

\begin{figure}[!ht]
\centering
\includegraphics[scale=0.75]{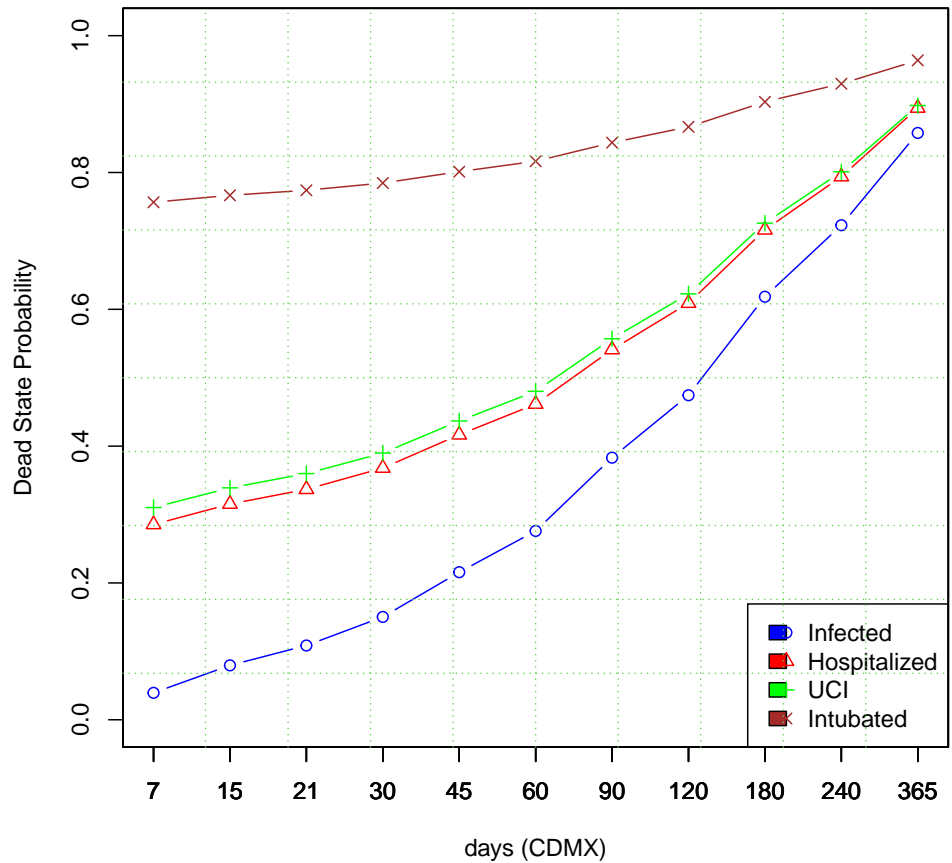}
\caption{Transition probabilities to dead state}\label{Fig.ToDeadState}
\end{figure}

The results achieved for Mexico City are shown in Figures \ref{Fig.ToRecoveredState},  \ref{Fig.ToSeveralStates}, and \ref{Fig.ToDeadState}. Figure \ref{Fig.ToDeadState} shows that, in the case of those who have been intubated, the probability of dying increases from 0.75 to 0.96. The probability of dying for people in the ICU increases from 0.31 to 0.81, this case is similar to that of hospitalized people. For infected persons, the probability of dying increases from 0.04 to 0.86. In Figure \ref{Fig.ToRecoveredState} can be observed that the probability of recovery for those who have been hospitalized goes from 0.37 to 0.0541 as time passes; the same results were found for people in the ICU. A special case is that of intubated patients, for whom the probability of recovery goes from 0.12 to 0.01, which means that as the disease evolves it is less likely to recover. These values are shown in Table \ref{Table.Prob.CDMX}. \\

For Mexico City, it can be seen that after 500 days since the pandemic arrived, large numbers of daily infections began to occur with values higher than four thousand infections per day. The same analysis was done for the sixteen delegations of Mexico City, the tables and graphs are presented in \textbf{Appendix A}.\\

\begin{center}
\begin{table}[!ht]
\scalebox{0.85}{
\begin{tabular}{|c||cccc|ccc|ccc|}
\hline 
Days & $E\curvearrowright F$ & $H\curvearrowright F$ & $U\curvearrowright F$ & $I\curvearrowright F$ & $H\curvearrowright U$ & $U\curvearrowright I$ & $H\curvearrowright I$ & $H\curvearrowright S$ & $U\curvearrowright S$ & $I\curvearrowright S$\\\hline\hline
\textbf{7}  & 0.0394 & 0.2855 & 0.3102 & 0.7565 & 2.858e-04 & 3.433e-04 & 3.456e-04 & 0.3662 & 0.3539 & 0.1248\\\hline
\textbf{15} & 0.0795 & 0.3153 & 0.3390 & 0.7667 & 2.803e-04 & 3.273e-04 & 3.390e-04 & 0.3505 & 0.3384 & 0.1195\\\hline
\textbf{21} & 0.1085 & 0.3369 & 0.3598 & 0.7740 & 2.715e-04 & 3.169e-04 & 3.283e-04 & 0.3395 & 0.3277 & 0.1157\\\hline
\textbf{30} & 0.1503 & 0.3679 & 0.3899 & 0.7846 & 2.587e-04 & 3.021e-04 & 3.129e-04 & 0.3236 & 0.3124 & 0.1103\\\hline
\textbf{45} & 0.2157 & 0.4166 & 0.4368 & 0.8012 & 2.388e-04 & 2.788e-04 & 2.889e-04 & 0.2987 & 0.2883 & 0.1018\\\hline
\textbf{60} & 0.2760 & 0.4615 & 0.4801 & 0.8165 & 2.205e-04 & 2.574e-04 & 2.666e-04 & 0.2757 & 0.2661 & 0.0939\\\hline
\textbf{90} & 0.3831 & 0.5411 & 0.5570 & 0.8436 & 1.878e-04 & 2.193e-04 & 2.272e-04 & 0.2349 & 0.2268 & 0.0801\\\hline
\textbf{120} & 0.4744 & 0.6090 & 0.6226 & 0.8668 & 1.601e-04 & 1.869e-04 & 1.936e-04 & 0.2002 & 0.1932 & 0.0682\\\hline
\textbf{180} & 0.6184 & 0.7161 & 0.7259 & 0.9033 & 1.162e-04 & 1.357e-04 & 1.405e-04 & 0.1453 & 0.1403 & 0.0495\\\hline
\textbf{240} & 0.7229 & 0.7939 & 0.8010 & 0.9298 & 8.437e-05 & 9.850e-05 & 1.020e-04 & 0.1055 & 0.1018 & 0.0359\\\hline
\textbf{365} & 0.8578 & 0.8942 & 0.8979 & 0.9639 & 4.329e-05 & 5.055e-05 & 5.237e-05 & 0.0541 & 0.0523 & 0.0185\\\hline
\end{tabular} }
\caption{Transition probabilities between states for several times}\label{Table.Prob.CDMX}
\end{table}
\end{center}

\section{Conclusion}
\label{Sec:3}
The pandemic worldwide, despite the efforts of scientists, governments, health authorities, and of course the population that has complied with sanitary and containment measures for almost a year, including vaccination, this enormous health problem does not seem to be over. The graphs show that as a consequence of vaccination, despite the increase in infections, deaths have no increased in the same way as in the first wave of the pandemic.  

Based on the data,  it can be seen that for Mexico City there are a total of $11920$ infected persons with COVID19 who are hospitalized, of these, $3545$ who were in intensive care and intubated lost their lives. With respect to the persons who were intubated (regions $III$, $IV$, $VI$ and $VII$, in figure \ref{fig:Regiones}),  those who were intubated and in the intensive care unit were $5516$, and, of the intubated persons, $12631$, unfortunately, passed away. From Figures \ref{Fig.ToRecoveredState},  \ref{Fig.ToSeveralStates} and  \ref{Fig.ToDeadState}, can be concluded the following

\begin{Obs}

The probability of a hospitalized person being intubated in the intensive care unit is low, and the same is true for moving from the intensive care unit to intubation. 

People in the intensive care unit, although they have a probability close to 0.3, as the days go by, it can grow to values close to 0.48 in the first 60 days.

The same occurs for people who are hospitalized; the probability of death in the first 7 days is 0.29, which increases to 0.46 in the first 60 days.

For intubated persons, the probability of death is 0.76 from the beginning and grows relatively low to 0.81 within 60 days. 

For an infected person, if he/she does not manage to recover in the average time, 14 days, the probability of losing his/her life could become very high, when the probability of losing his/her life in the first days is very low.

In hospitalized persons and those in the intensive care unit, the probability of recovery goes from values close to 0.36 to 0.28 in the first 60 days, while intubated persons have a probability of recovery of 0.12 and decreases to 0.09 in the first 60 days. This means that the people who arrive at the hospital are either in delicate or very serious conditions. 

At least in Mexico City, people do not consider attending hospitals as an option to receive care and prefer to stay at home even at the risk of losing their lives due to complications derived from the disease.

\end{Obs}

The COVID pandemic is not over, the consequences of the third wave in Mexico City have not been as devastating as it was in the first wave, i.e., hospitals have not been overwhelmed, the demand for medical oxygen has not reached the high levels observed months ago.  However, local authorities have decided that the activities of a big city like Mexico City should be resumed, considering that the vaccination campaign is gradually advancing in all age groups and delegations. Notwithstanding, they also continue to emphasize the need to continue with the sanitary measures that were made known to the entire population from the beginning.

The delegations with higher number of infections and deaths are Iztapalapa, Gustavo A Madero, and Alvaro Obregon, this information can be corroborated in Table \ref{Table.Datos.CDMX}, also the corresponding tables for some of the delegations can be found  in \textbf{Appendix A}.

For future work, it is planned to include in the analysis comorbidity conditions such as obesity, asthma, diabetes, pulmonary diseases, pneumonia, hypertension, among others, and determine the effects on COVID19 patients and their evolution in the states considered (hospitalized, intensive care unit, intubated, deaths). It is also important to know the current situation of infected persons and their relationship with age, i.e. young, adult, elderly and old. 
Finally, another issue being studied is the development of the pandemic taking into account the delegations with which it is surrounded. 

\section*{Acknowledgment}
I acknowledge the support of my institution, the Universidad Autónoma de la Ciudad de México,  for the realization of this work.


\bibliographystyle{unsrtnat}
\bibliography{references}  






\section*{Appendix A: Tables and graphs for delegations}\label{Sect.Apendice.A}

The delegations with the highest number of cases and deaths are Iztalapapa (517,663 and 8,810), Alvaro Obregon (281,778 and 3,809), Gustavo A Madero (345,557 and 6,759), followed by Tlalpan (226,018 and 2,457), although the Coyoacan delegation has a record of 3,119 deaths due to covid19.\\

Analogous to the analysis provided  for Mexico City, it can be observed that for the Iztapalapa delegation, there are $21947$ hospitalized; those who were in intensive care and intubated were $686$, intensive care and who died $740$, intubated and who died $2744$. The following can be observed

\begin{Obs}
For hospitalized patients, the likelihood of being admitted to the intensive care unit or being intubated, as well as the likelihood that those already in the intensive care unit will be intubated,  for some delegations are 

\begin{itemize}
\item Tlalpan: the probability of going from hospitalized to uni ranges from 0.00005 to 0.00015, from ICU to Intubated from 0.0001 to 0.00020, hospitalized to incubated from 0.0001 to 0.00025.
\item Gustavo A. Madero: the probability of moving from hospitalized to ICU, from ICU to Intubated, and from inpatient to incubated ranges from 0.00005 to 0.00020, i.e., it is unlikely to move from one state to another.
\item Alvaro O.: The probability of moving from hospitalized to ICU ranges from 0.00005 to 0.00007 with a maximum of 0.00015, the same is true for patients to be hospitalized as well as intubated.
\end{itemize}

\end{Obs}

\begin{Obs}
For the most affected delegations, those patients who are in some of the following states: Hospitalized, Intensive Care Unit, or intubated, the probability of death is
\begin{itemize}
\item Alvaro O.: of intubated patients from 0.78 to 0.9, ICU from 0.3 to 0.75, hospitalized patients from 0.3 to 0.75, and those infected from 0.1 to 0.7.
\item Gustavo A. Madero: of intubated and hospitalized patients, the probability of death rises from 0.3 to 0.85, intubated patients rise from 0.8 to 0.9, while those who are infected, the probability of death rises from 0.1 in the first days and increases to 0.8 as time goes by.
\item Iztapalapa: has a behavior similar to that of the Alvaro O delegation, but in this case, the probability of death exceeds 0.9, those hospitalized and in the ICU goes from 0.3 in the first 7 days to 0.9 as time goes by. Those infected ranges from 0.1 in the first days to a maximum of 0.8.
\item Tlalpan: intubated persons range from a value close to 0.7 and grow to a value higher than 0.8, those in ICU grow from 0.3 to 0.65, hospitalized persons range from 0.2 to 0.6, while infected persons range from 0.1 in the first 7 days and grow to a value close to 0.6.
\end{itemize}

\end{Obs}

Finally, 
\begin{Obs}
For the same delegations as above, for those patients who are in one of the following states: Hospitalized, Intensive Care Unit, or intubated, the probability of recovery is.
\begin{itemize}
\item Tlalpan: of hospitalized from 0.3 drops to 0.1, for those in ICU from 0.6 drops to values close to 0.2, while those who are only hospitalized drops from 0.7 to 0.3.  
\item Iztapalapa: of hospitalized from 0.2 drops to 0.05, for those in ICU from 0.5 drops to values close to 0.1, while those only hospitalized drops from 0.5 to 0.1.  
\item Gustavo A. Madero: for patients from 0.2 drops to 0.05, for ICU from 0.5 drops to values close to 0.1, while those only inpatients drop from 0.5 to 0.1  
\item Alvaro O.: Of hospitalized and ICU falls from 0.6 to 0.2, while those intubated falls from 0.2. to 0.1
\end{itemize}
\end{Obs}

This information can be corroborated in the graphs and tables below.

\begin{figure}[h!]
\centering
\subfigure[Azcapotzalco]{\includegraphics[width=5cm, height=5cm]{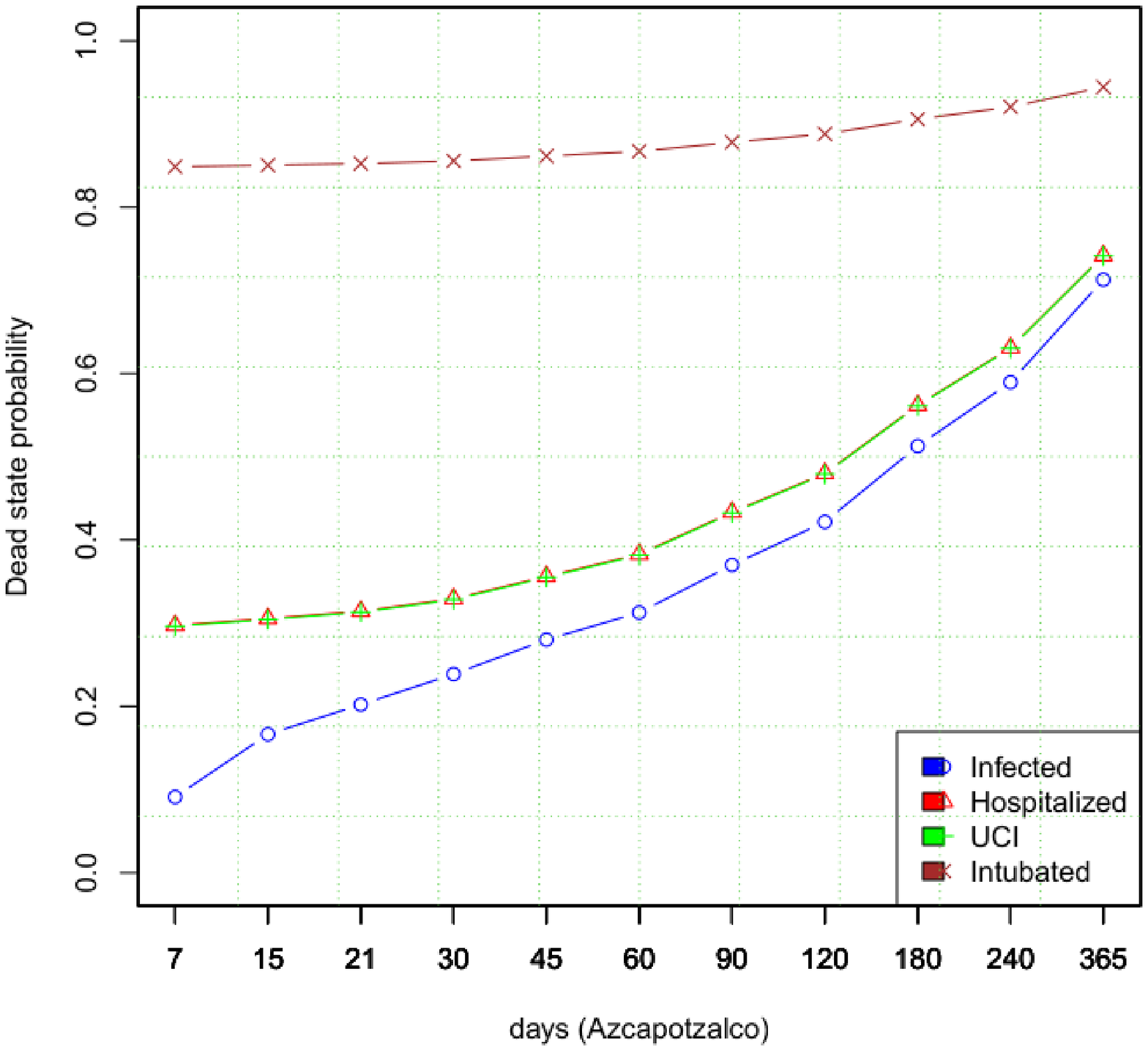}}\hspace{3mm}
\subfigure[Coyoacan]{\includegraphics[width=5cm, height=5cm]{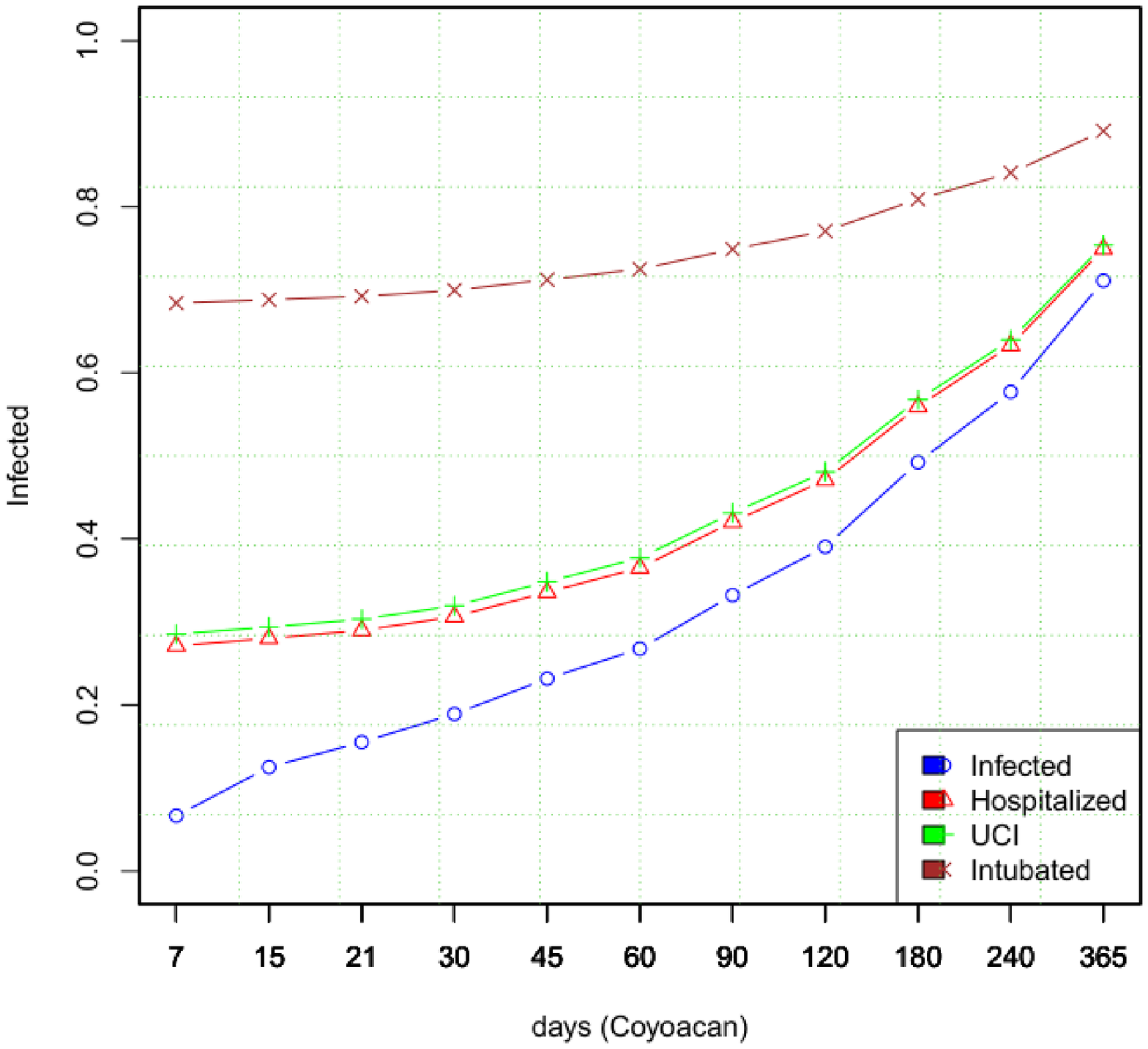}}\hspace{3mm}
\subfigure[Cuajimalpa]{\includegraphics[width=5cm, height=5cm]{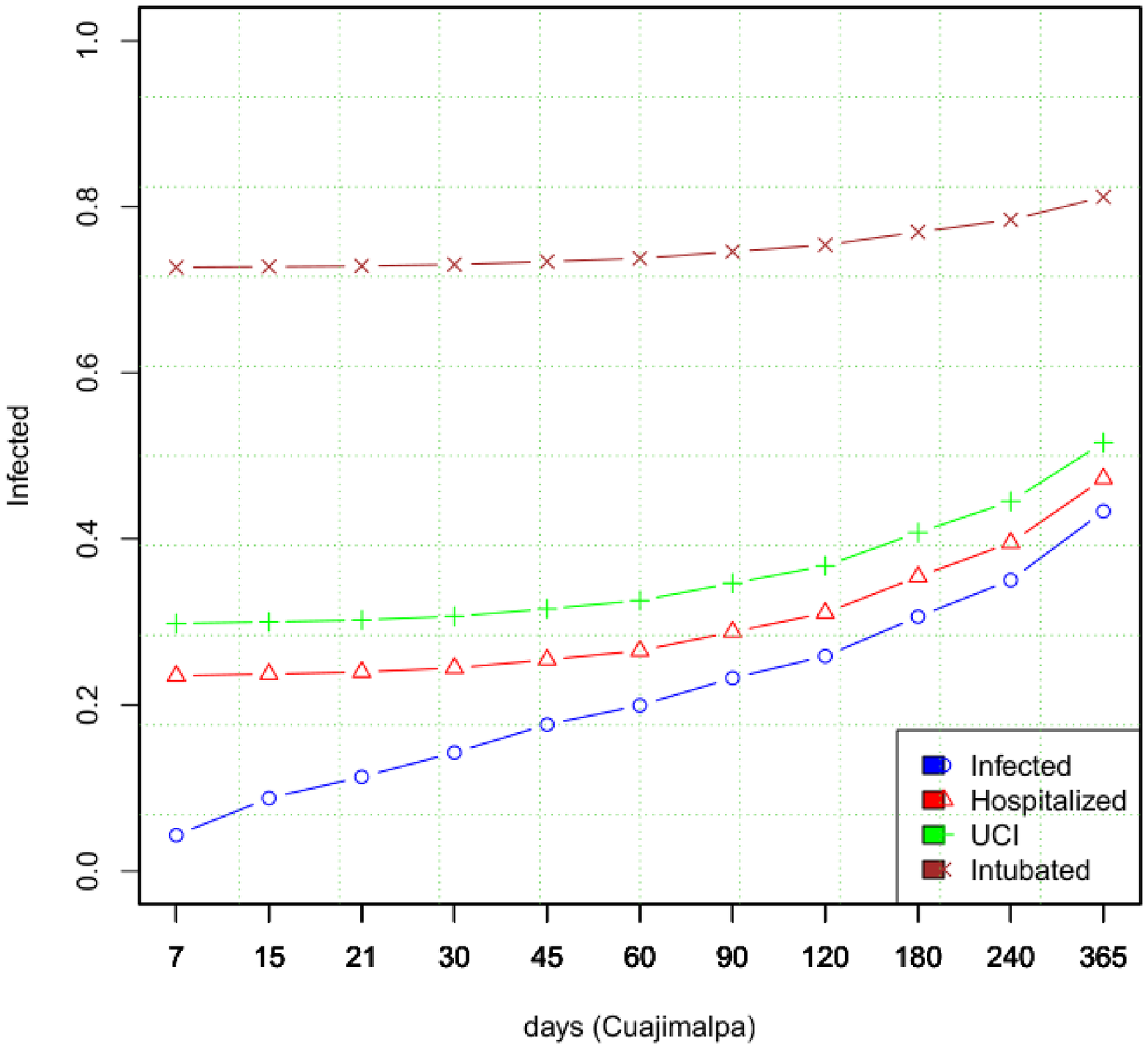}}\hspace{3mm}
\subfigure[Gustavo A.]{\includegraphics[width=5cm, height=5cm]{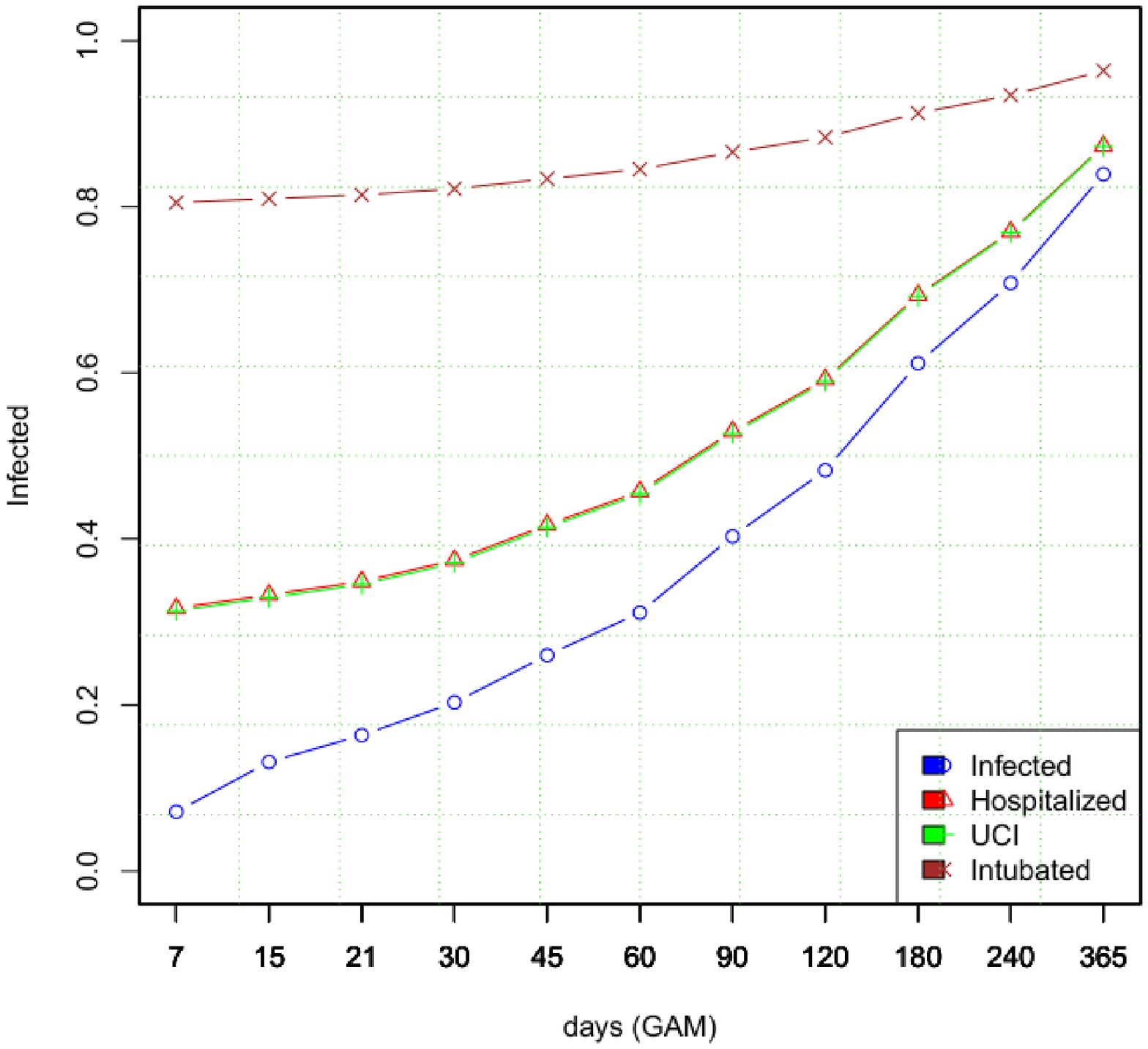}}
\hspace{3mm}
\subfigure[Iztacalco]{\includegraphics[width=5cm, height=5cm]{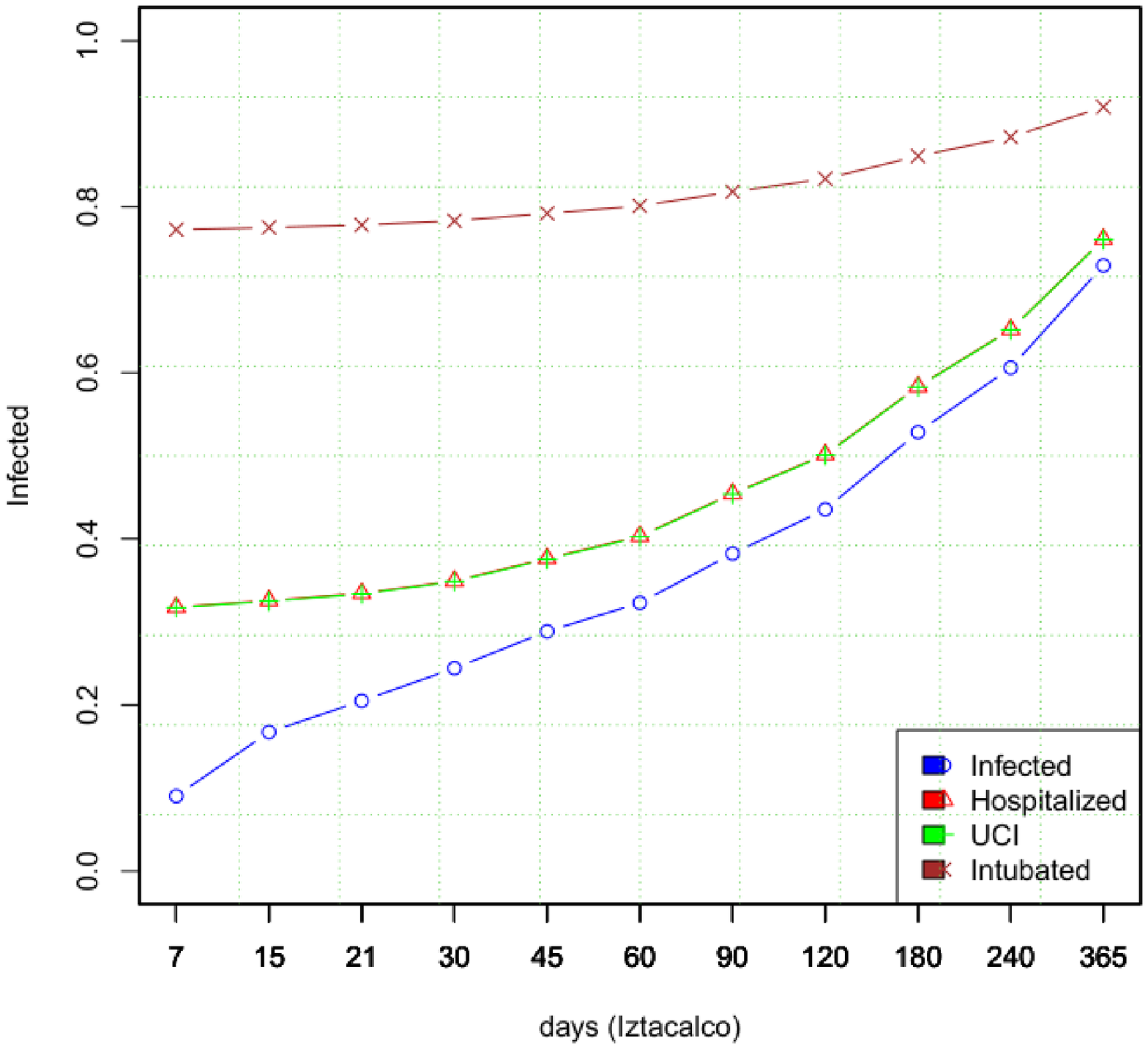}}\hspace{3mm}
\subfigure[Iztapalapa]{\includegraphics[width=5cm, height=5cm]{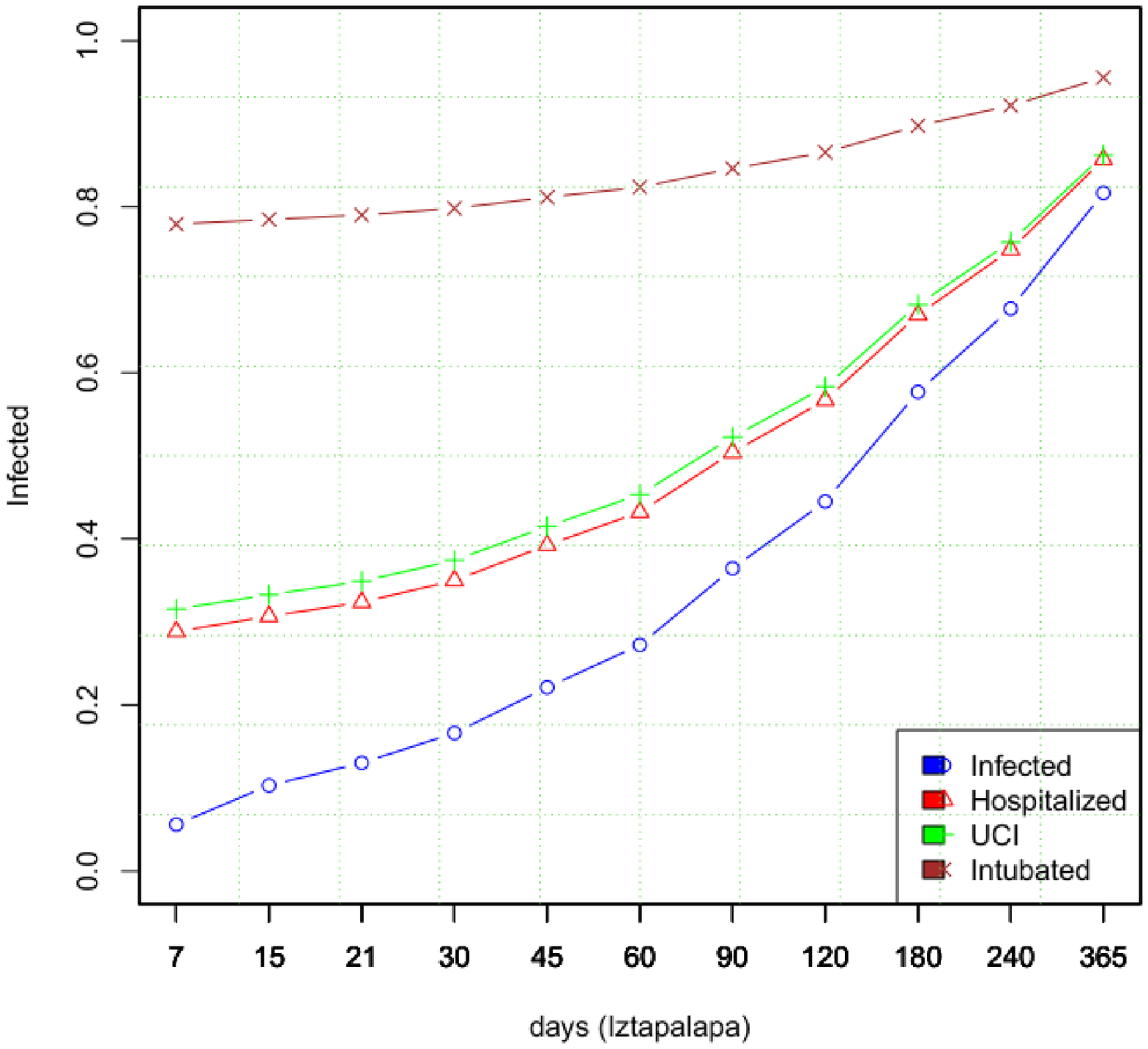}}\hspace{3mm}
\subfigure[Magdalena C]{\includegraphics[width=5cm, height=5cm]{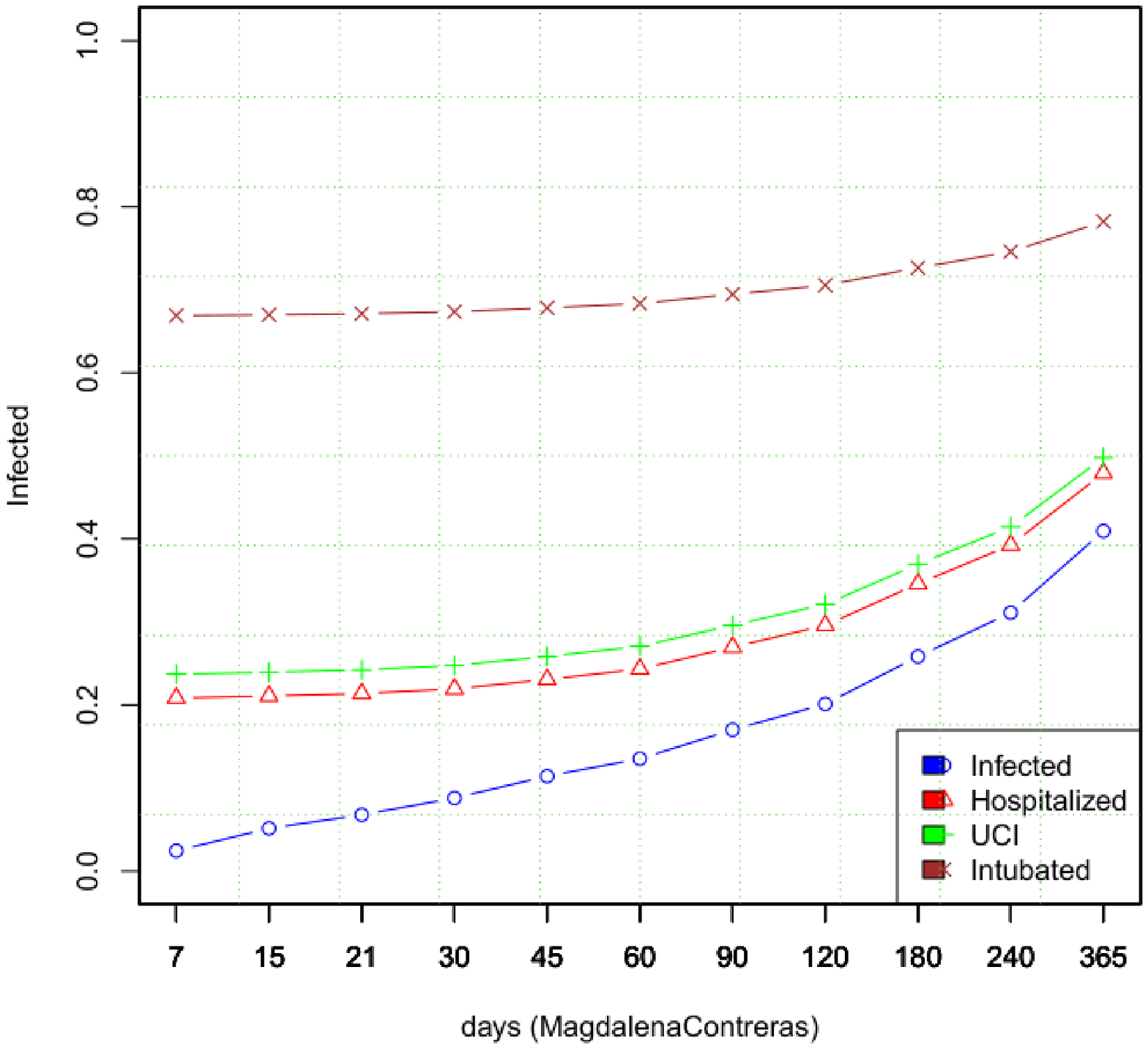}}
\hspace{3mm}
\subfigure[Alvaro O]{\includegraphics[width=5cm, height=5cm]{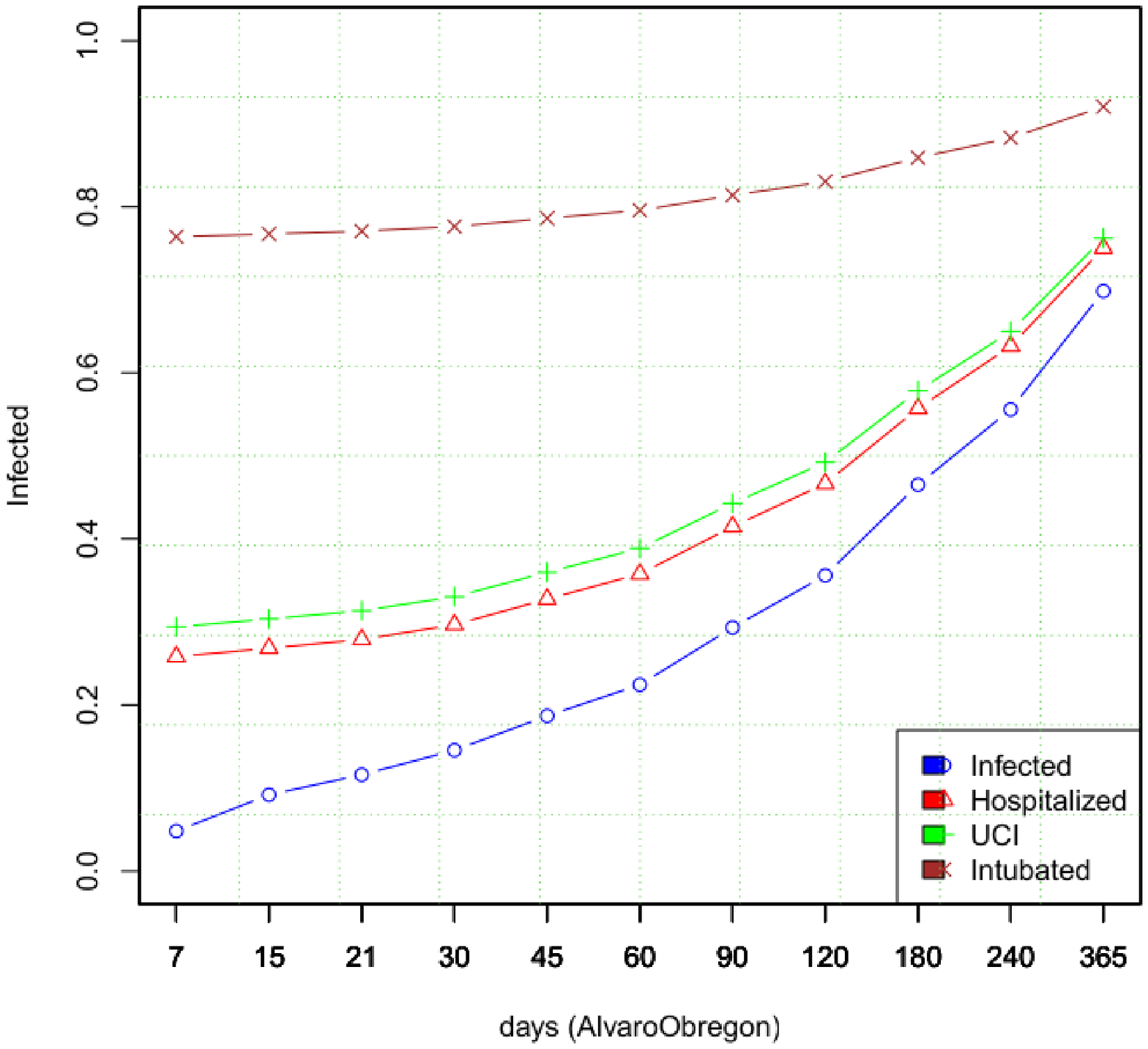}}\hspace{3mm}
\caption{Transition probabilities from Infected,  Hospitalized,  UCI and Intubated to Death state for delegations: Azcapotzalco,  Coyoacan, Cuajimalpa, Gustavo Madero,  Iztacalco, Iztapalapa,  Magdalena C.  and Alvaro O.} \label{fig:PrimerasOcho}
\end{figure}
\begin{figure}[h!]
\centering
\subfigure[Benito J]{\includegraphics[width=5cm, height=5cm]{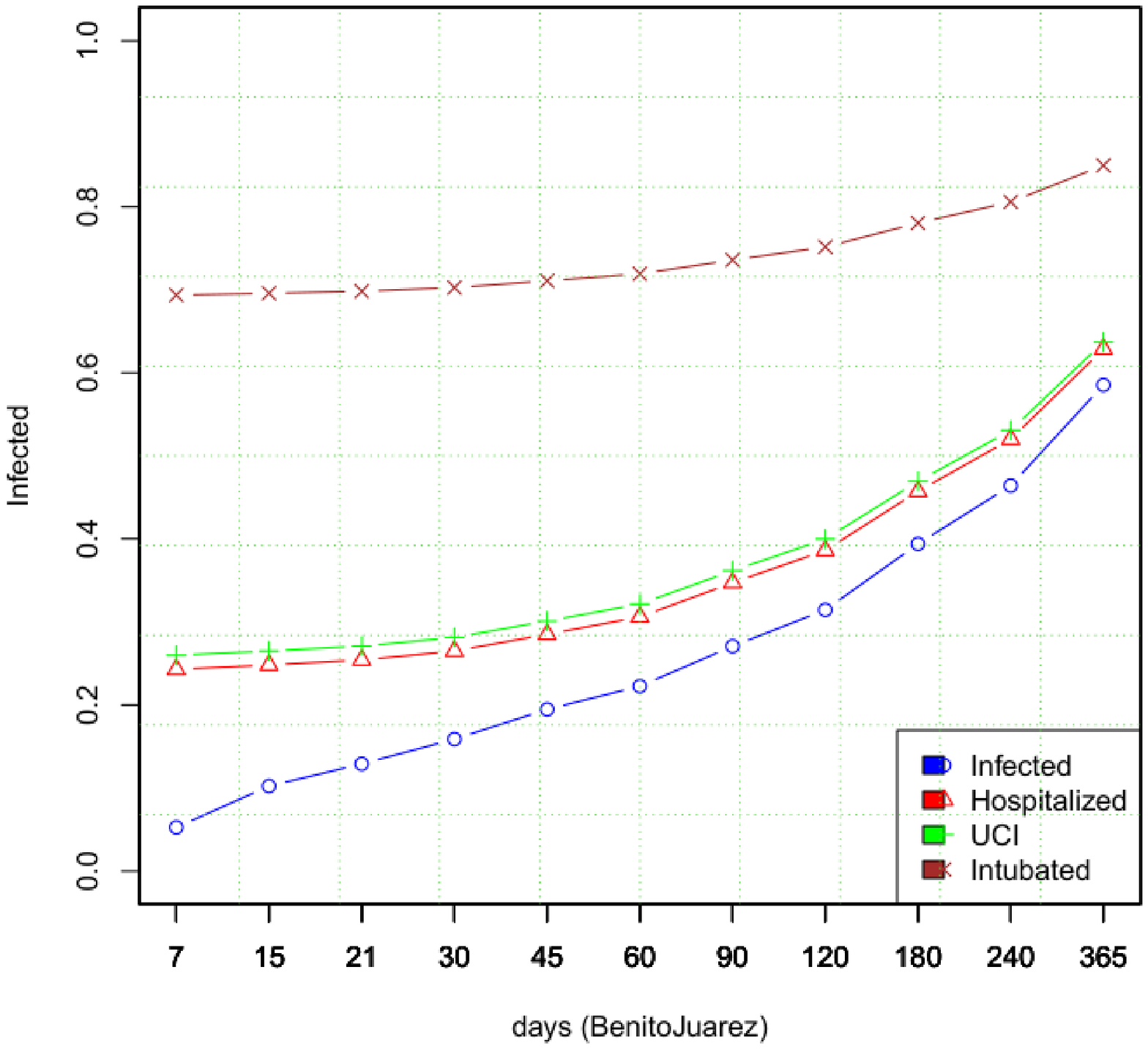}}\hspace{3mm}
\subfigure[Miguel H]{\includegraphics[width=5cm, height=5cm]{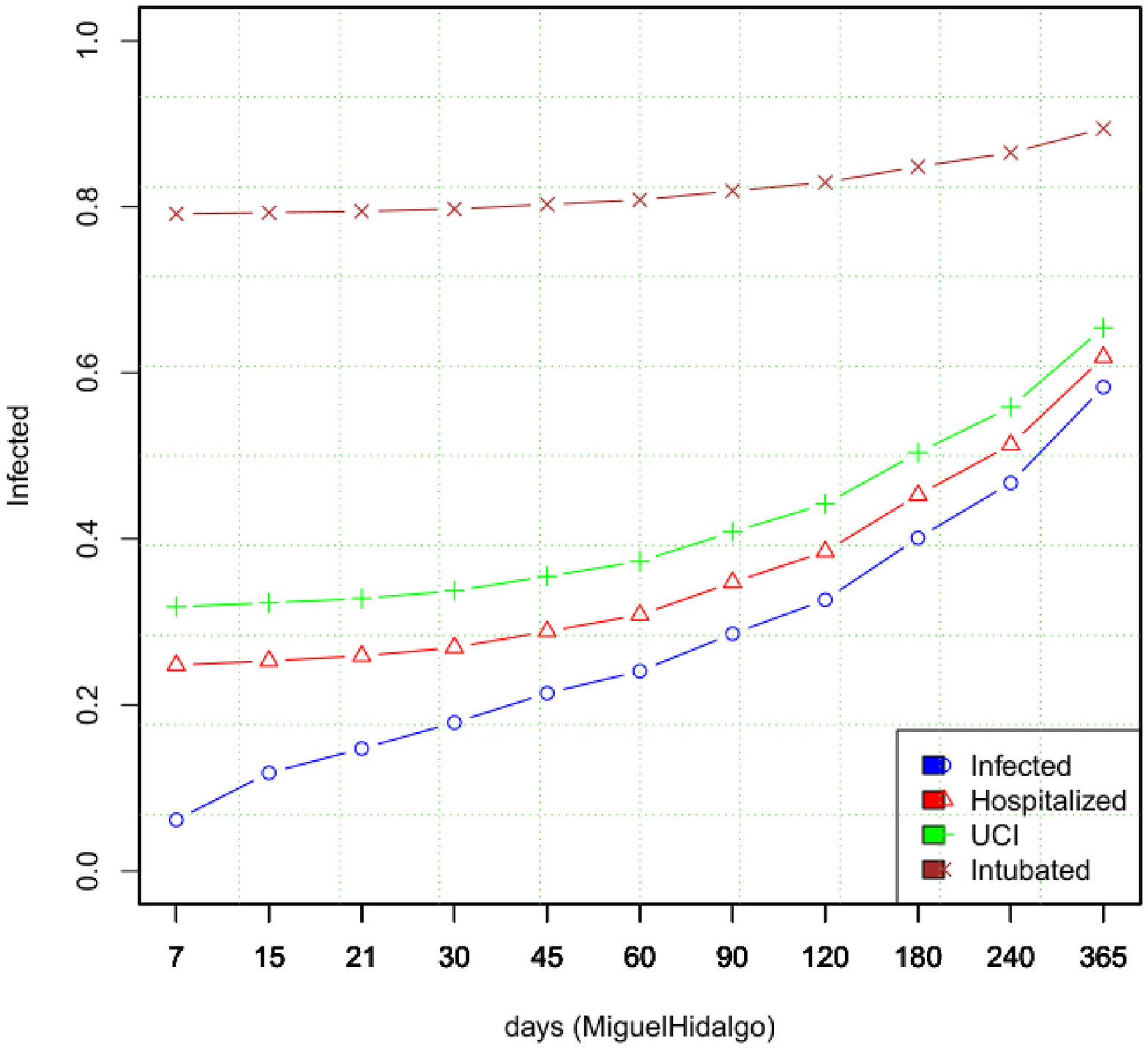}}\hspace{3mm}
\subfigure[Milpa A]{\includegraphics[width=5cm, height=5cm]{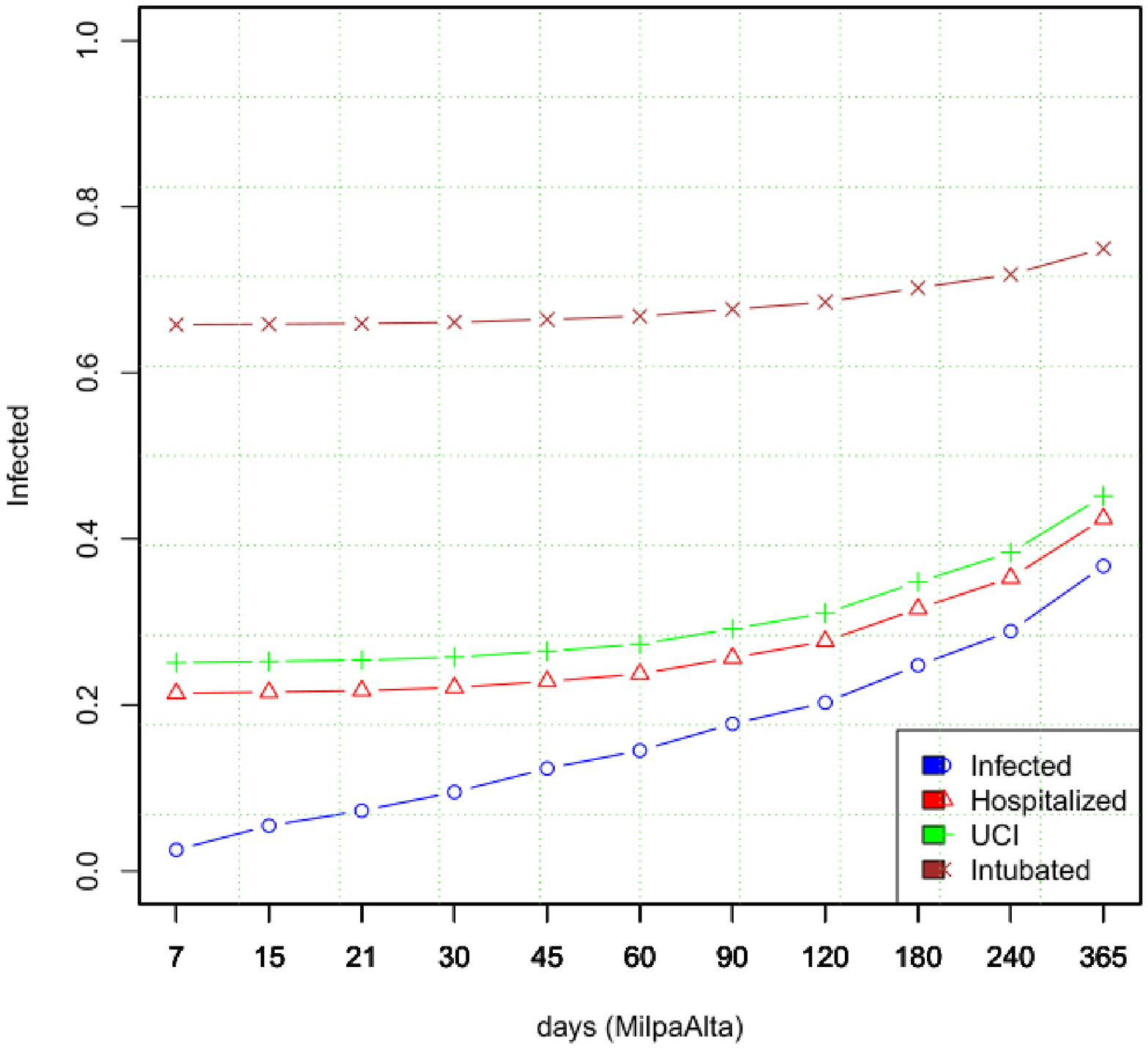}}
\hspace{3mm}
\subfigure[Tlahuac]{\includegraphics[width=5cm, height=5cm]{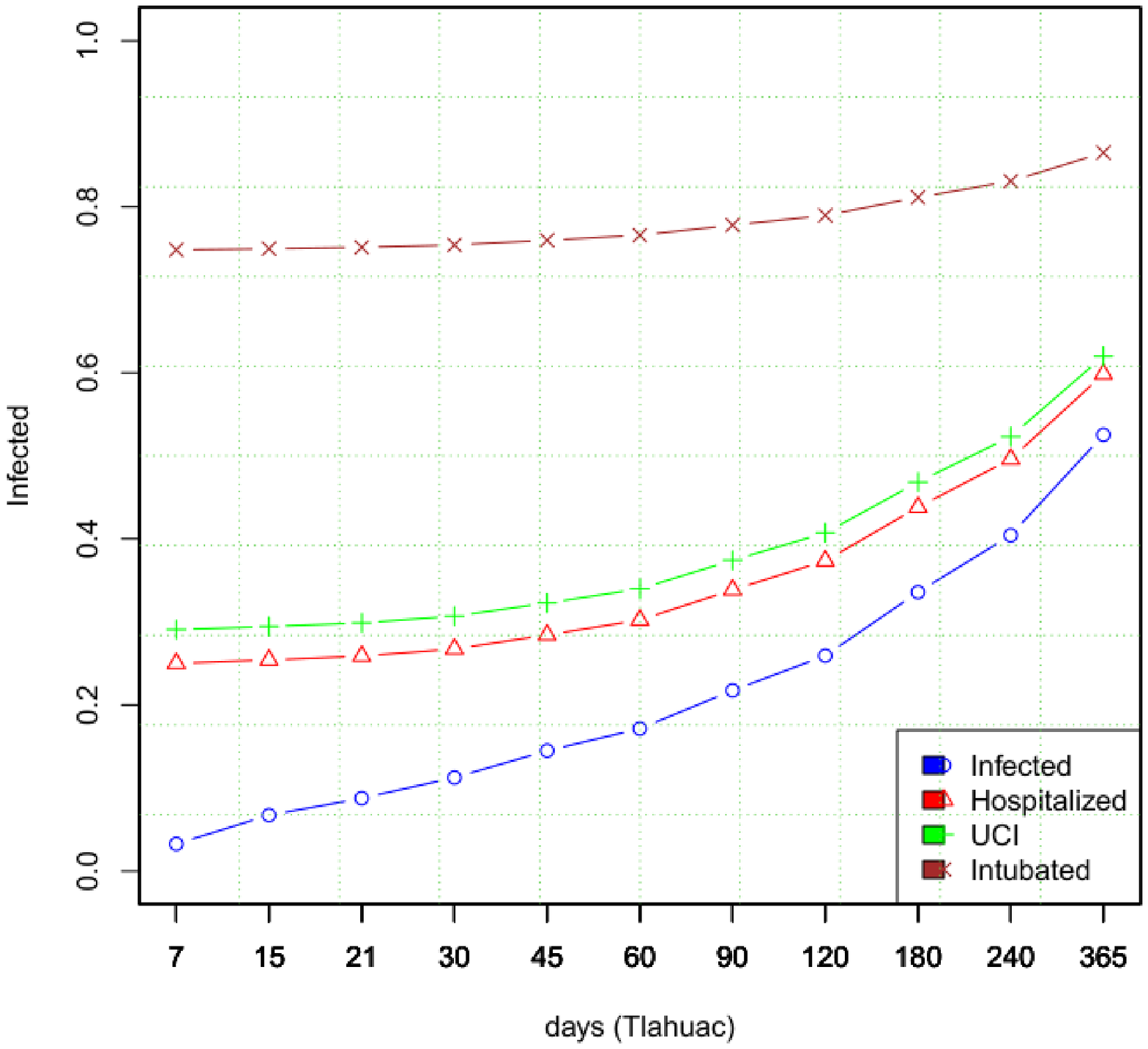}}
\hspace{3mm}
\subfigure[Tlalpan]{\includegraphics[width=5cm, height=5cm]{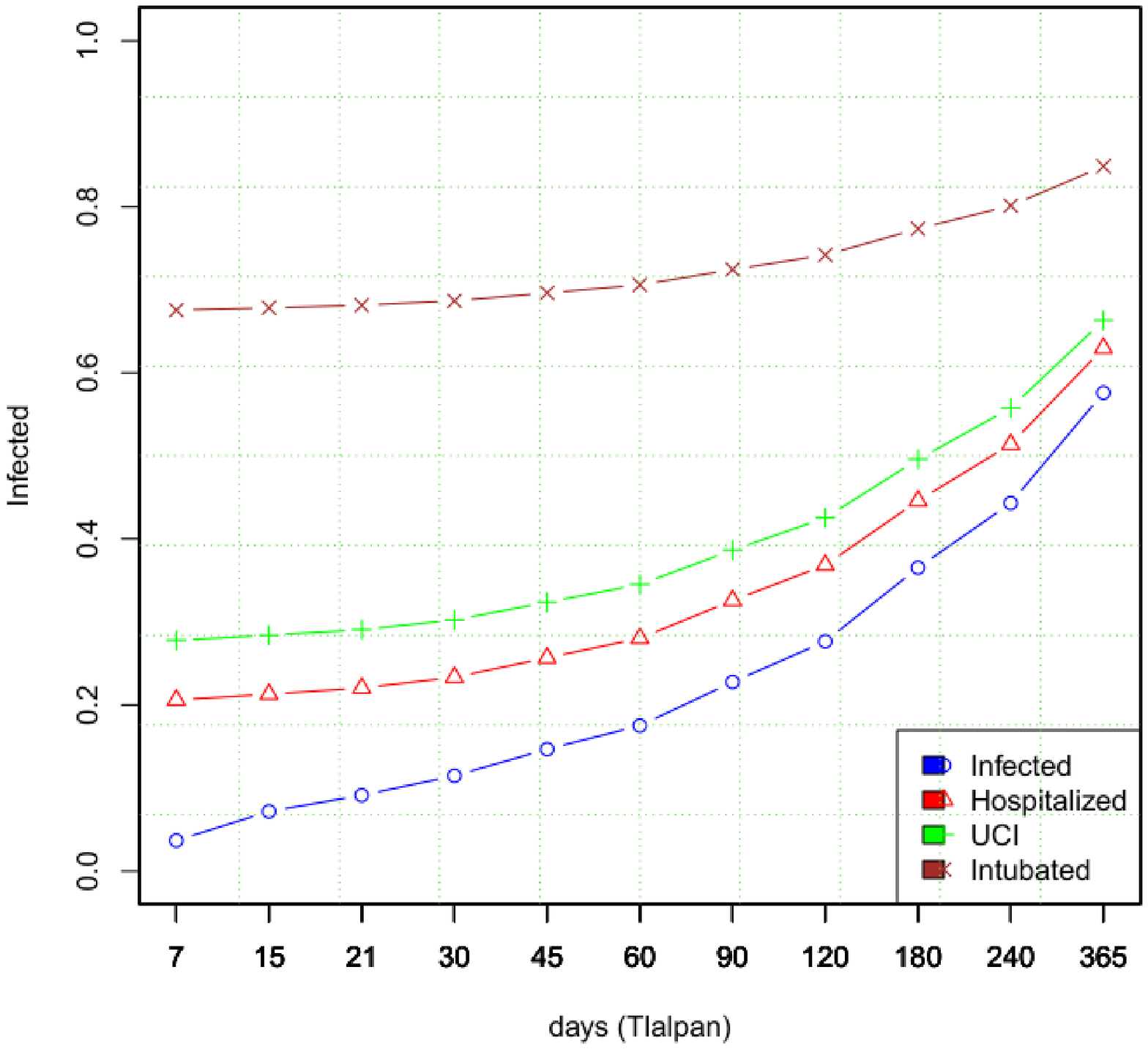}}
\hspace{3mm}
\subfigure[Xochimilco]{\includegraphics[width=5cm, height=5cm]{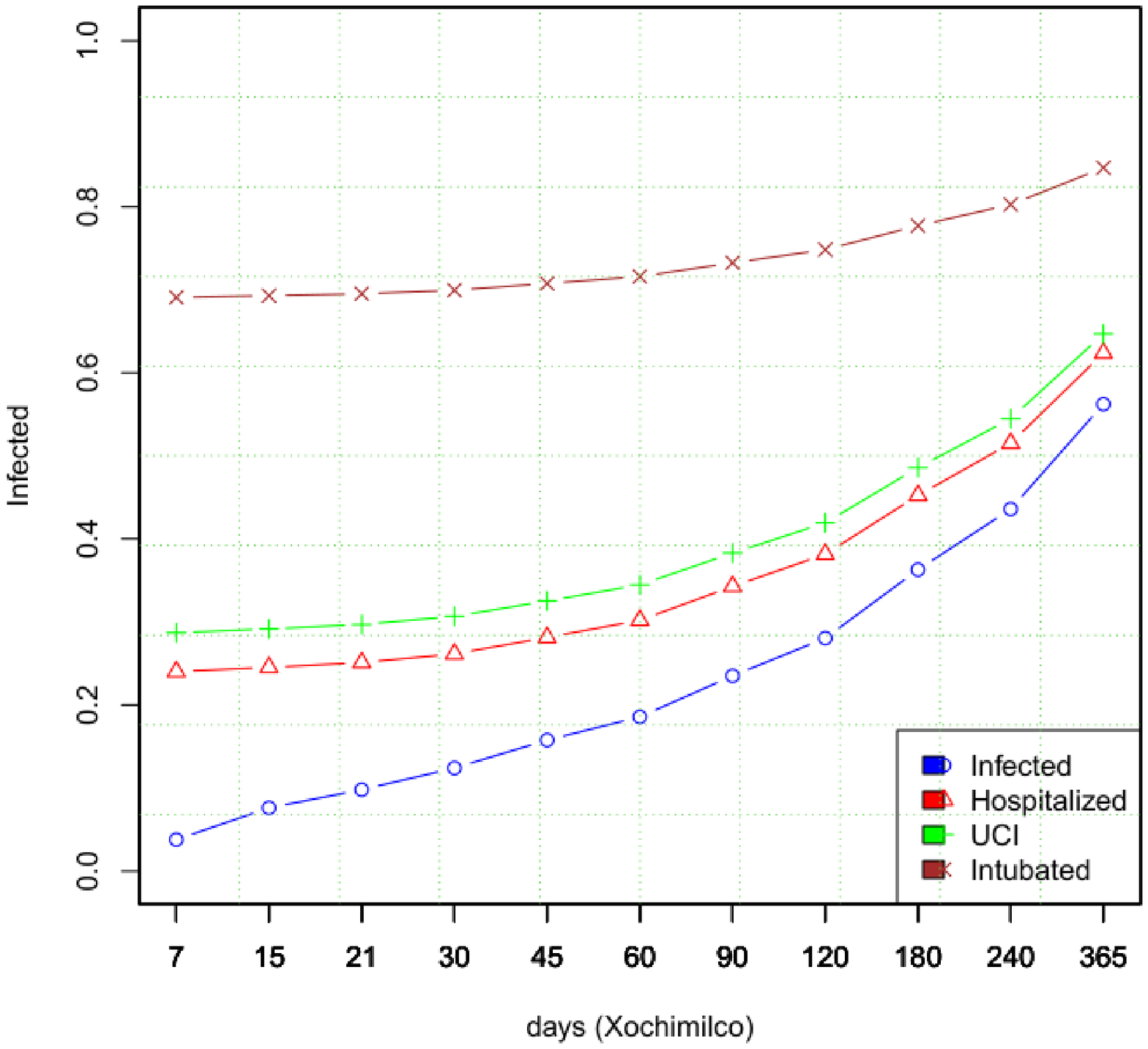}}\hspace{3mm}
\subfigure[Venustiano Carranza]{\includegraphics[width=5cm, height=5cm]{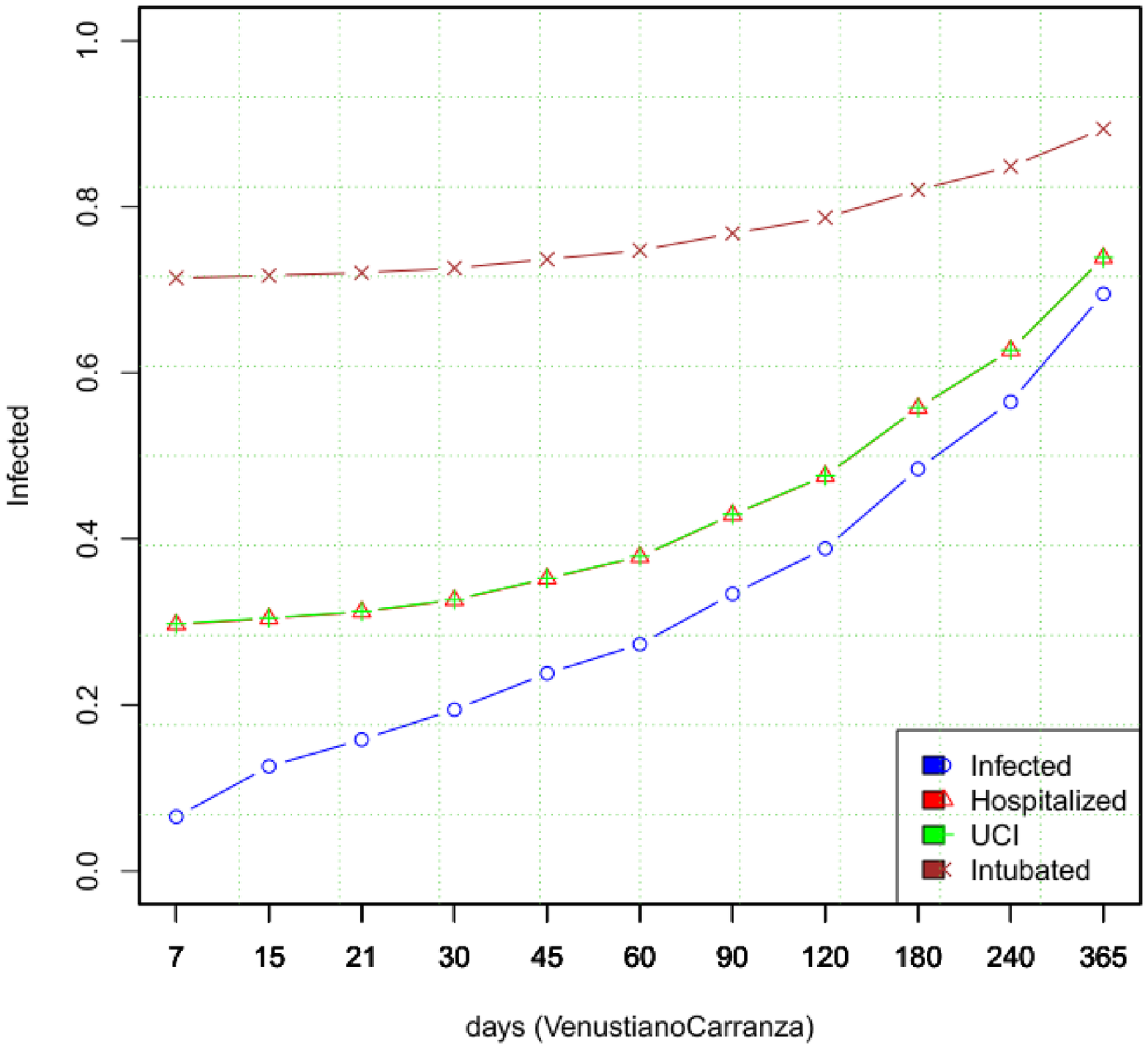}}
\hspace{3mm}
\subfigure[Cuauhtemoc]{\includegraphics[width=5cm, height=5cm]{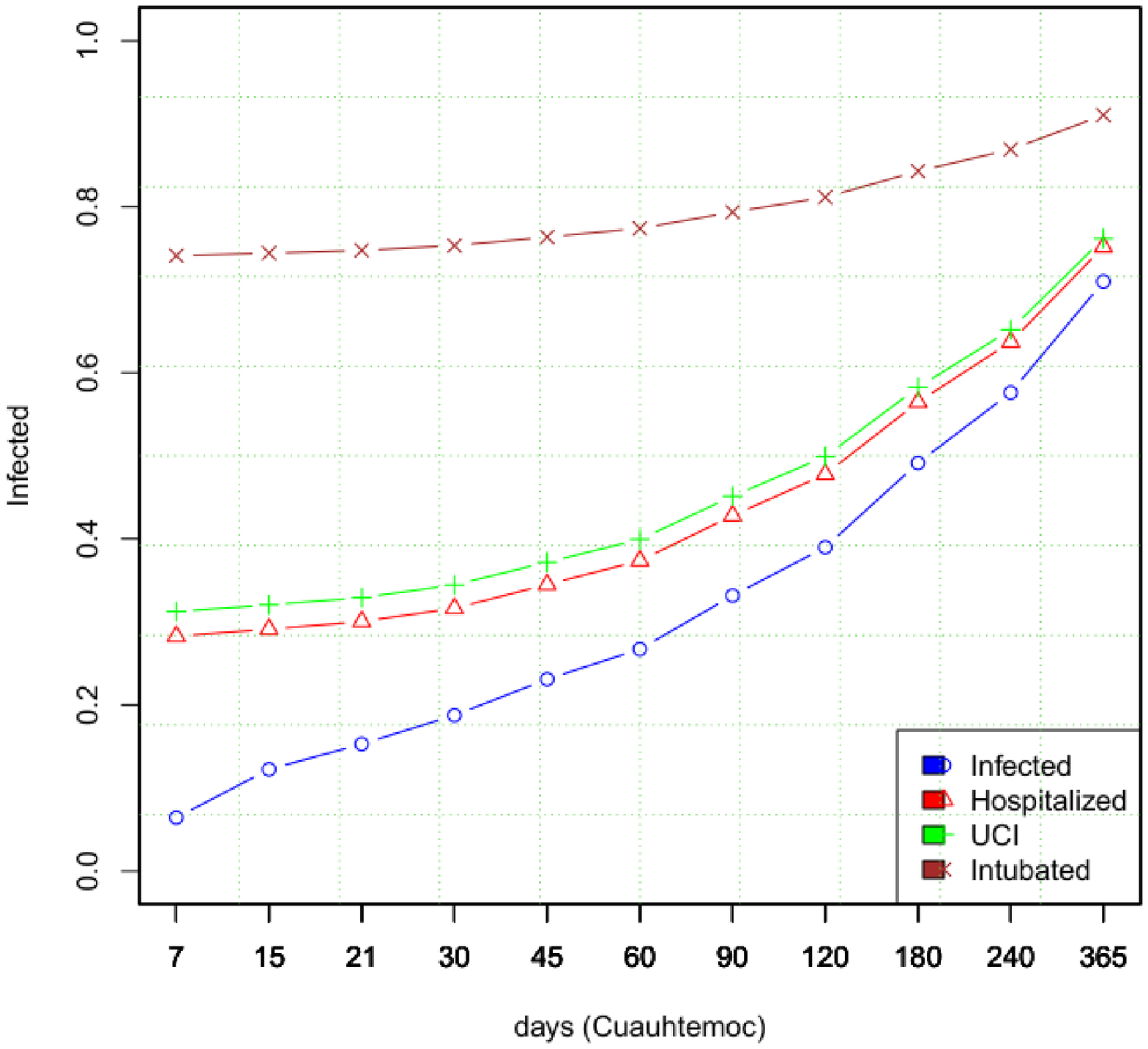}}\hspace{3mm}
\caption{Transition probabilities from Infected,  Hospitalized,  UCI and Intubated to Death state for delegations: Benito J., Miguel H.,Milpa A.,  Tlahuac,  Tlalpan,  Xochimilco, Venustiano C.  and Cuauhtemoc } \label{fig:SegundasOcho}
\end{figure}
\begin{center}
\begin{table}[!ht]
\scalebox{0.75}{
\begin{tabular}{|c||cccc|ccc|ccc|}
\hline 
Days & $E\curvearrowright F$ & $H\curvearrowright F$ & $U\curvearrowright F$ & $I\curvearrowright F$ & $H\curvearrowright U$ & $U\curvearrowright I$ & $H\curvearrowright I$ & $H\curvearrowright S$ & $U\curvearrowright S$ & $I\curvearrowright S$\\\hline\hline
\textbf{7}  &   0.03671  & 0.2060 &  0.2778  & 0.6754 &  5.451e-05 &  7.889e-05  &   8.227e-05 &  0.6955  & 0.6360  & 0.2837\\\hline
\textbf{15} &   0.0718  & 0.2129  & 0.2839  & 0.6783  & 1.250e-04 &  1.692e-04  & 1.888e-04 &  0.6166  & 0.5626 &  0.2517\\\hline
\textbf{21}  &  0.0913  & 0.2204  & 0.2907  & 0.6814  & 1.526e-04  & 2.081e-04  &  2.304e-04 &  0.5816  & 0.5301  & 0.2375\\\hline
\textbf{30} &   0.1147  & 0.2334  & 0.3024 &  0.6867 &  1.724e-04 &  2.363e-04 &  2.604e-04  & 0.5492  & 0.5002  & 0.2244\\\hline
\textbf{45}  &  0.1467 &  0.2568  & 0.3237 &  0.6962  & 1.806e-04 &  2.481e-04  &  2.728e-04  & 0.5188  & 0.4722  & 0.2120\\\hline
\textbf{60}  &  0.1751 &  0.2803 &  0.3451  & 0.7059&  1.785e-04 &  2.453e-04  & 2.696e-04  & 0.4988 & 0.4539 & 0.2038\\\hline
\textbf{90}   & 0.2276 &  0.3258 &  0.3865 &  0.7244 &  1.684e-04 &  2.314e-04  &  2.543e-04  & 0.4661 &  0.4241  & 0.1905\\\hline
\textbf{120}  & 0.2766 &  0.3685 &  0.4254  & 0.7419  & 1.578e-04  & 2.169e-04 &  2.386e-04 &  0.4365  & 0.3972  & 0.1784\\\hline
\textbf{180} &  0.3653  & 0.4459  & 0.4958  & 0.7735 &  1.385e-04 &  1.903e-04  & 2.091e-04 &  0.3829 &  0.3485 & 0.1565\\\hline
\textbf{240}  & 0.4431 &  0.5139 &  0.5576 &  0.8013  & 1.215e-04 &  1.669e-04 &  1.835e-04  & 0.3360  & 0.3057  & 0.1373\\\hline
\textbf{365}  & 0.5760 &  0.6298 &  0.6632  & 0.8487  & 9.251e-05 &  1.271e-04  & 1.397e-04 &  0.2558  & 0.2328  & 0.1046\\\hline
\end{tabular} }
\caption{Transition probabilities between states for several times}\label{Table.Prob.Tlalpan}
\end{table}
\end{center}
\begin{figure}[h!]
\centering
\subfigure[Azcapotzalco]{\includegraphics[width=5cm, height=5cm]{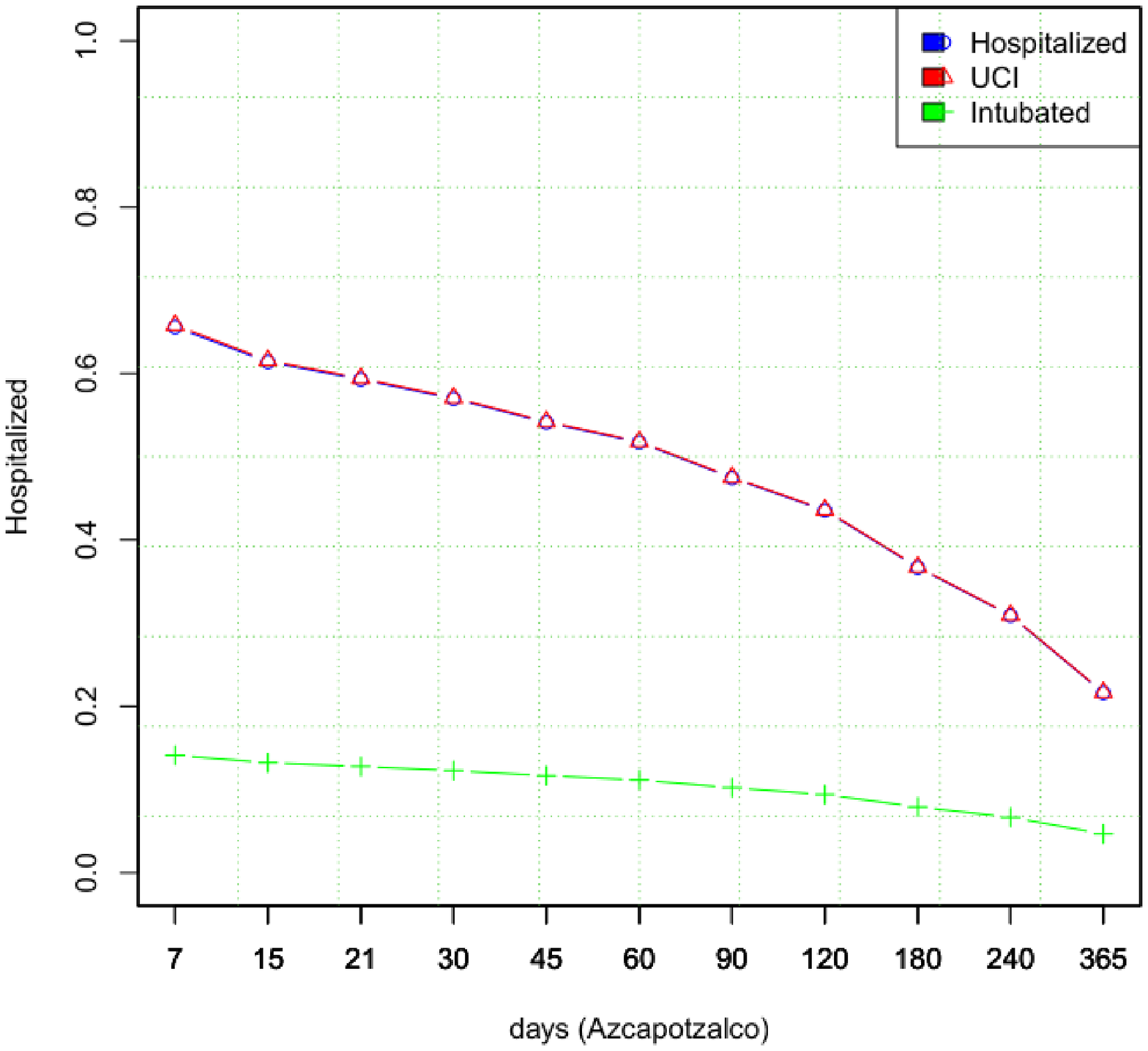}}\hspace{3mm}
\subfigure[Coyoacan]{\includegraphics[width=5cm, height=5cm]{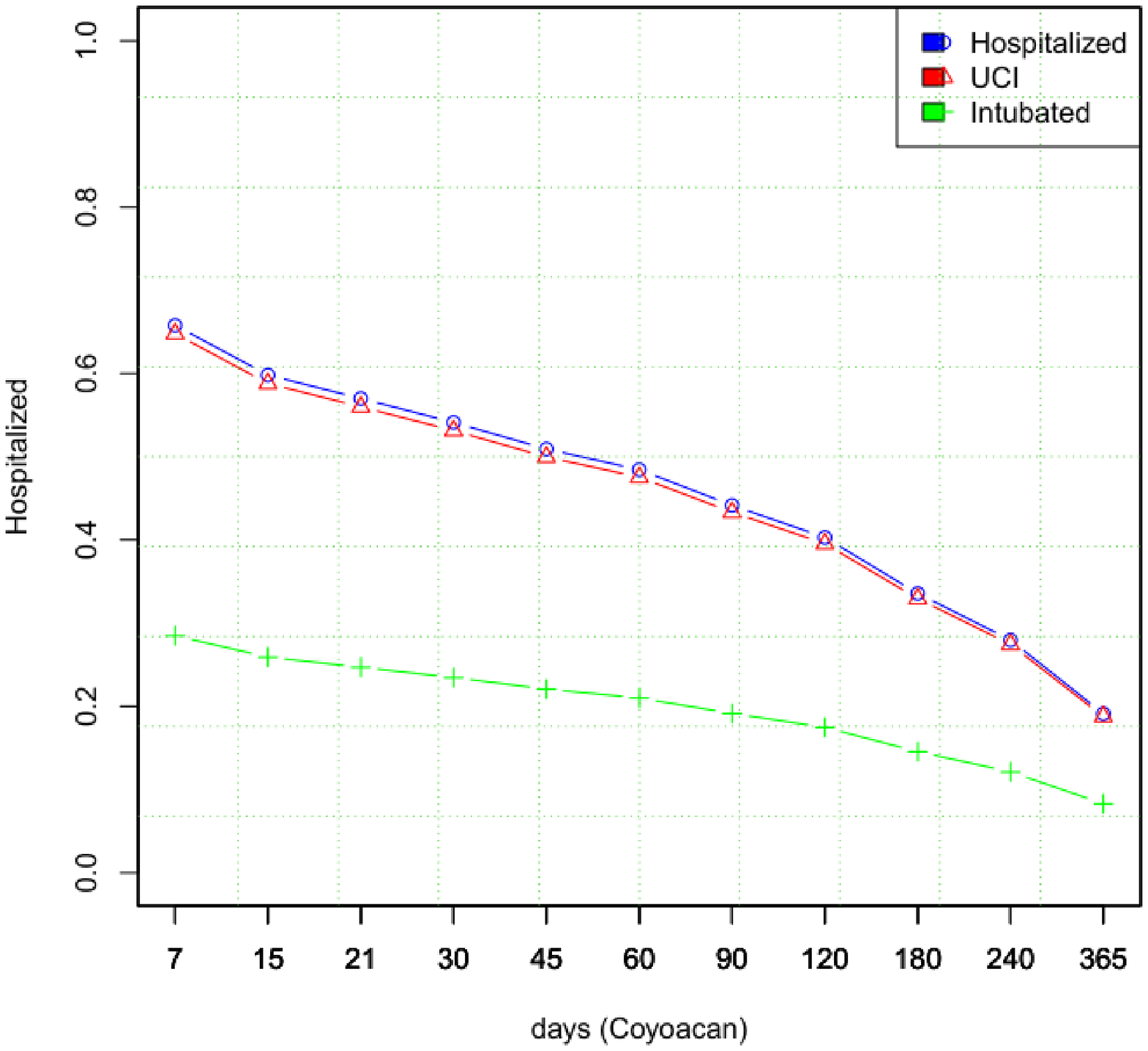}}\hspace{3mm}
\subfigure[Cuajimalpa]{\includegraphics[width=5cm, height=5cm]{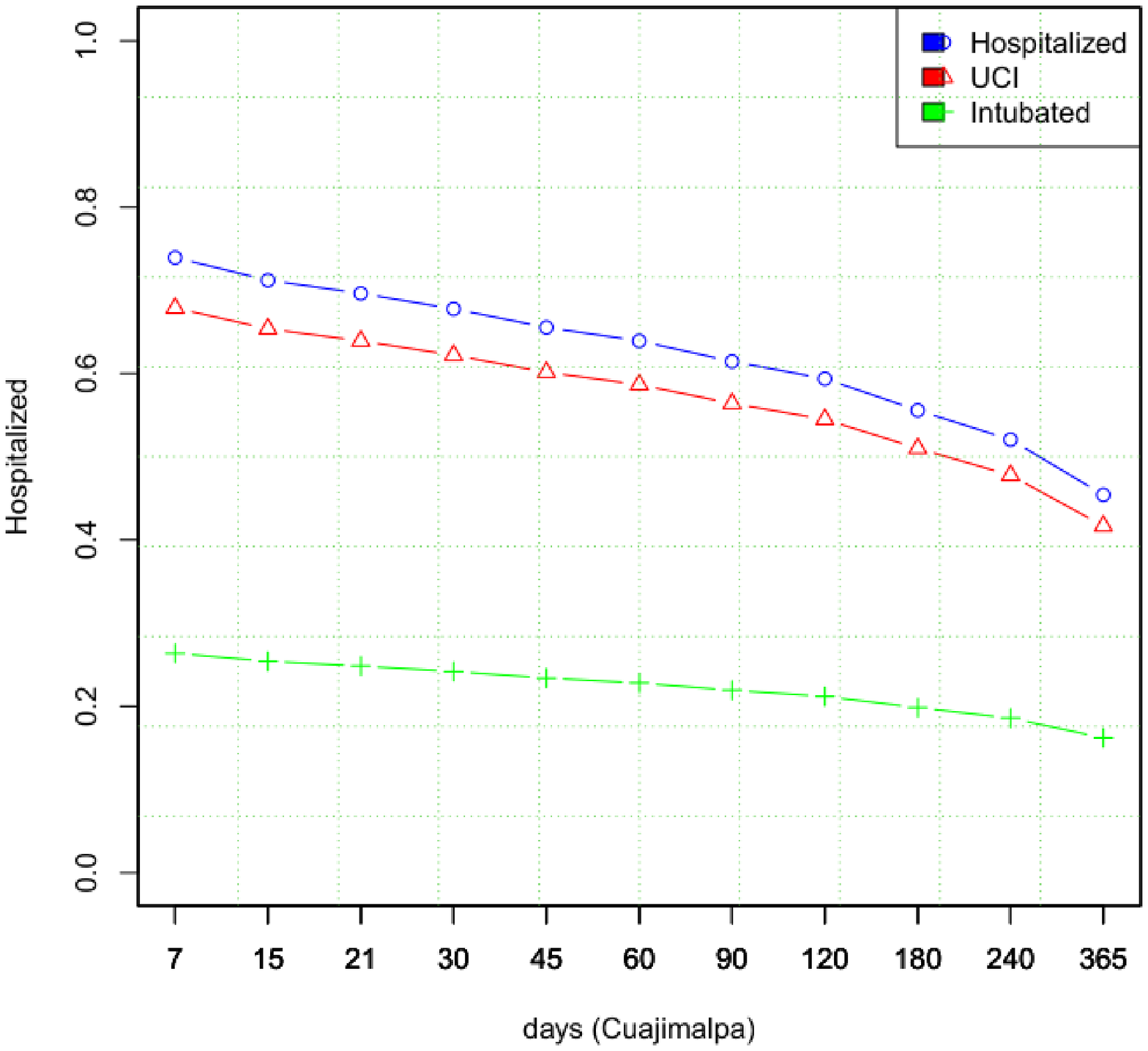}}\hspace{3mm}
\subfigure[Gustavo A.]{\includegraphics[width=5cm, height=5cm]{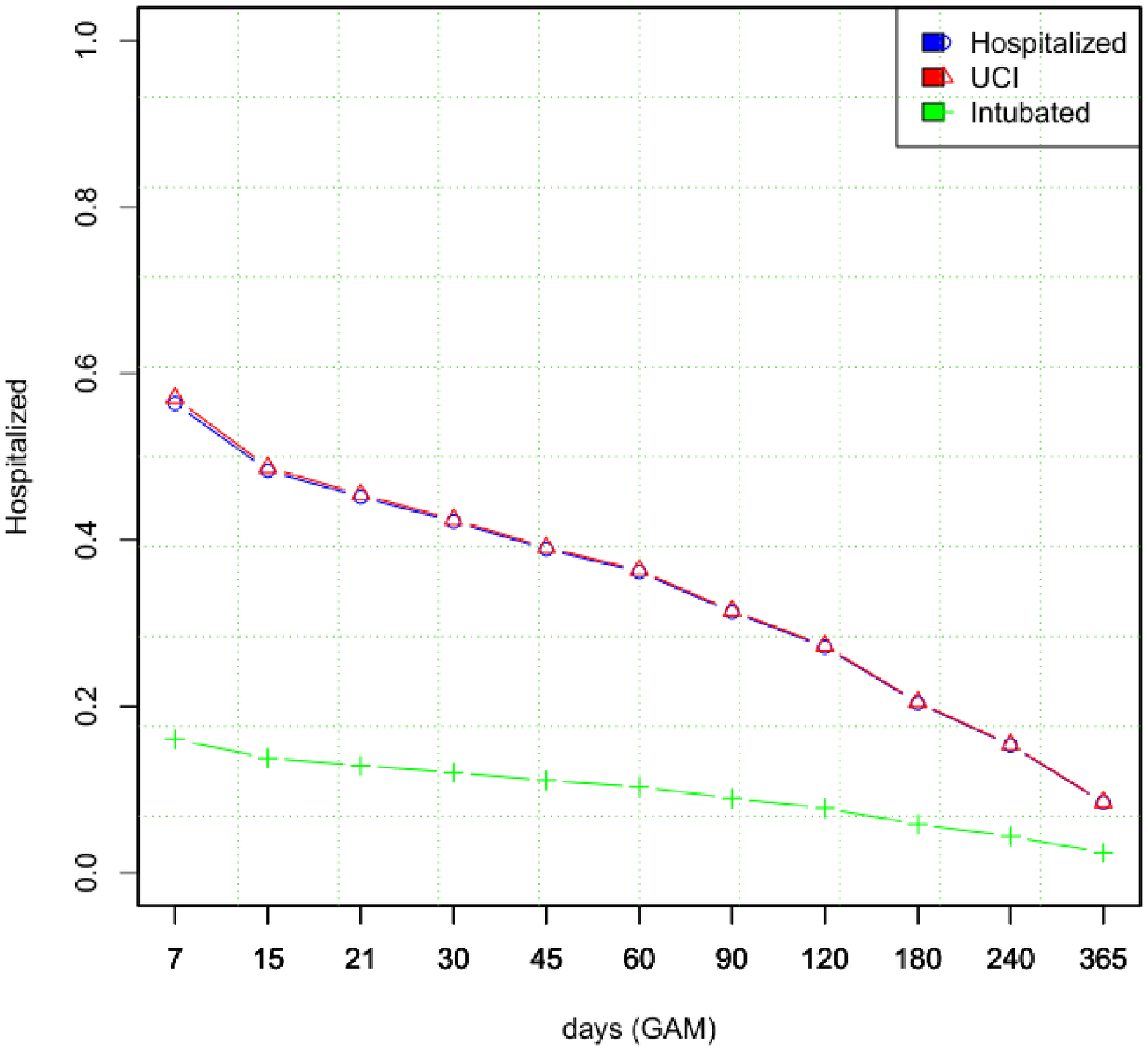}}\hspace{3mm}
\subfigure[Iztacalco]{\includegraphics[width=5cm, height=5cm]{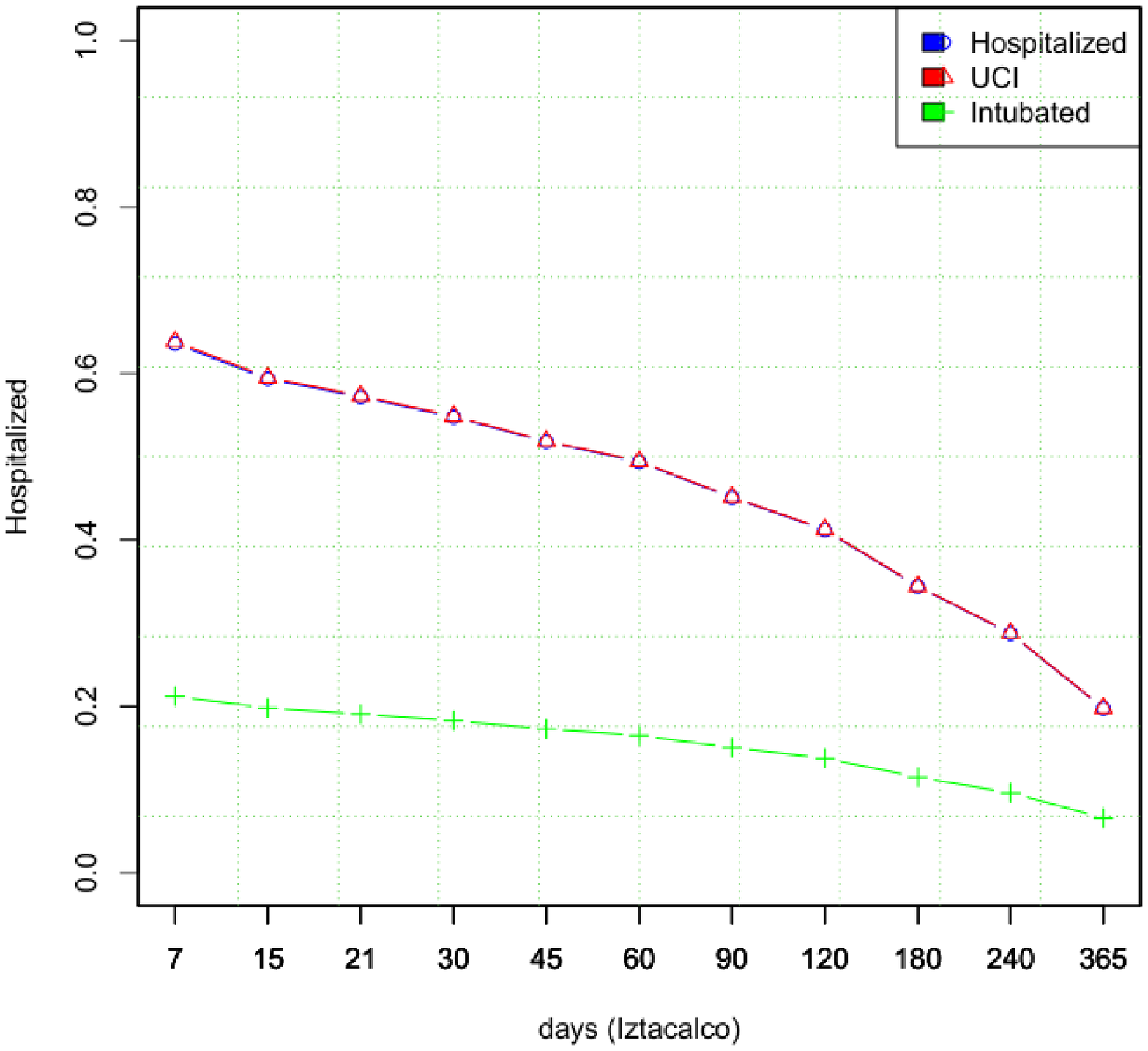}}\hspace{3mm}
\subfigure[Iztapalapa]{\includegraphics[width=5cm, height=5cm]{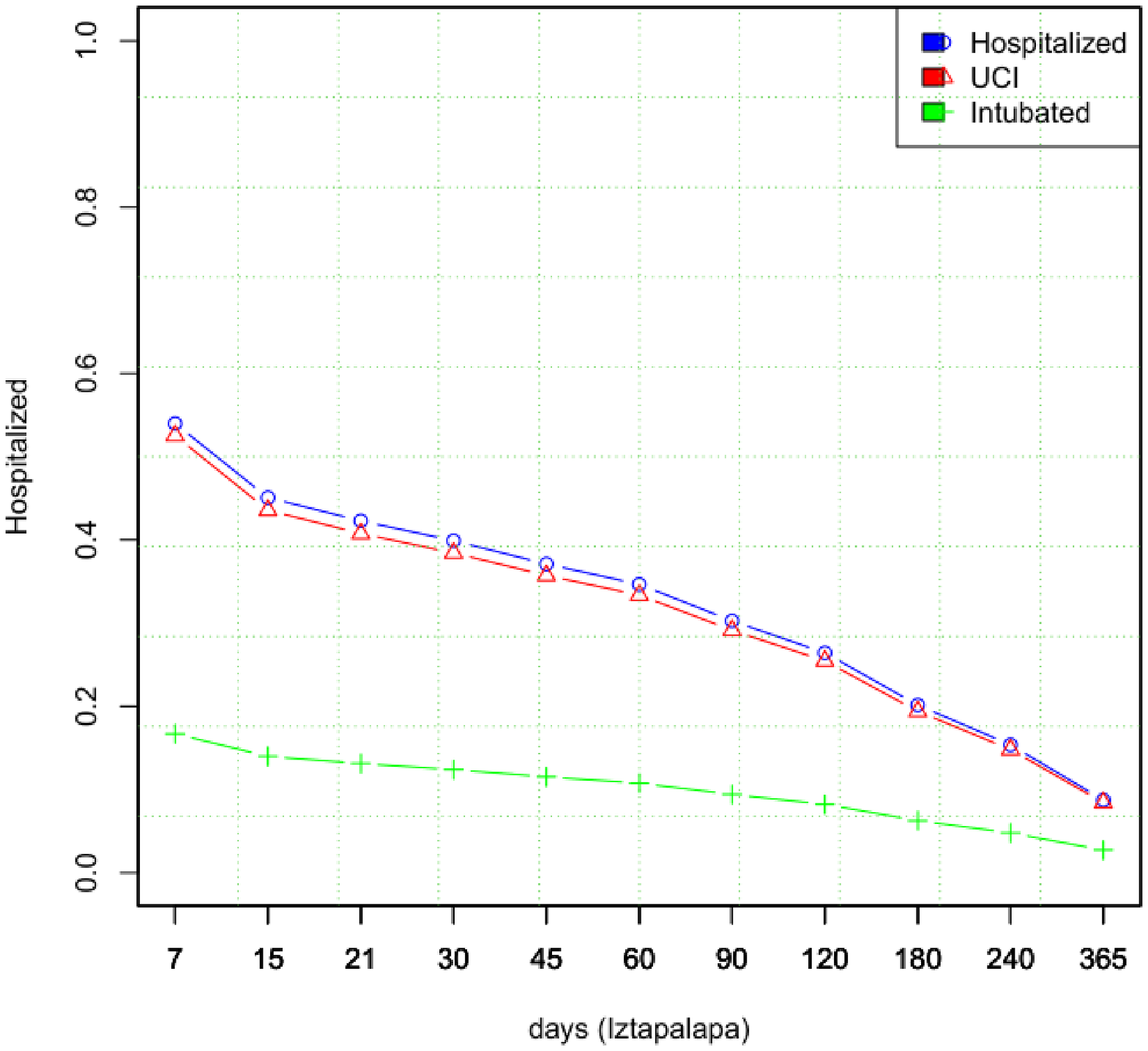}}\hspace{3mm}
\subfigure[Magdalena C]{\includegraphics[width=5cm, height=5cm]{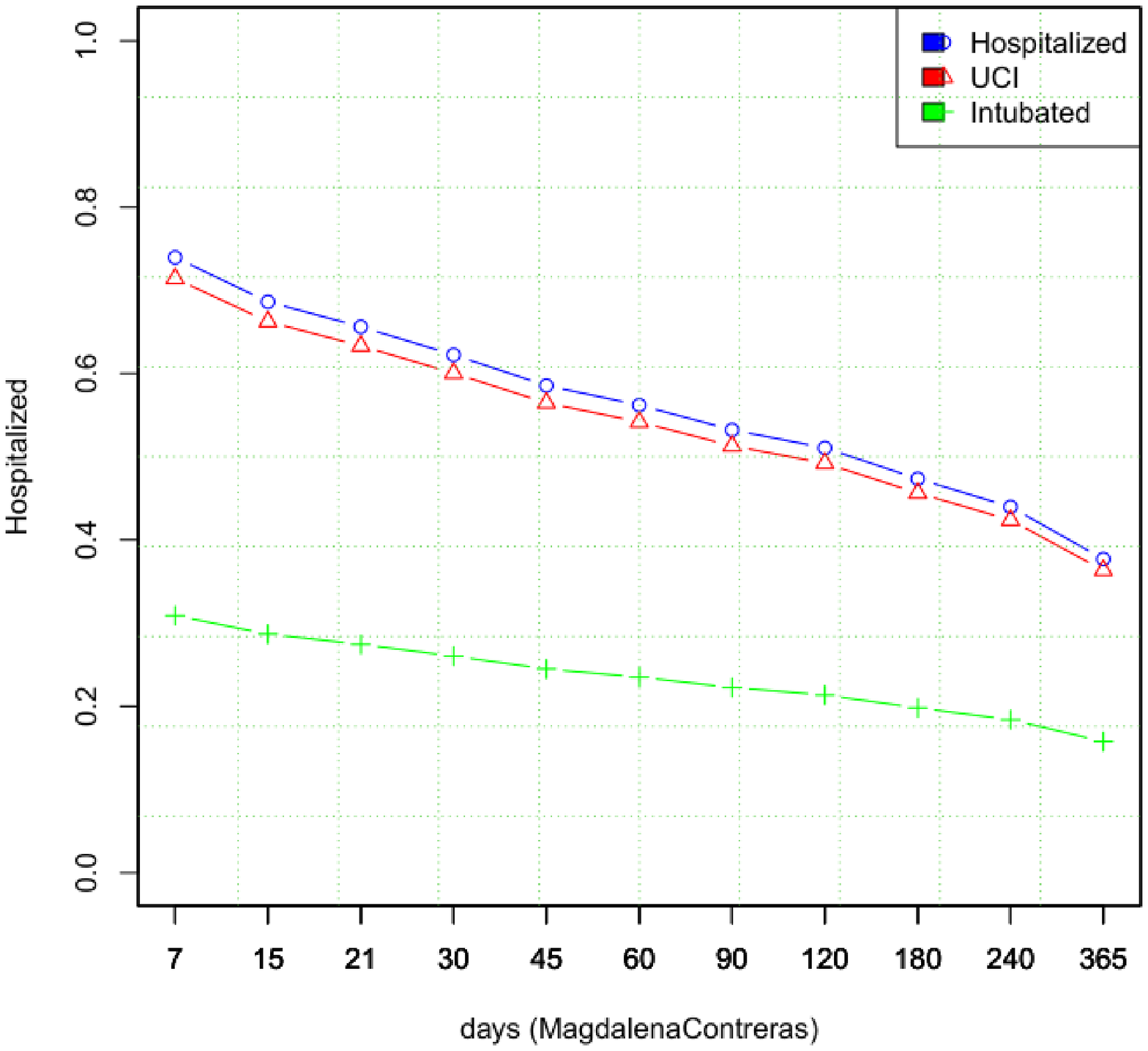}}\hspace{5mm}
\subfigure[Alvaro O]{\includegraphics[width=5cm, height=5cm]{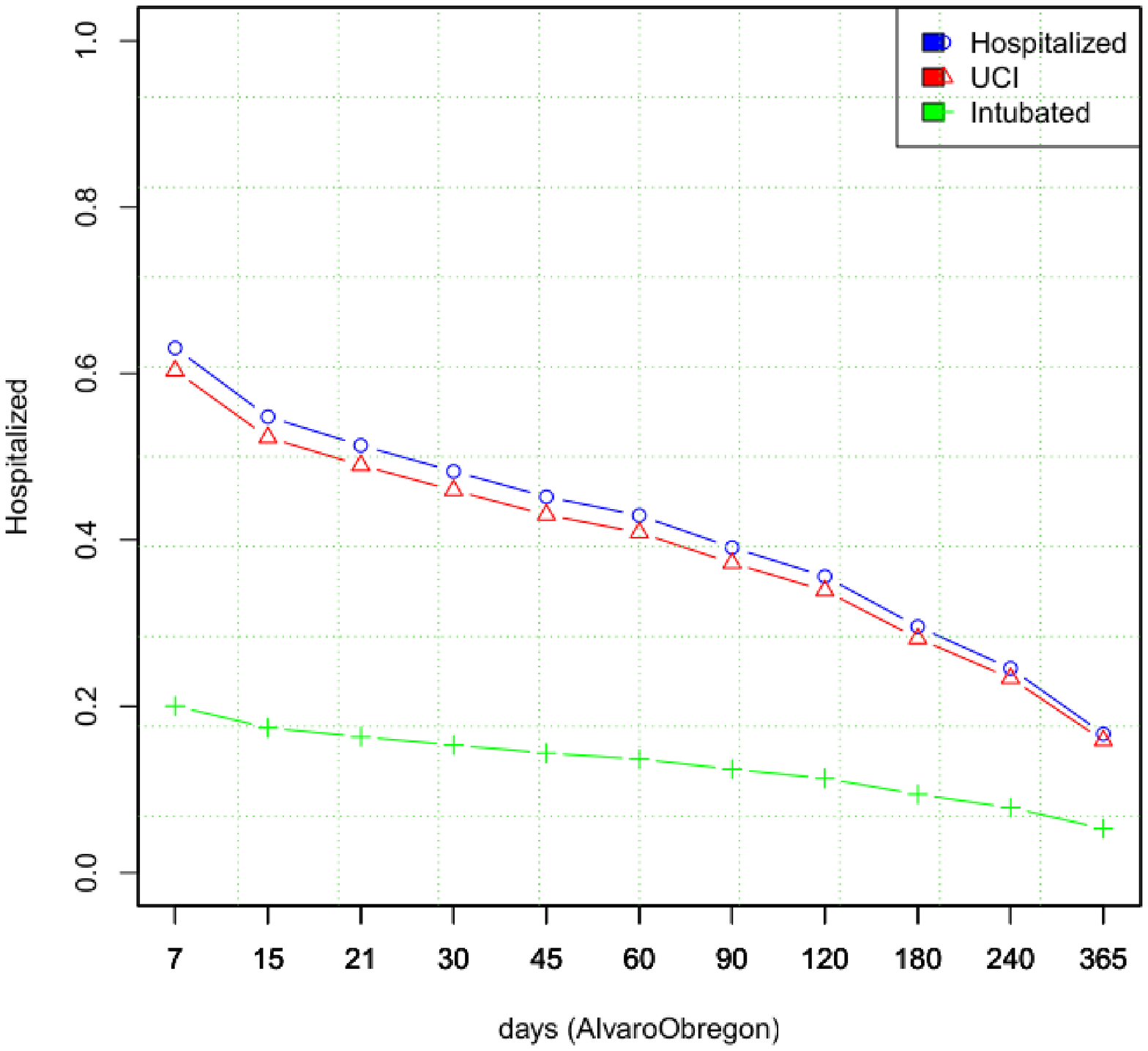}}\hspace{3mm}
\caption{Transition probabilities from hospitalized,  UCI and Intibated  to Recovered state for delegations: Azcapotzalco,  Coyoacan, Cuajimalpa, Gustavo Madero,  Iztacalco, Iztapalapa,  Magdalena C., and Alvaro O.} \label{fig:TercerasOcho}
\end{figure}
\begin{center}
\begin{table}[!ht]
\scalebox{0.75}{
\begin{tabular}{|c||cccc|ccc|ccc|}
\hline 
Days & $E\curvearrowright F$ & $H\curvearrowright F$ & $U\curvearrowright F$ & $I\curvearrowright F$ & $H\curvearrowright U$ & $U\curvearrowright I$ & $H\curvearrowright I$ & $H\curvearrowright S$ & $U\curvearrowright S$ & $I\curvearrowright S$\\\hline\hline
\textbf{7} &    0.0714 &  0.3165 &  0.3133562 &  0.8050 &  7.894e-05 &  1.076e-04 &   7.875e-05 &  0.5639 &  0.5706 &  0.16036\\\hline
\textbf{15} &   0.1316  & 0.3328  & 0.3293483 &  0.8097 &  1.583e-04  & 1.568e-04  &  1.579e-04 &  0.4832 &  0.4871 &  0.1376\\\hline
\textbf{21} &   0.1635 &  0.3488  & 0.3453244  & 0.8143 &  1.798e-04  & 1.793e-04  &  1.792e-04  & 0.4516 &  0.4546 &  0.1287\\\hline
\textbf{30} &   0.2031 &  0.3745 &  0.3711081 &  0.8216 &  1.874e-04 &  1.875e-04 &   1.867e-04 &  0.4221 &  0.4246 &  0.1203\\\hline
\textbf{45}  &  0.2598&  0.4170 &  0.4138643 &  0.8338 &  1.803e-04 &  1.806e-04 &  1.797e-04 &  0.3889 &  0.3910 &  0.1109\\\hline
\textbf{60}  &  0.3112 &  0.4572 &  0.4542353 &  0.8452 &  1.687e-04 &  1.691e-04  &  1.682e-04 &  0.3614 &  0.3634 &  0.1031\\\hline
\textbf{90}  &  0.4031 &  0.5295  & 0.5269798 &  0.8658 &  1.464e-04 &  1.467e-04 &   1.459e-04 &  0.3132 &  0.3149 &  0.0893\\\hline
\textbf{120} &  0.4826 &  0.5922 &  0.5900412 &  0.8837 &  1.269e-04  & 1.271e-04 &  1.264e-04 &  0.2714 &  0.2729 &  0.0774\\\hline
\textbf{180}  & 0.6114 &  0.6937 &  0.6920639 &  0.9127 &  9.529e-05  & 9.548e-05 &  9.497e-05 &  0.2039 &  0.2050 &  0.0581\\\hline
\textbf{240}  & 0.7081  & 0.7699  & 0.7686971 &  0.9344 &  7.158e-05 &  7.172e-05  & 7.133e-05  & 0.1531  & 0.1540 &  0.04367\\\hline
\textbf{365}  & 0.8392 &  0.8733  & 0.8725719 &  0.9639 &  3.944e-05  & 3.951e-05 &  3.929e-05 &  0.0844 &  0.0848 &  0.0241\\\hline
\end{tabular} }
\caption{Transition probabilities between states for several times,  Gustavo A Madero}\label{Table.Prob.GAM}
\end{table}
\end{center}
\begin{figure}[h!]
\centering
\subfigure[Benito J]{\includegraphics[width=5cm, height=5cm]{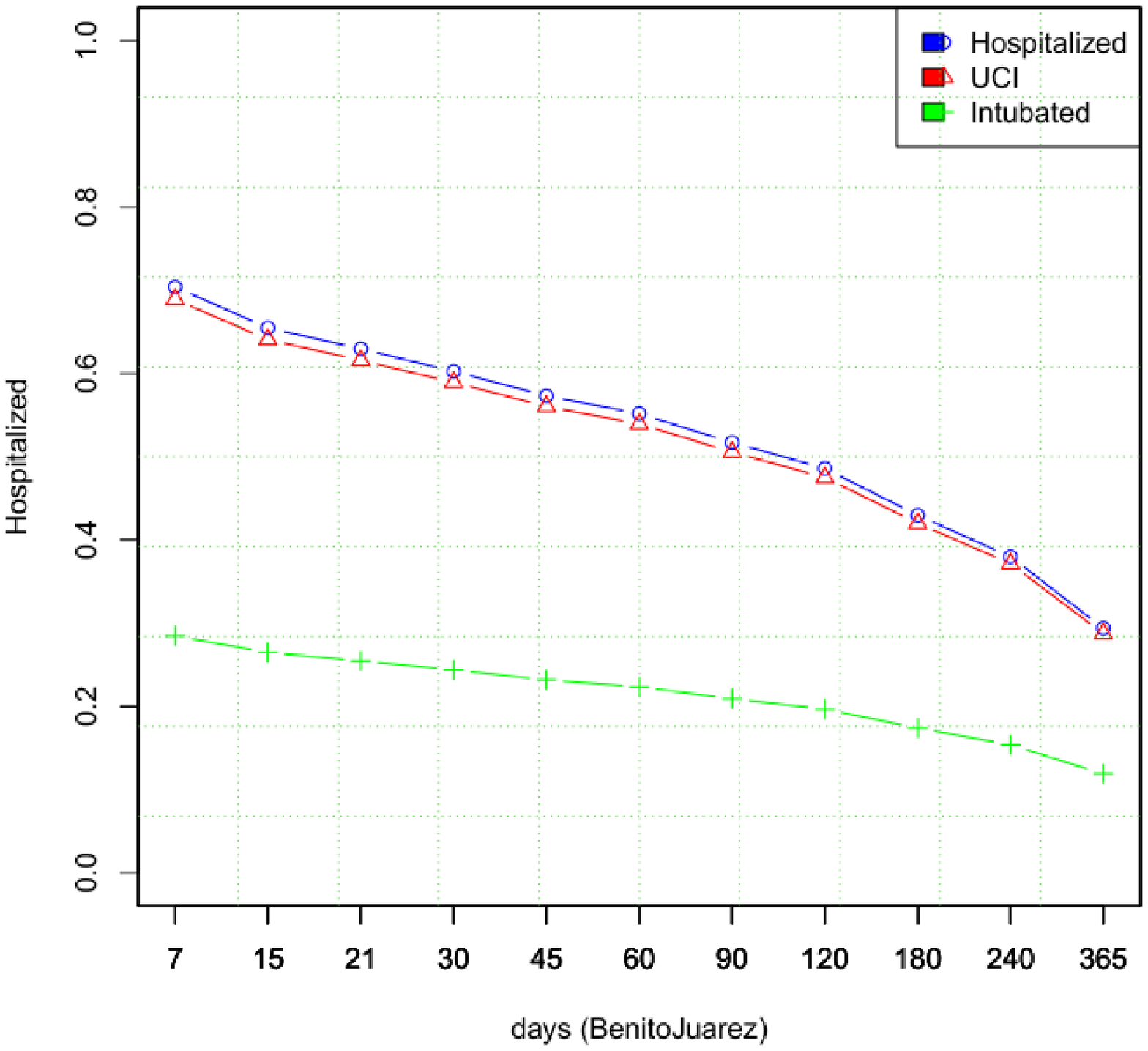}}\hspace{3mm}
\subfigure[Miguel H]{\includegraphics[width=5cm, height=5cm]{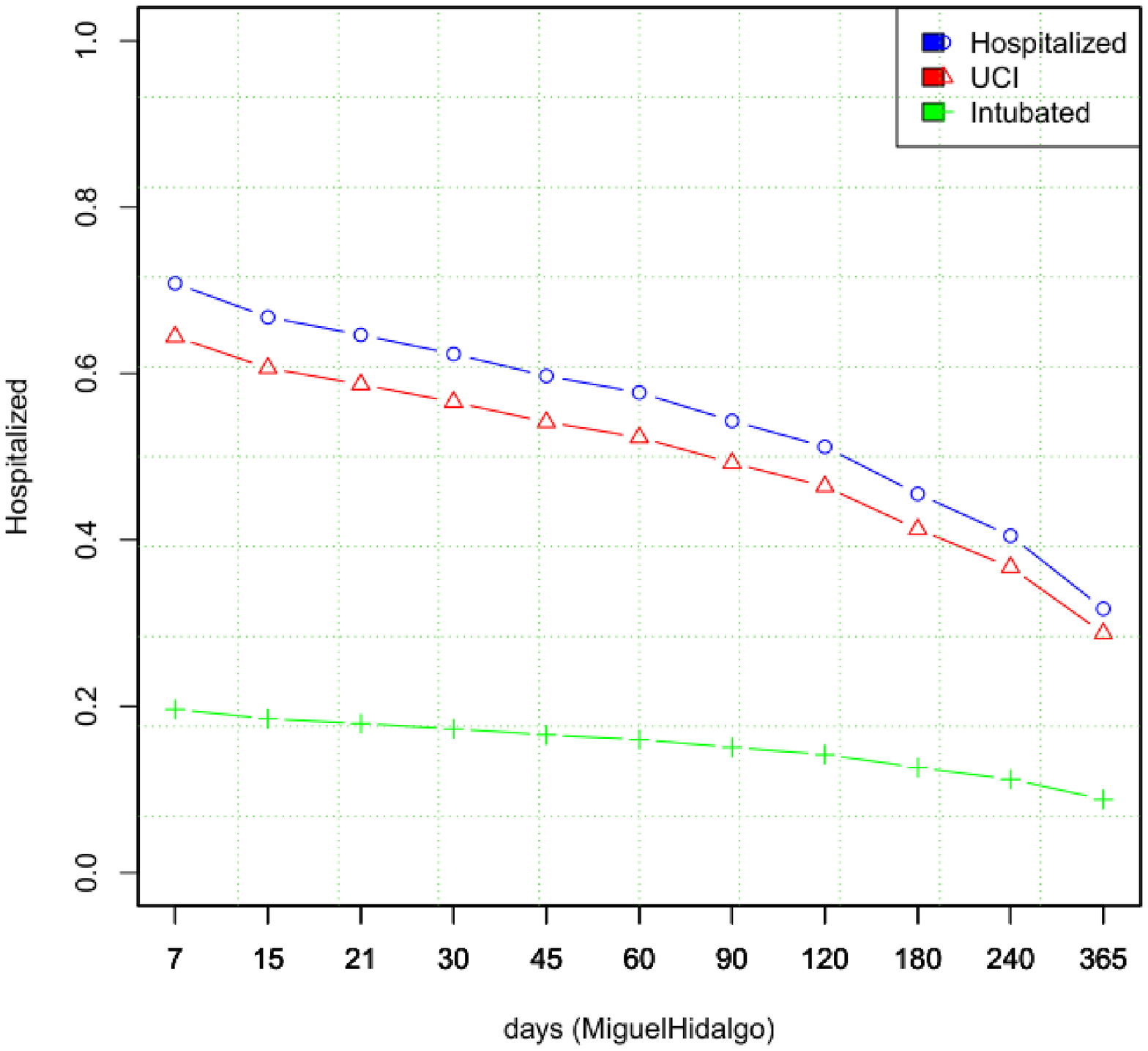}}\hspace{3mm}
\subfigure[Milpa A]{\includegraphics[width=5cm, height=5cm]{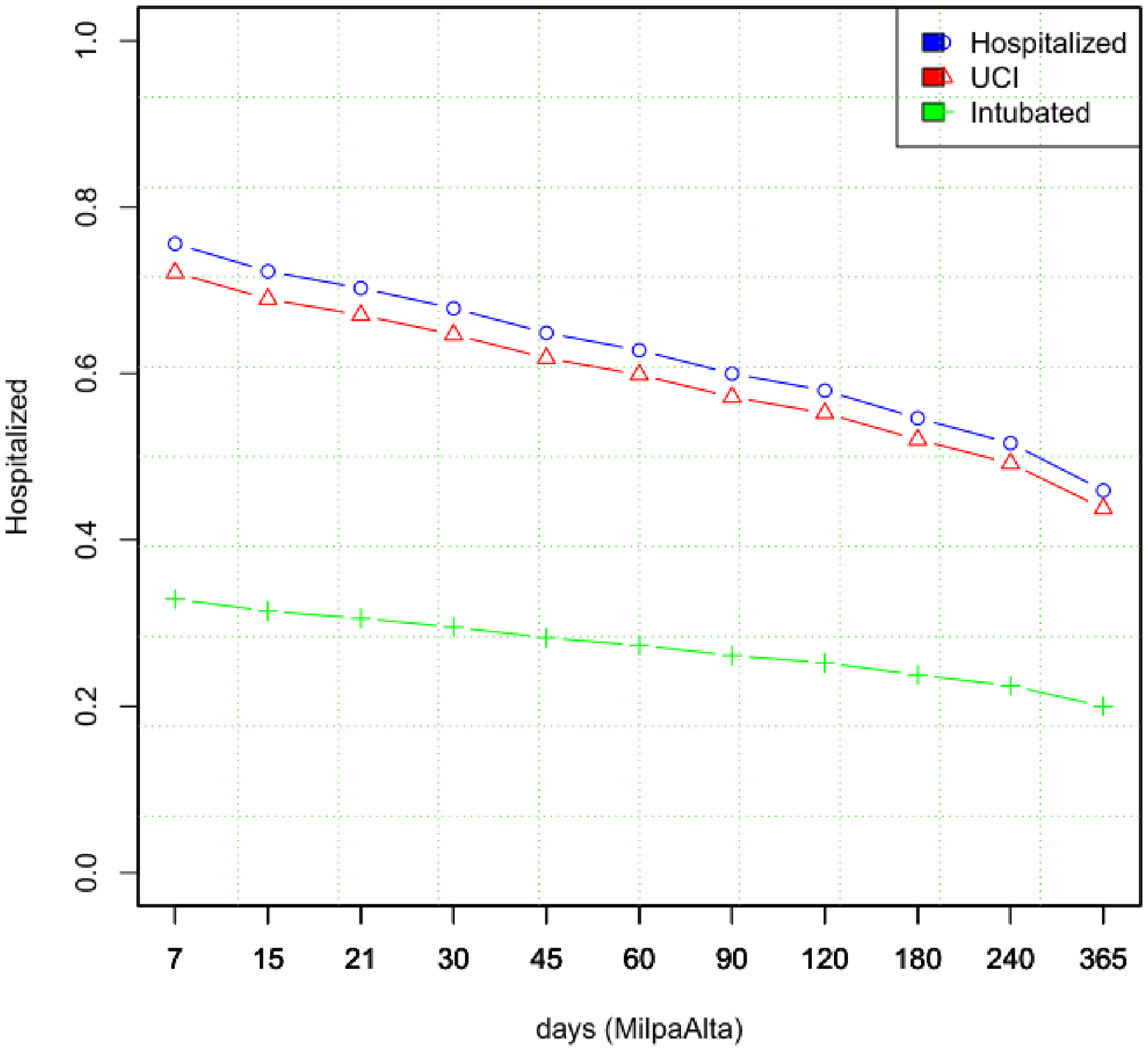}}\hspace{3mm}
\subfigure[Tlahuac]{\includegraphics[width=5cm, height=5cm]{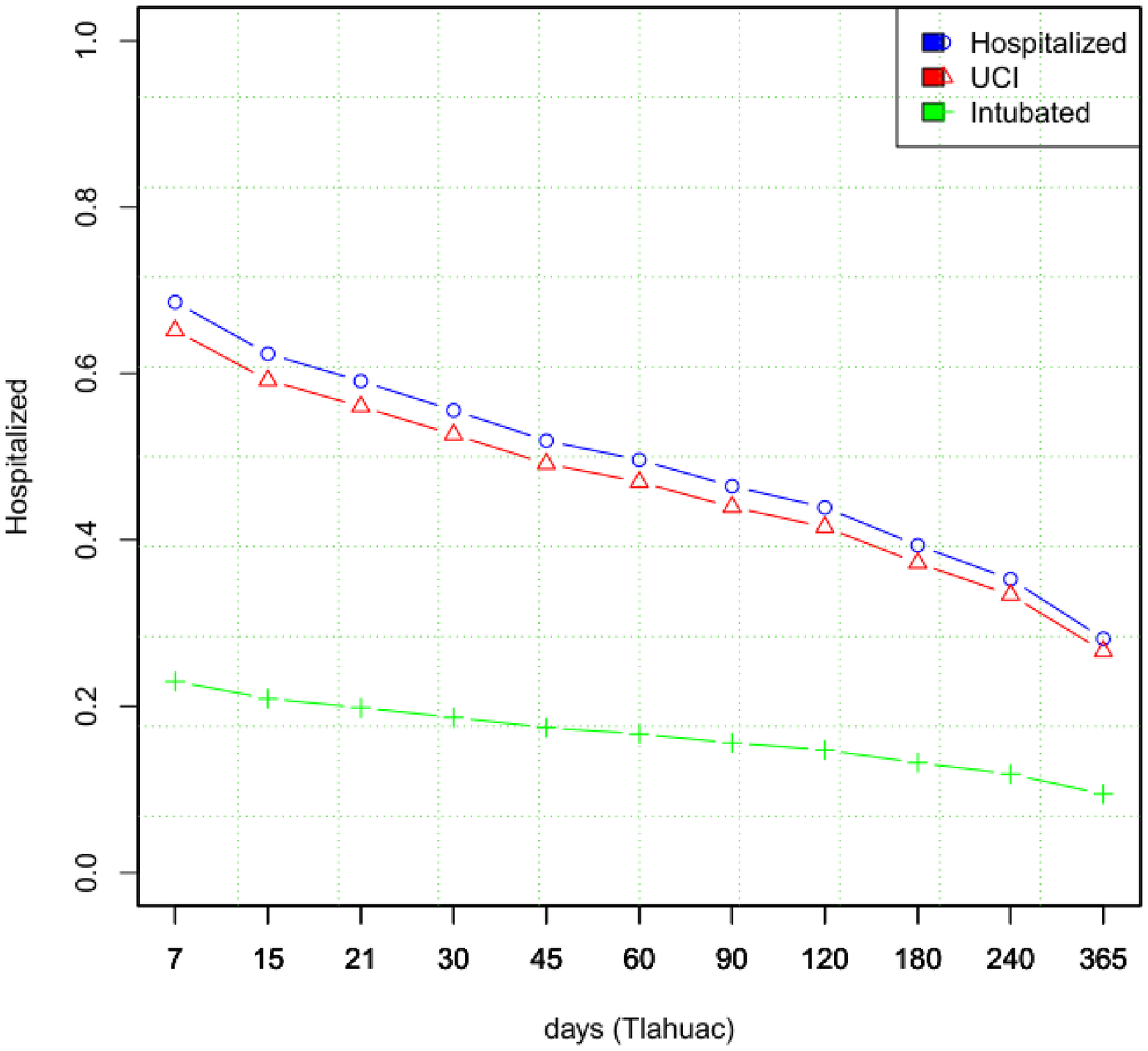}}\hspace{3mm}
\subfigure[Tlalpan]{\includegraphics[width=5cm, height=5cm]{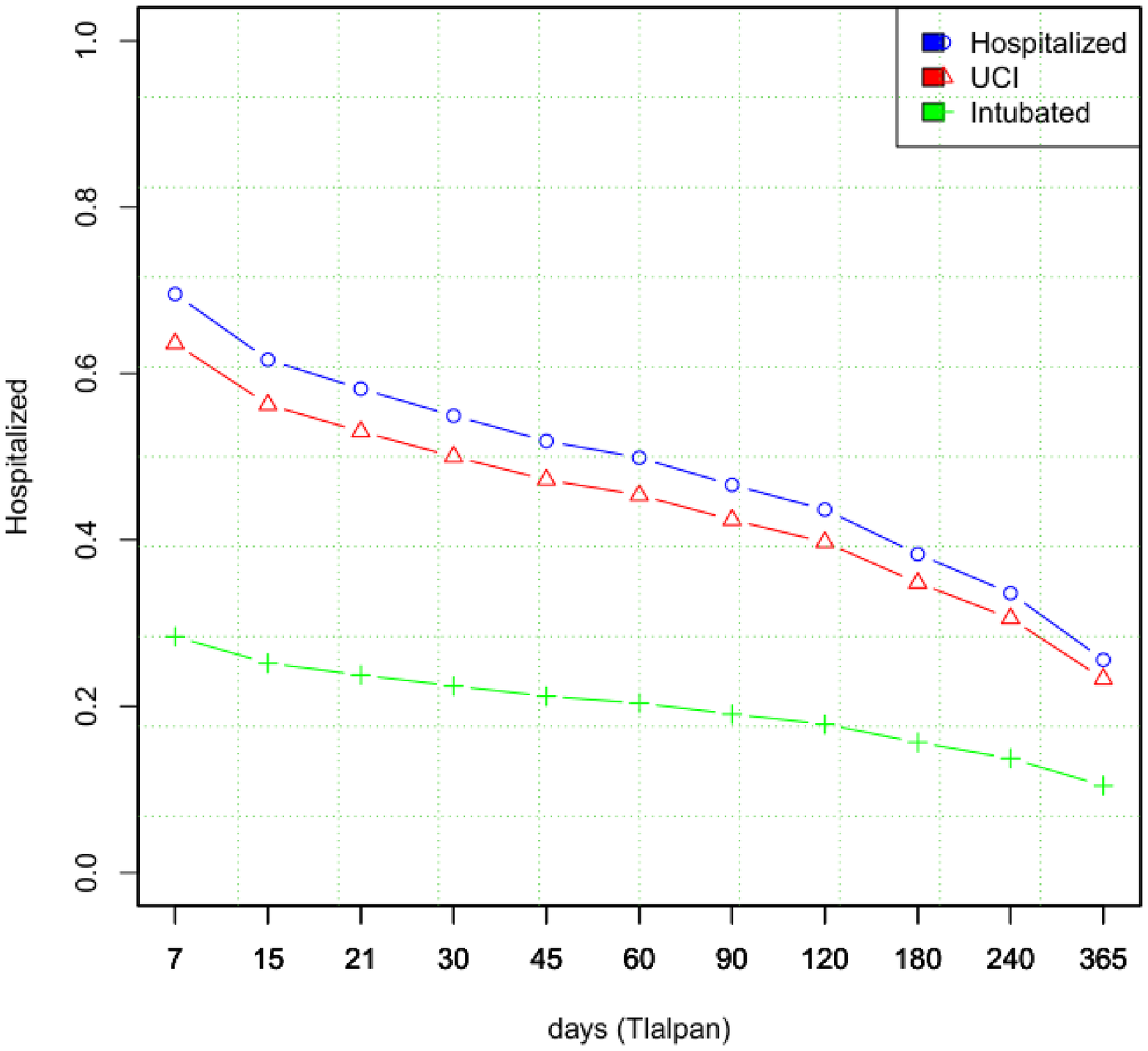}}\hspace{3mm}
\subfigure[Xochimilco]{\includegraphics[width=5cm, height=5cm]{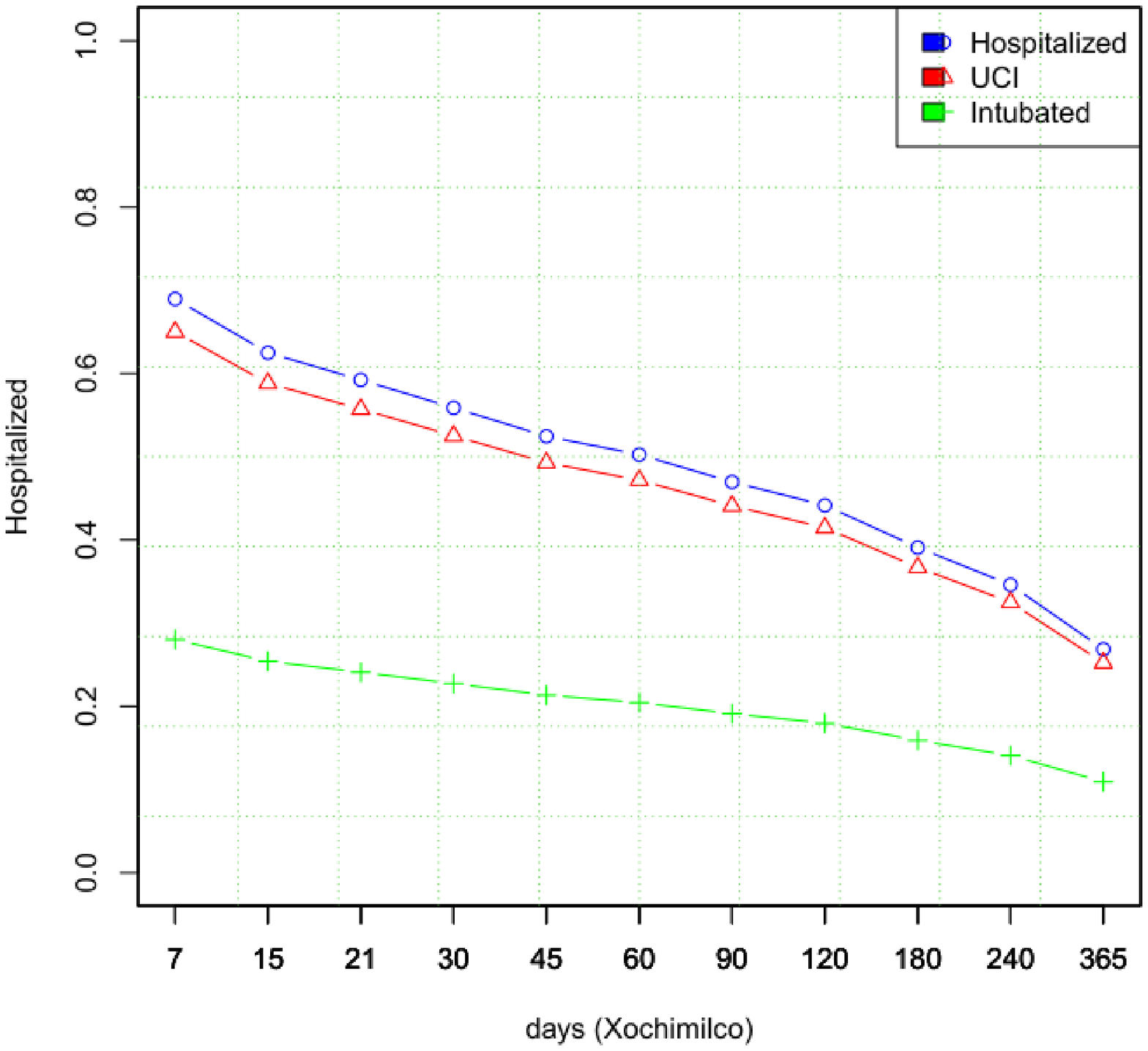}}\hspace{3mm}
\subfigure[Venustiano Carranza]{\includegraphics[width=5cm, height=5cm]{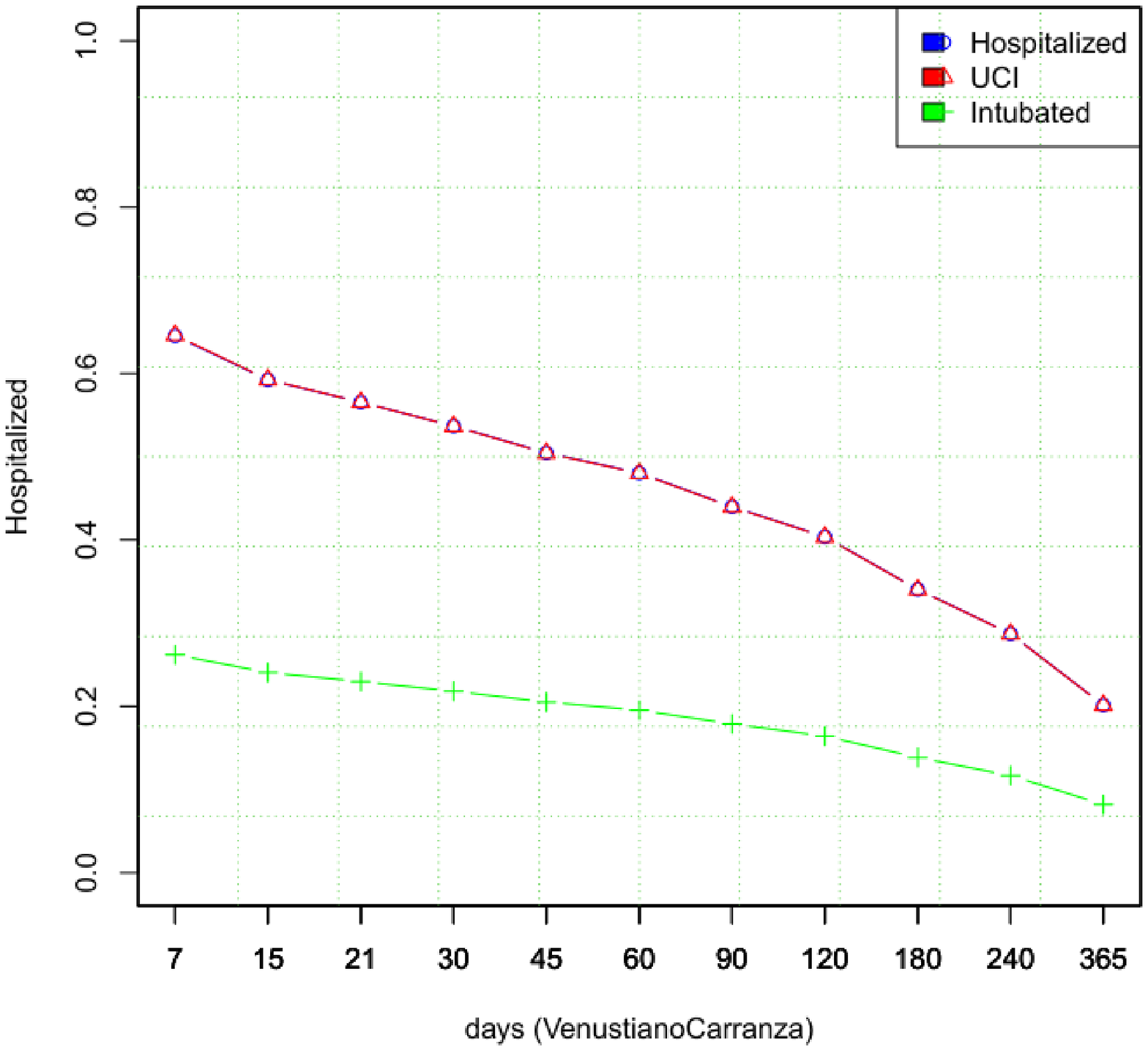}}\hspace{5mm}
\subfigure[Cuauhtemoc]{\includegraphics[width=5cm, height=5cm]{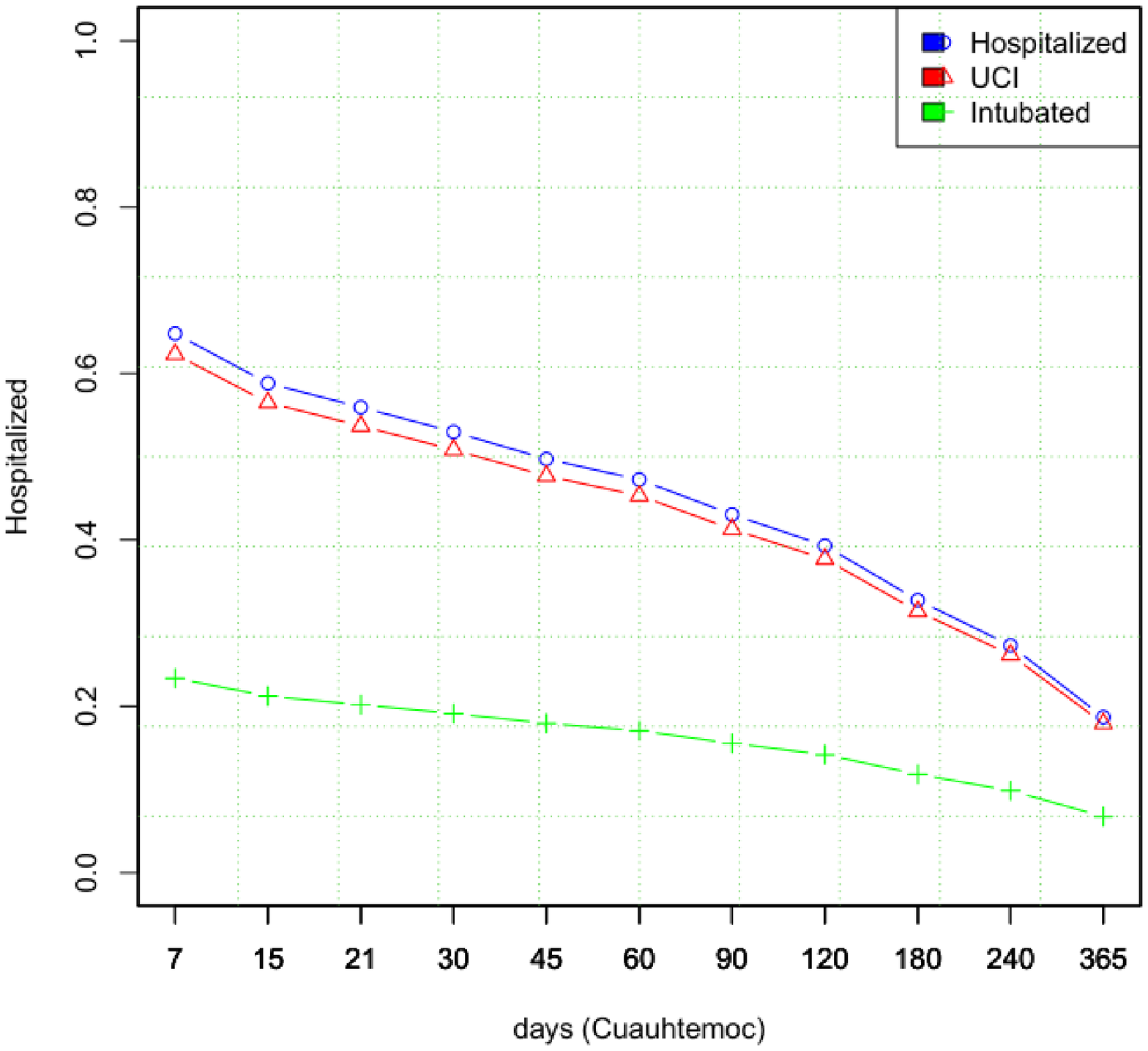}}\hspace{3mm}
\caption{Transition probabilities from hospitalized,  UCI and Intibated  to Recovered state for delegations: Benito J., Miguel H., Milpa A.,  Tlahuac,  Tlalpan,  Xochimilco, Venustiano C.  and Cuauhtemoc } \label{fig:CuartasOcho}
\end{figure}
\begin{center}
\begin{table}[!ht]
\scalebox{0.75}{
\begin{tabular}{|c||cccc|ccc|ccc|}
\hline 
Days & $E\curvearrowright F$ & $H\curvearrowright F$ & $U\curvearrowright F$ & $I\curvearrowright F$ & $H\curvearrowright U$ & $U\curvearrowright I$ & $H\curvearrowright I$ & $H\curvearrowright S$ & $U\curvearrowright S$ & $I\curvearrowright S$\\\hline\hline
\textbf{7}  &     0.05599  &   0.2884  &   0.3155 &    0.7790  &   9.866e-05 &    1.873e-04  &    0.00014  &   0.5399  &   0.5254  &   0.1669\\\hline
\textbf{15} &     0.1031 &    0.3068  &   0.3327 &    0.7847 &    1.795e-04 &    2.422e-04  &    0.00025&    0.4508 &    0.4355  &   0.1398\\\hline
\textbf{21}  &    0.1301 &    0.3237 &    0.3489 &    0.7900 &    1.951e-04 &    2.645e-04 &     0.00027 &    0.4229  &   0.4077 &    0.1312\\\hline
\textbf{30}  &    0.1662 &    0.3499 &    0.3742 &    0.7982 &    1.964e-04 &    2.668e-04 &     0.00028  &   0.3989 &    0.3842 &    0.1239\\\hline
\textbf{45}  &    0.2212 &   0.3924 &    0.4150  &   0.8113  &   1.858e-04 &    2.526e-04  &    0.00026 &    0.3710  &   0.3572 &    0.1152\\\hline
\textbf{60} &     0.2722 &    0.4322  &   0.4533  &   0.8237  &   1.738e-04 &    2.363e-04  &    0.00025  &   0.3466  &   0.3336  &   0.1076\\\hline
\textbf{90}  &    0.3645  &   0.5041  &   0.5226  &   0.8460 &    1.518e-04 &    2.064e-04 &    0.00021  &   0.3026 &    0.2913 &    0.0939\\\hline
\textbf{120}  &   0.4451  &   0.5670  &   0.5832 &    0.8656  &   1.326e-04 &    1.802e-04 &    0.00019 &    0.2642 &    0.2544  &   0.0821\\\hline
\textbf{180}  &   0.5768  &   0.6698 &    0.6821  &   0.8975  &   1.011e-04 &    1.374e-04 &    0.00014  &   0.2015  &   0.1939  &   0.0626\\\hline
\textbf{240}  &   0.6773 &    0.7482  &   0.7576 &    0.9218 &    7.708e-05 &    1.048e-04  &   0.00019  &   0.1536  &   0.1479 &    0.0477\\\hline
\textbf{365}  &   0.8166  &   0.8568  &   0.8622 &    0.9556  &   4.382e-05 &    5.957e-05 &  0.00006  &   0.0873  &   0.0841  &   0.0271\\\hline
\end{tabular} }
\caption{Transition probabilities between states for several times, Iztapalapa}\label{Table.Prob.Iztapalapa}
\end{table}
\end{center}
\begin{figure}[h!]
\centering
\subfigure[Azcapotzalco]{\includegraphics[width=5cm, height=5cm]{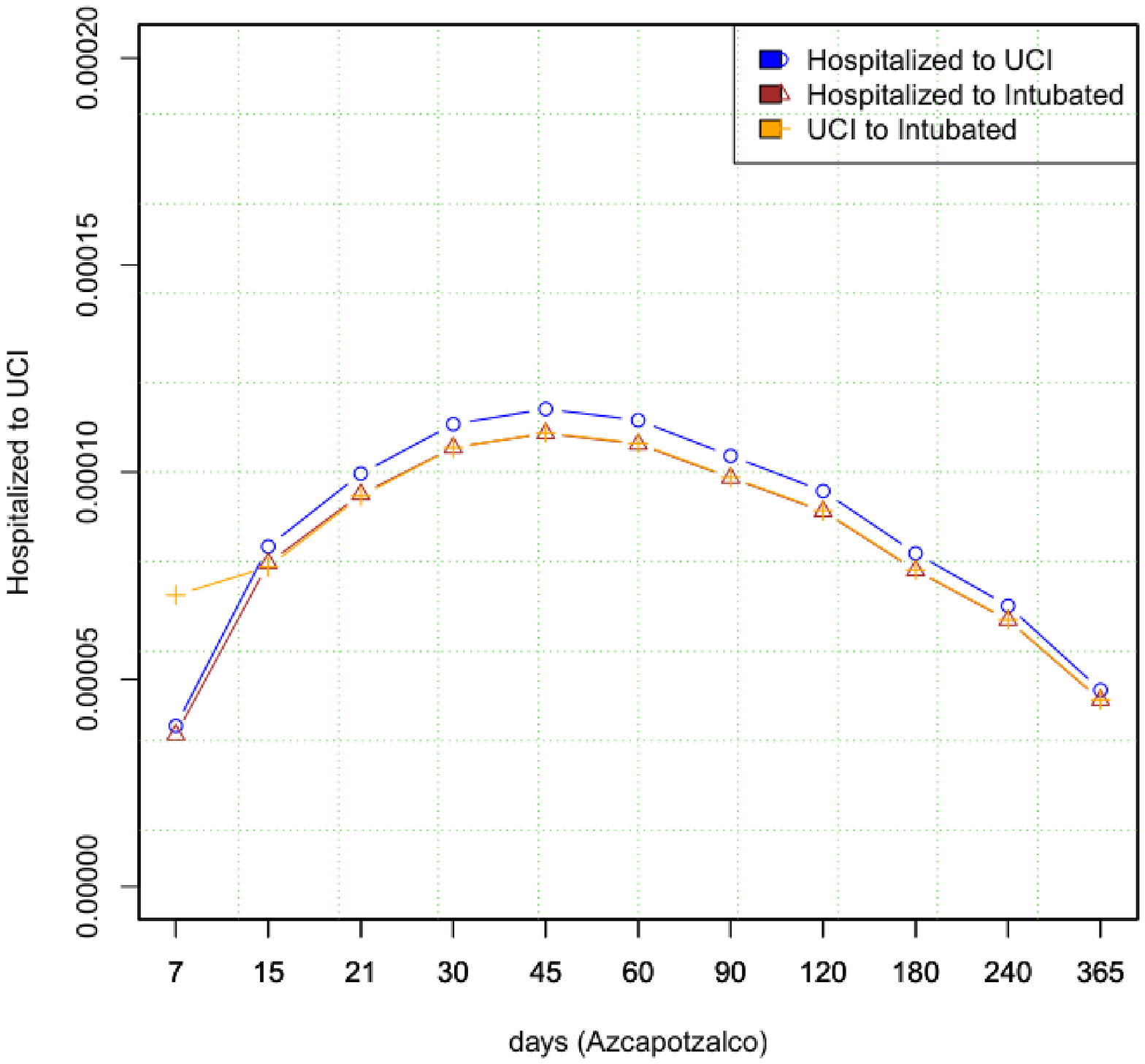}}\hspace{3mm}
\subfigure[Coyoacan]{\includegraphics[width=5cm, height=5cm]{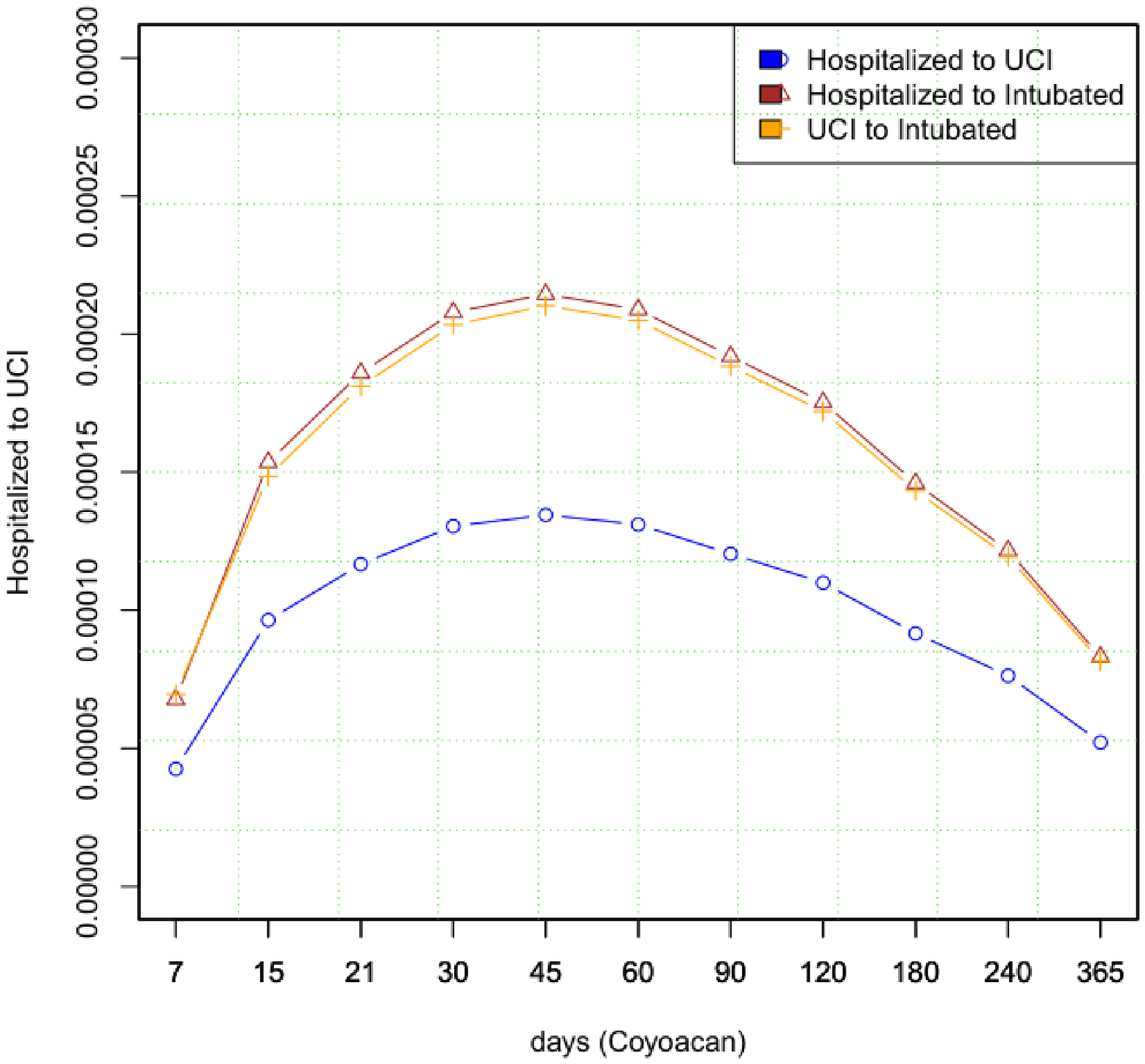}}\hspace{3mm}
\subfigure[Cuajimalpa]{\includegraphics[width=5cm, height=5cm]{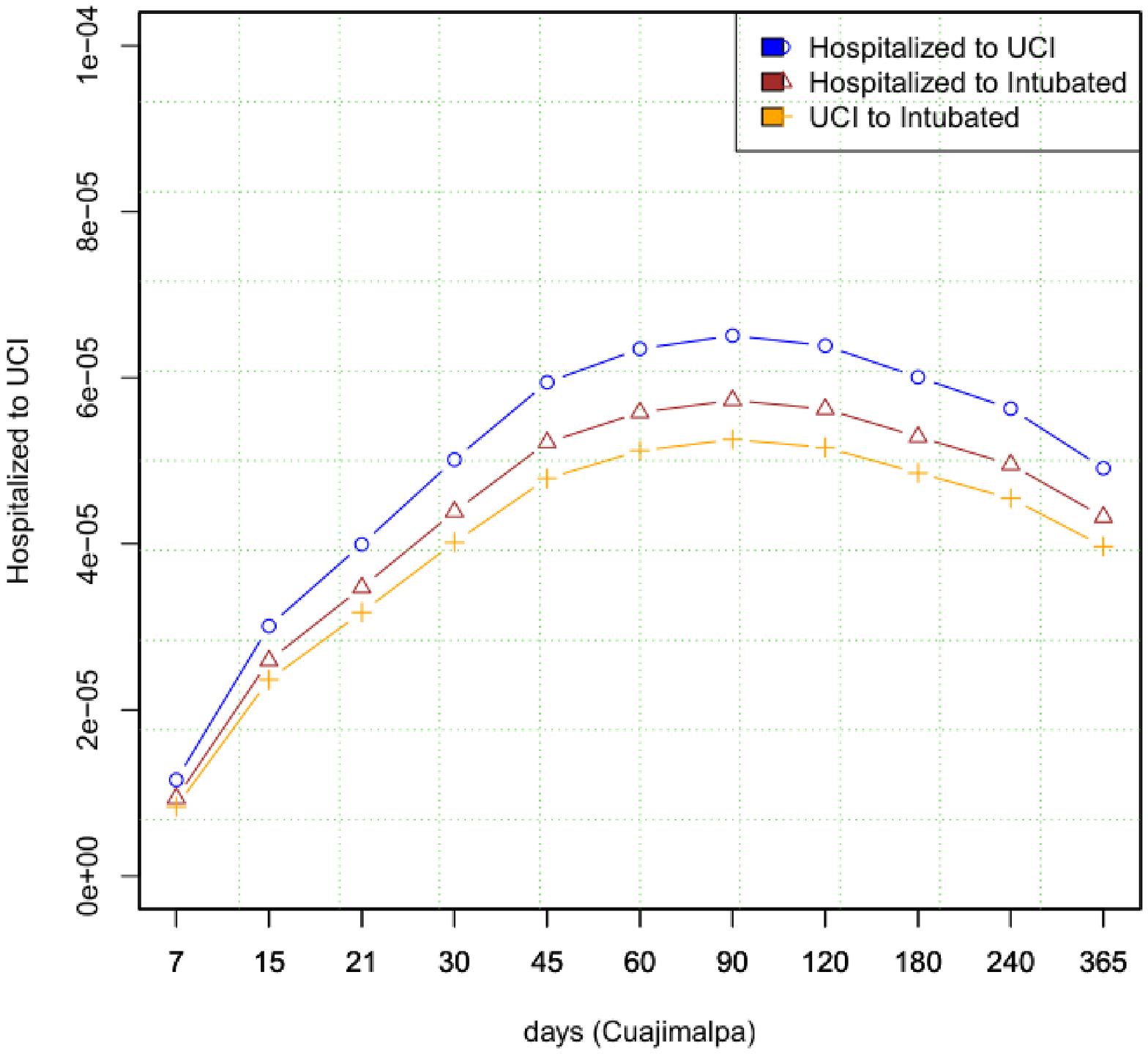}}\hspace{3mm}
\subfigure[Gustavo A.]{\includegraphics[width=5cm, height=5cm]{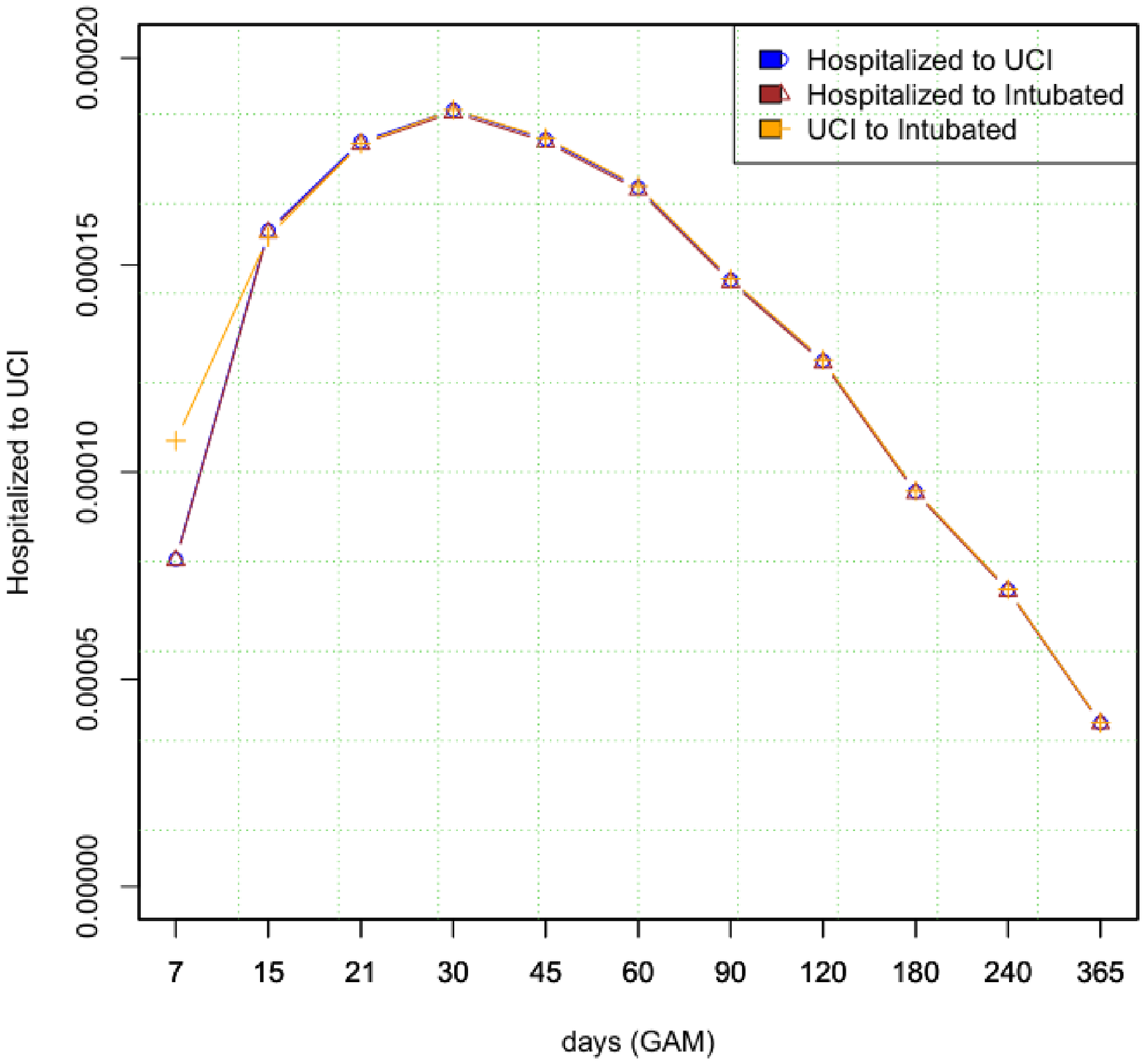}}\hspace{5mm}
\subfigure[Iztacalco]{\includegraphics[width=5cm, height=5cm]{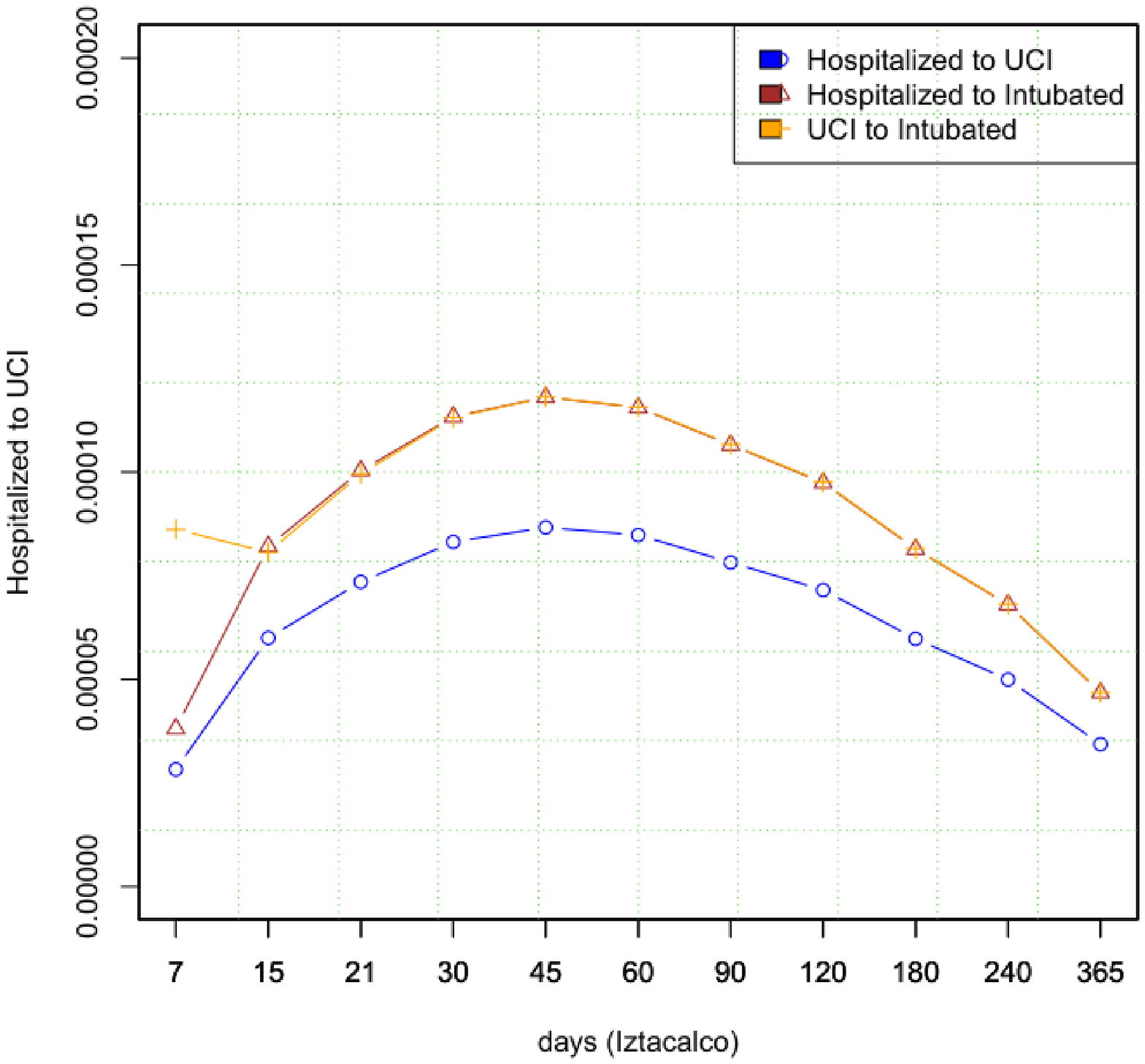}}\hspace{3mm}
\subfigure[Iztapalapa]{\includegraphics[width=5cm, height=5cm]{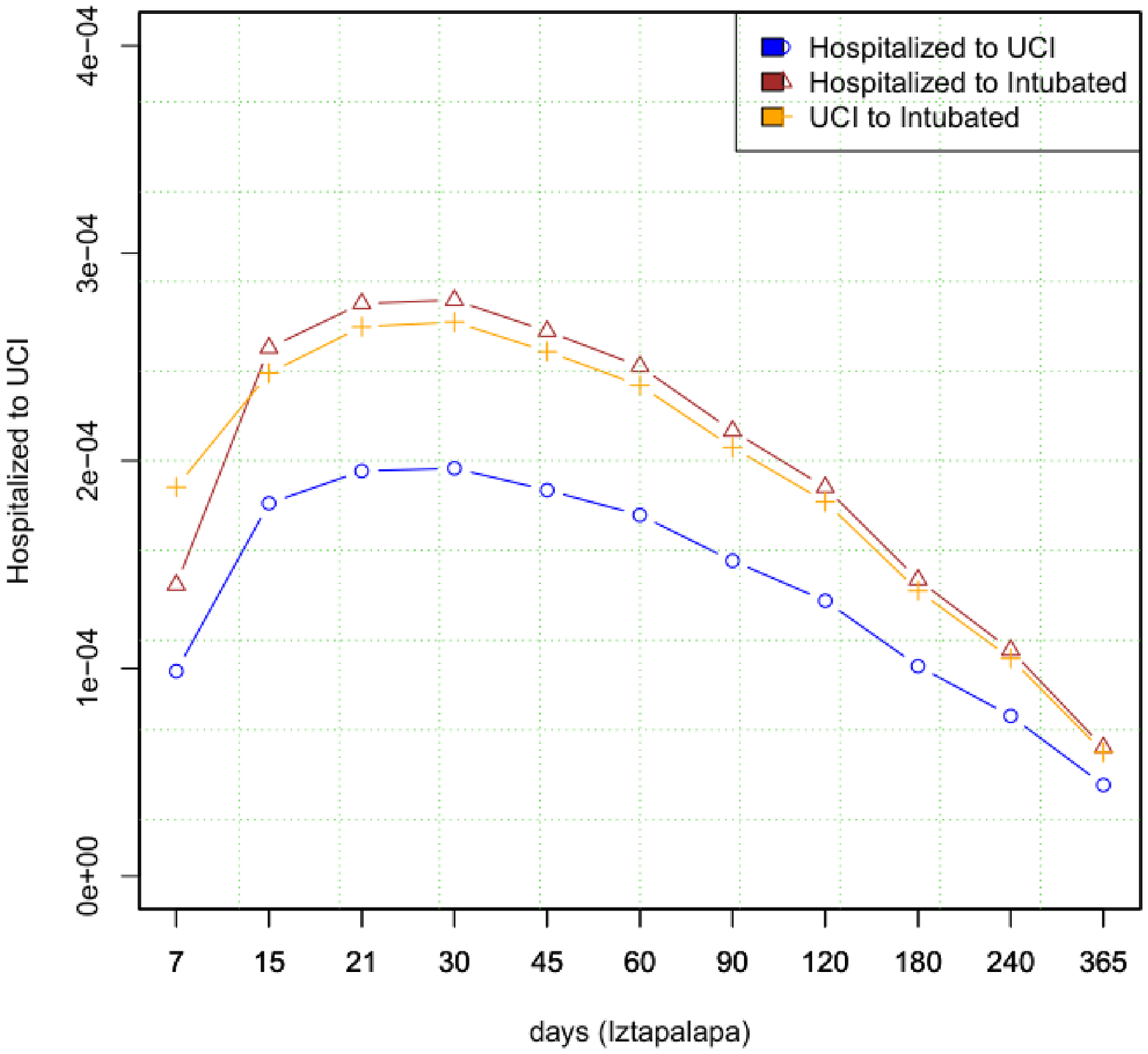}}\hspace{3mm}
\subfigure[Magdalena C]{\includegraphics[width=5cm, height=5cm]{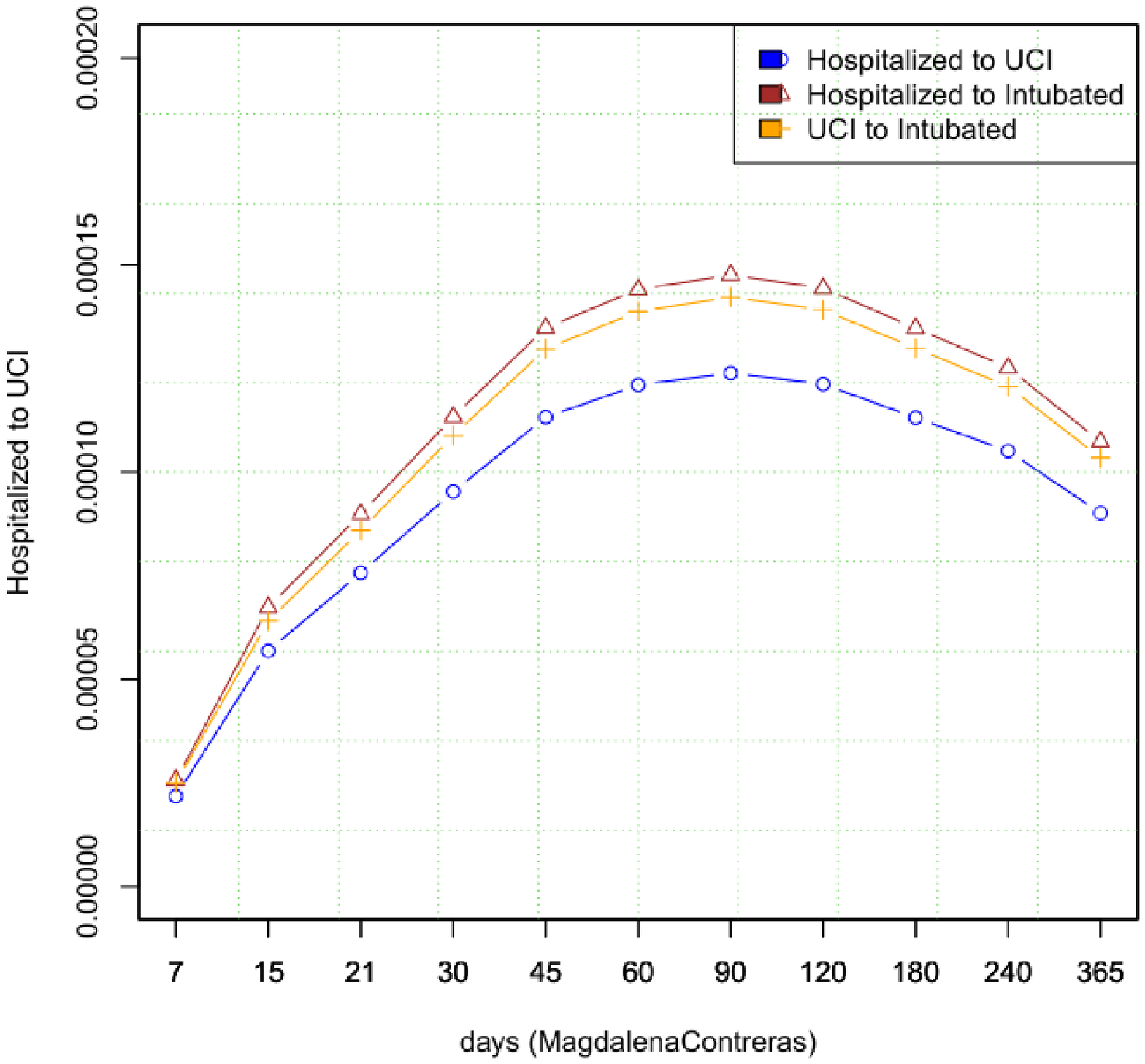}}\hspace{3mm}
\subfigure[Alvaro O]{\includegraphics[width=5cm, height=5cm]{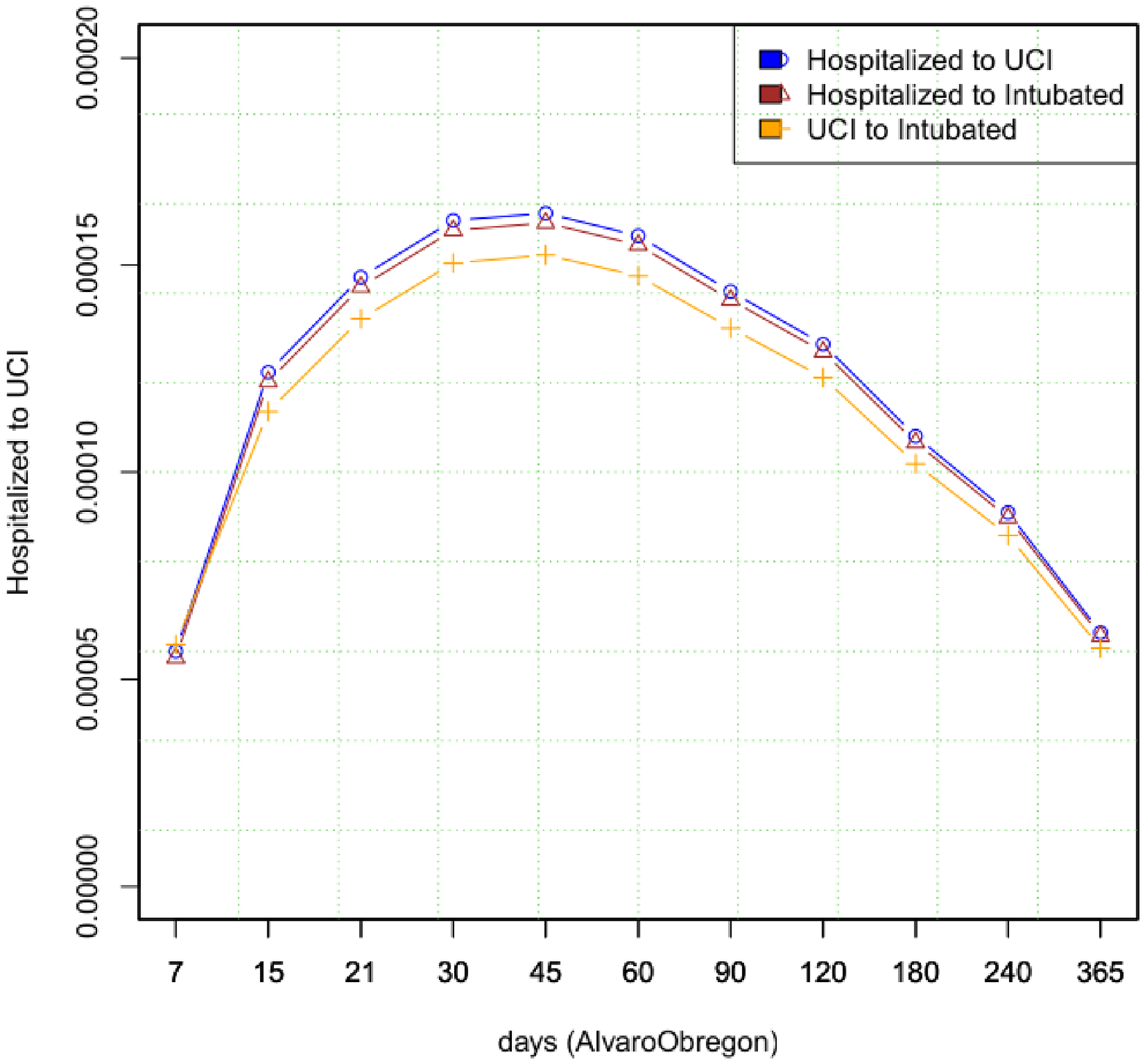}}\hspace{3mm}
\caption{Transition probabilities from hospitalized and UCI to intubated state for delegations: Azcapotzalco,  Coyoacan, Cuajimalpa, Gustavo Madero,  Iztacalco, Iztapalapa,  Magdalena C.  and Alvaro O.} \label{fig:QuintasOcho}
\end{figure}
\begin{center}
\begin{table}[!ht]
\scalebox{0.75}{
\begin{tabular}{|c||cccc|ccc|ccc|}
\hline 
Days & $E\curvearrowright F$ & $H\curvearrowright F$ & $U\curvearrowright F$ & $I\curvearrowright F$ & $H\curvearrowright U$ & $U\curvearrowright I$ & $H\curvearrowright I$ & $H\curvearrowright S$ & $U\curvearrowright S$ & $I\curvearrowright S$\\\hline\hline
\textbf{7} &    0.0481  & 0.2581 & 0.2941  & 0.7639  & 5.676e-05  & 5.839e-05   & 5.518e-05 &  0.6307 &  0.6034  & 0.2002\\\hline
\textbf{15} &   0.0918  & 0.2683  & 0.3036 &  0.7672  & 1.241e-04  & 1.146e-04 &   1.219e-04 &  0.5480  & 0.5230  & 0.1742\\\hline
\textbf{21}  &  0.1159  & 0.2789 &  0.3136  & 0.7706 &  1.471e-04 &  1.370e-04  & 1.447e-04  & 0.5136  & 0.4896  & 0.1633\\\hline
\textbf{30} &   0.1454 &  0.2966 &  0.3304  & 0.7762  & 1.608e-04  & 1.505e-04 &   1.584e-04  & 0.4824 &  0.4595&  0.1534\\\hline
\textbf{45}  &  0.1869 &  0.3275  & 0.3598 & 0.7861  & 1.625e-04 &  1.524e-04  &  1.602e-04  & 0.452  & 0.4301  & 0.1437\\\hline
\textbf{60} &   0.2243 &  0.3578  & 0.3886  & 0.7957 &  1.571e-04  & 1.474e-04   & 1.548e-04 &  0.4294 &  0.4088 &  0.1366\\\hline
\textbf{90}  &  0.2933  & 0.4147 &  0.4428 &  0.8138  & 1.436e-04  & 1.348e-04  & 1.4158e-04 &  0.3908  & 0.3721 &  0.1243\\\hline
\textbf{120} &  0.3559 &  0.4666  & 0.4923  & 0.8303  & 1.309e-04 &  1.228e-04  & 1.290e-04  & 0.3561  & 0.3390  & 0.1133\\\hline
\textbf{180}  & 0.4652 &  0.5571  & 0.5783  & 0.8591  & 1.087e-04 &  1.020e-04 & 1.072e-04 &  0.2958  & 0.2816  & 0.0941\\\hline
\textbf{240} &  0.5559 &  0.6322  & 0.6498 &  0.8829  & 9.028e-05 &  8.472e-05  & 8.899e-05  & 0.2456 &  0.2338 &  0.0781\\\hline
\textbf{365} &  0.6984 &  0.7502  & 0.7622  & 0.9205  & 6.13e-05 &  5.753e-05 &  6.043e-05 &  0.1668  & 0.1588  & 0.0531\\\hline
\end{tabular} }
\caption{Transition probabilities between states for several times, Alvaro Obregon}\label{Table.Prob.AlvaroO}
\end{table}
\end{center}
\begin{figure}[h!]
\centering
\subfigure[Benito J]{\includegraphics[width=5cm, height=5cm]{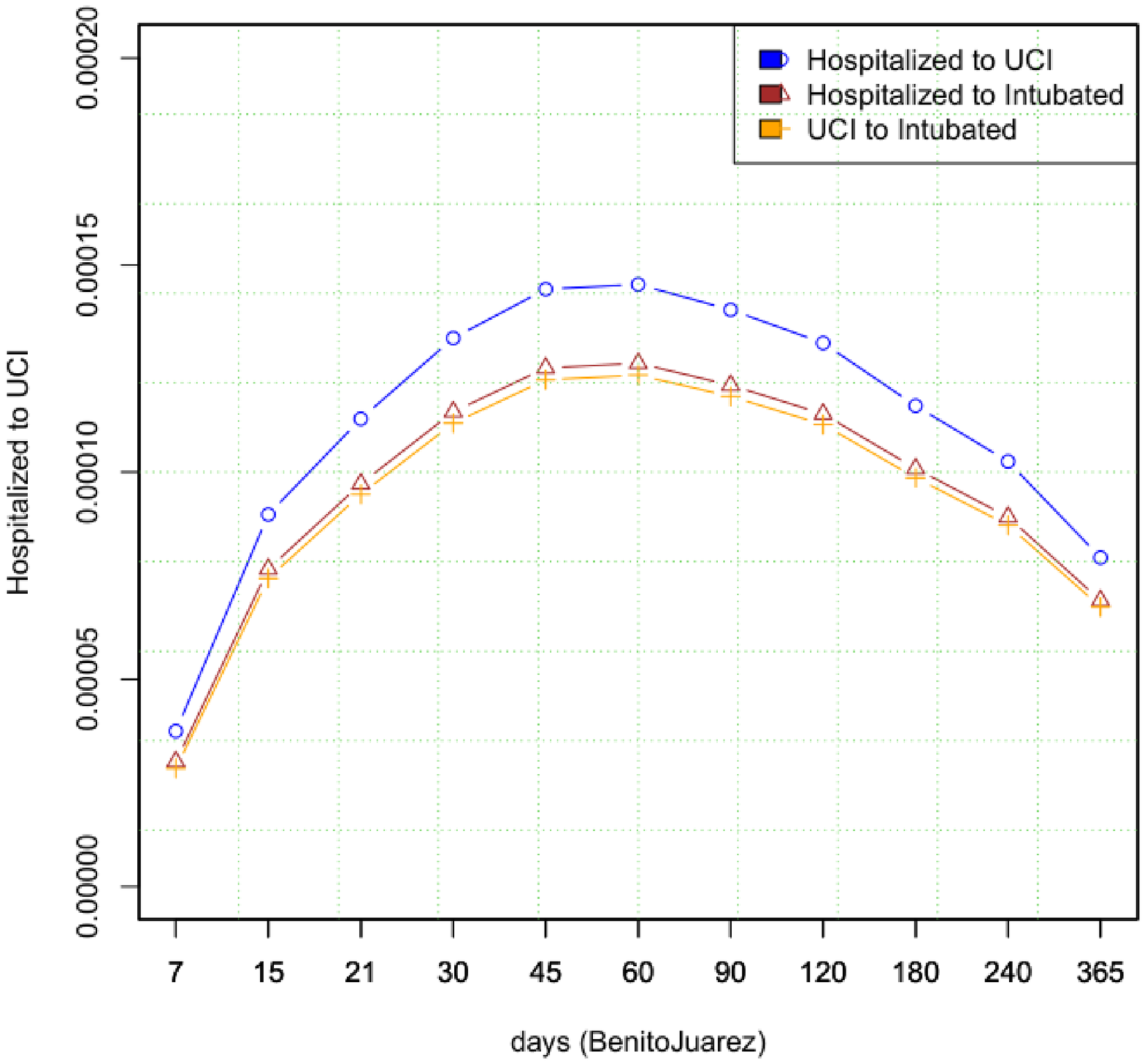}}\hspace{3mm}
\subfigure[Miguel H]{\includegraphics[width=5cm, height=5cm]{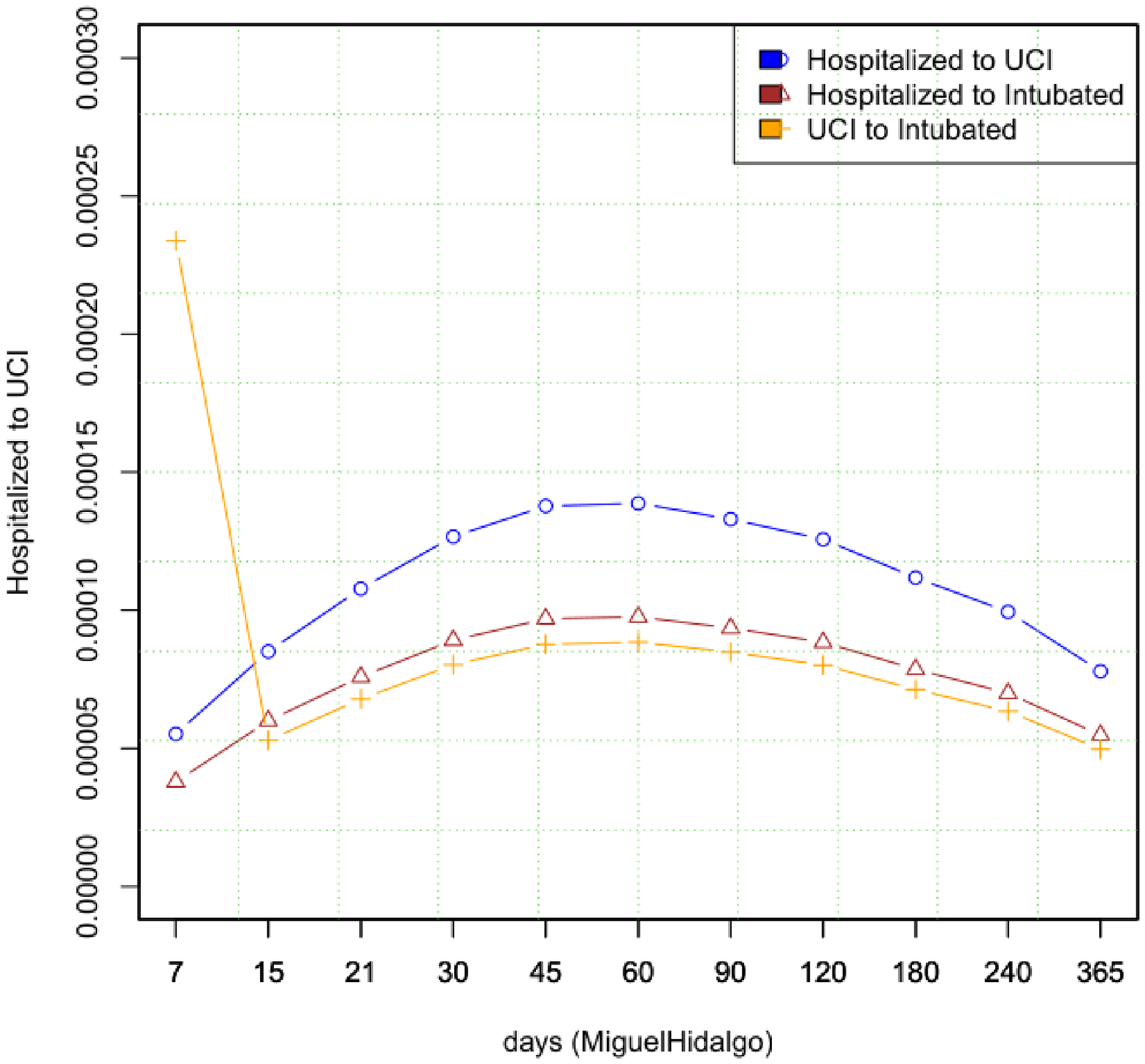}}\hspace{3mm}
\subfigure[Milpa A]{\includegraphics[width=5cm, height=5cm]{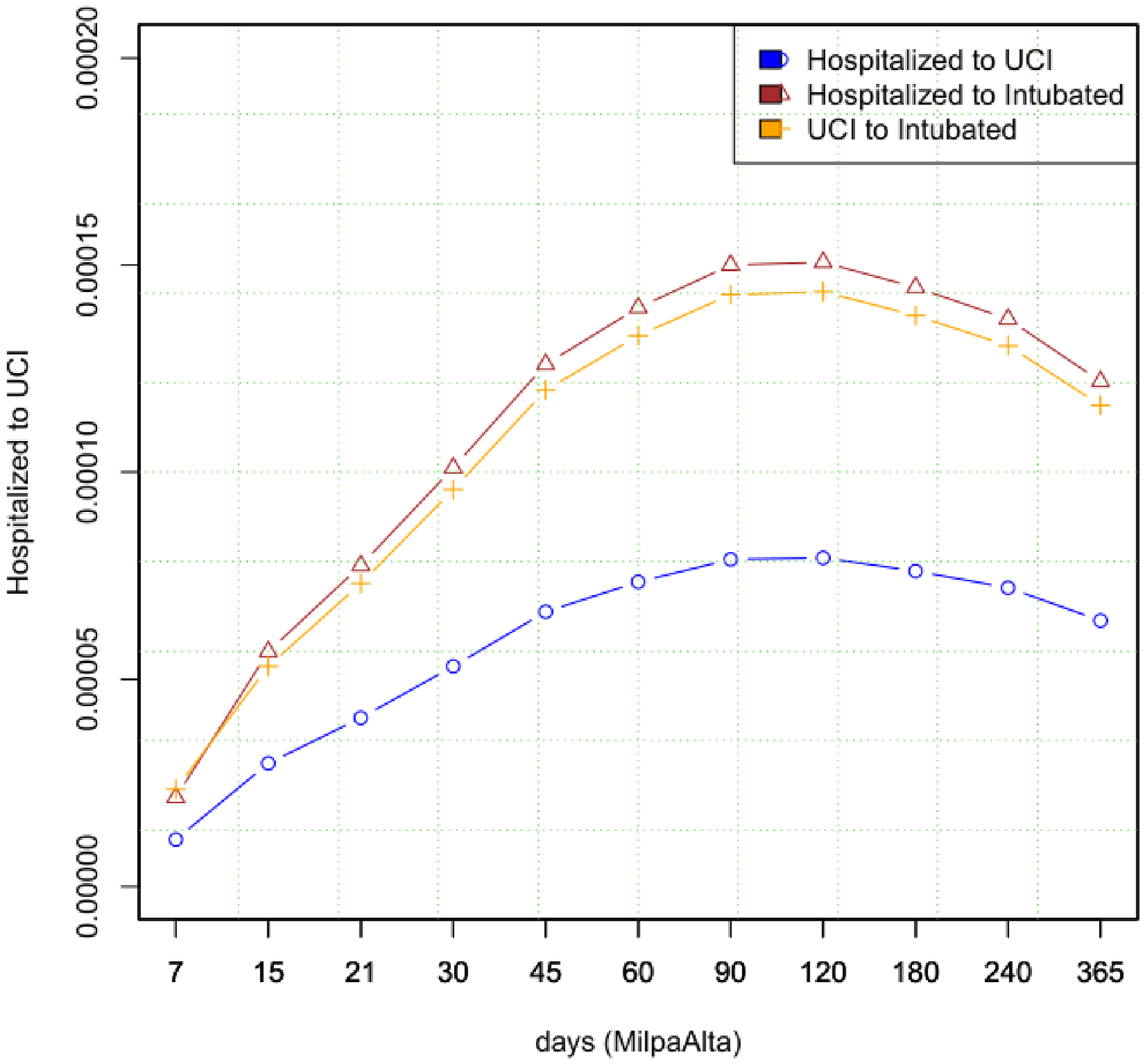}}\hspace{3mm}
\subfigure[Tlahuac]{\includegraphics[width=5cm, height=5cm]{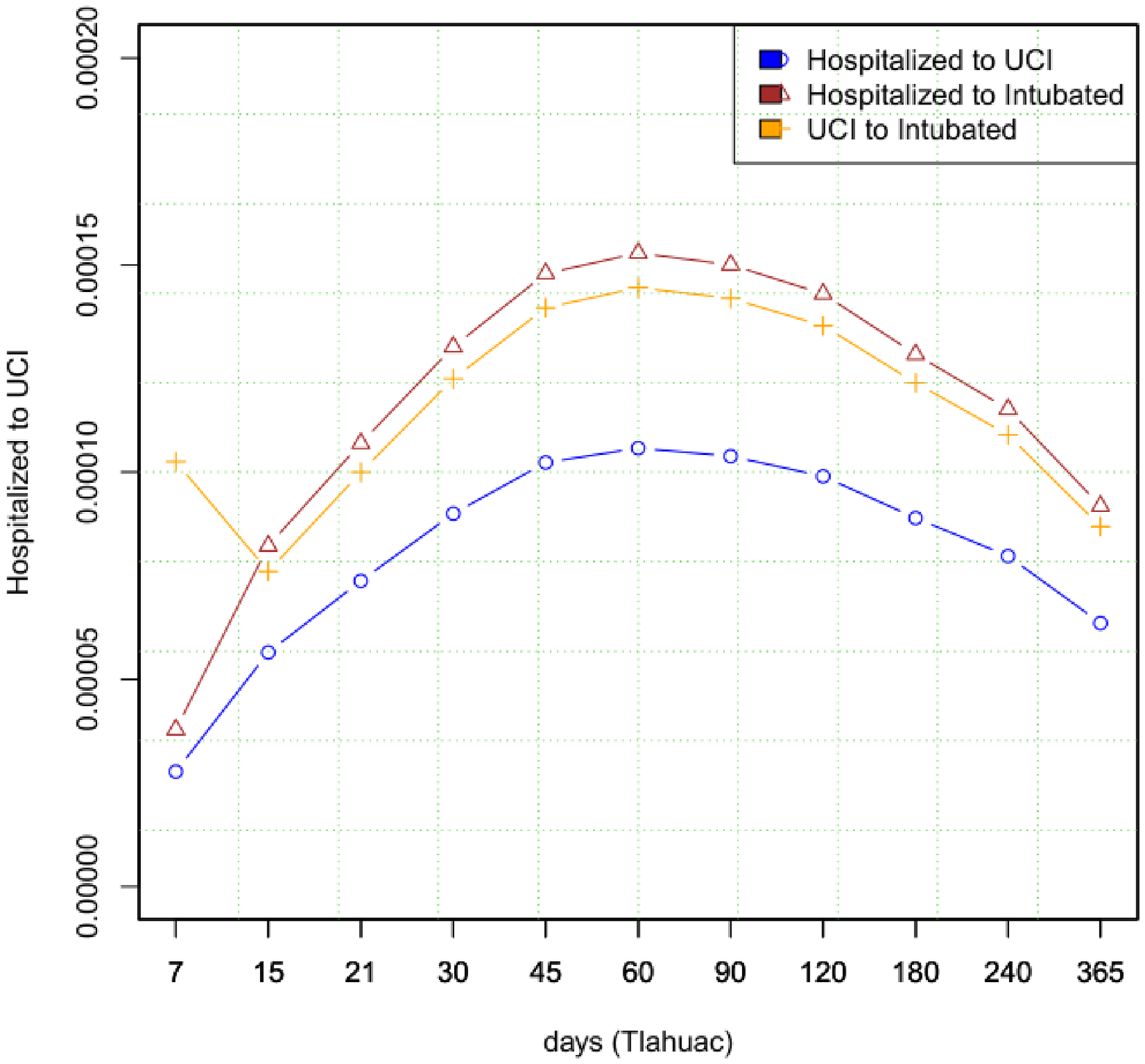}}\hspace{3mm}
\subfigure[Tlalpan]{\includegraphics[width=5cm, height=5cm]{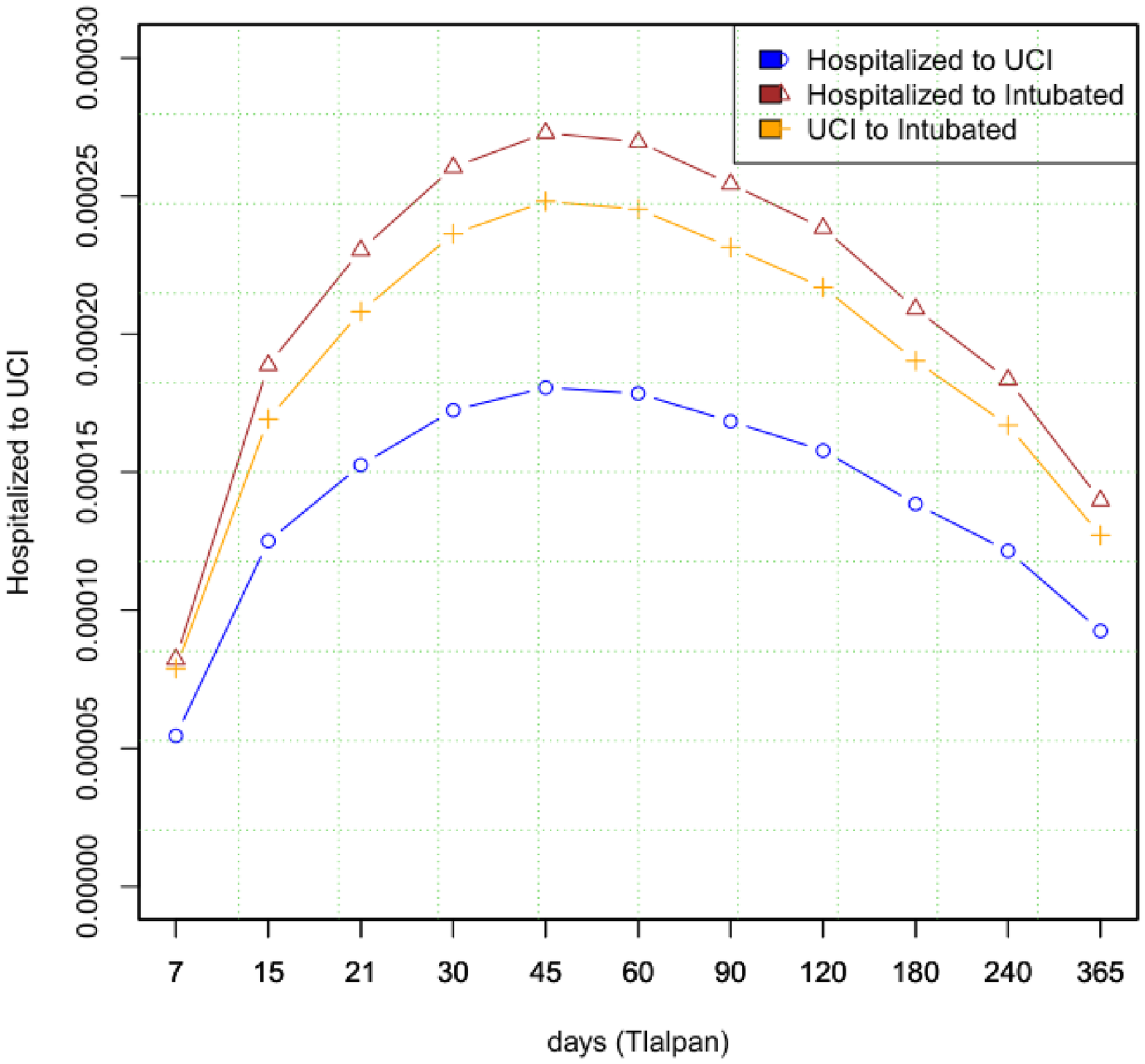}}\hspace{3mm}
\subfigure[Xochimilco]{\includegraphics[width=5cm, height=5cm]{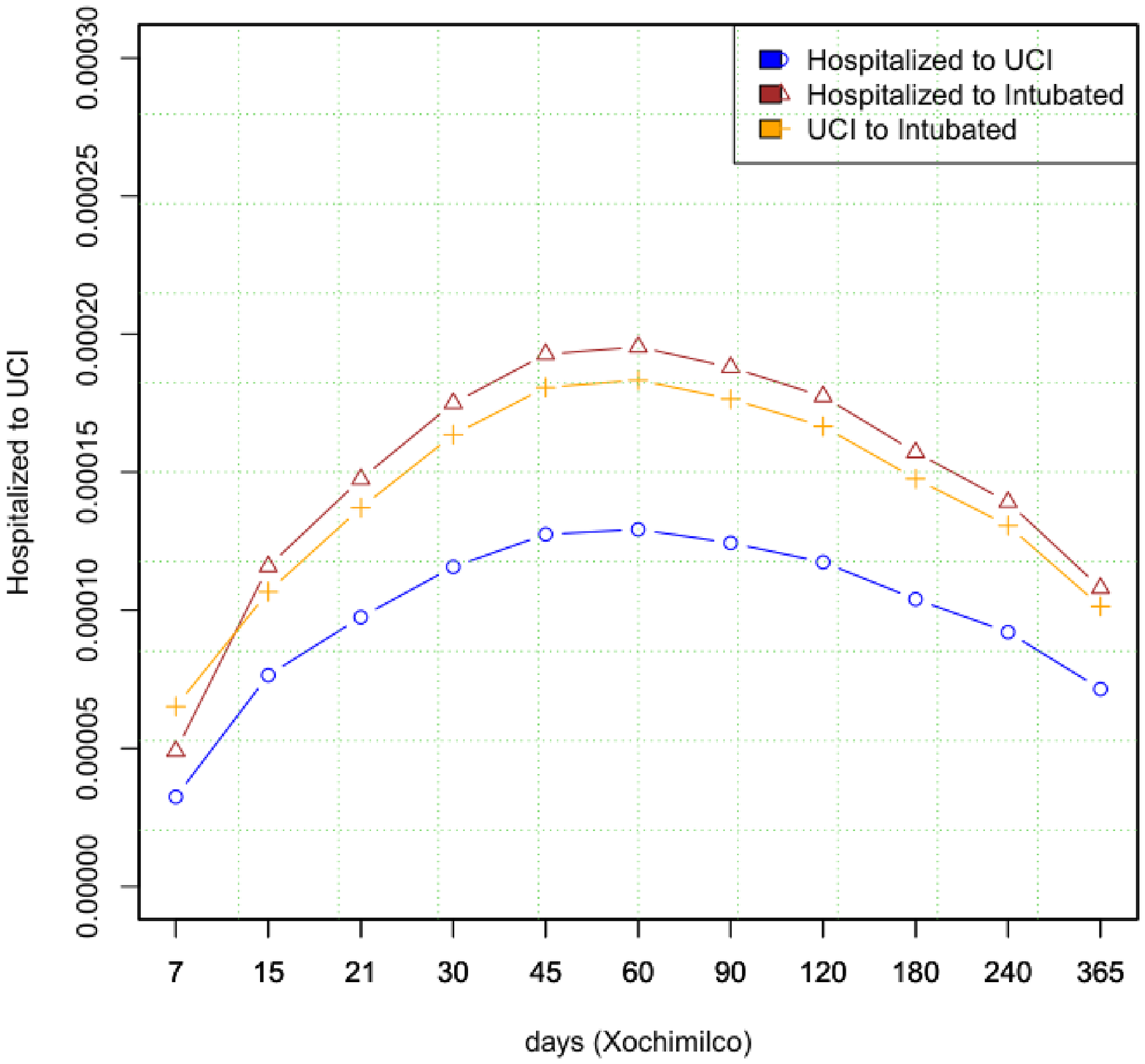}}\hspace{3mm}
\subfigure[Venustiano Carranza]{\includegraphics[width=5cm, height=5cm]{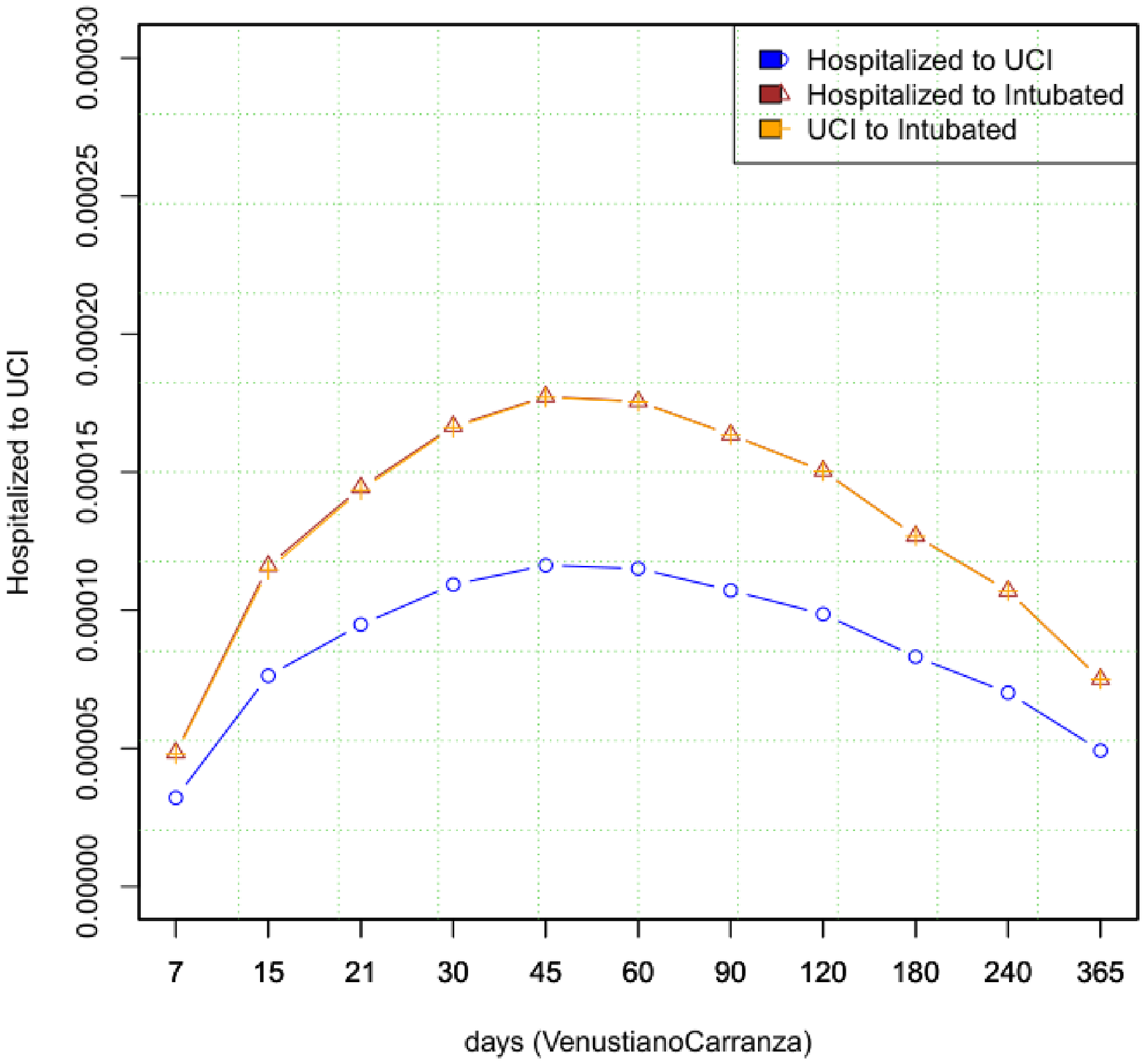}}\hspace{3mm}
\subfigure[Cuauhtemoc]{\includegraphics[width=5cm, height=5cm]{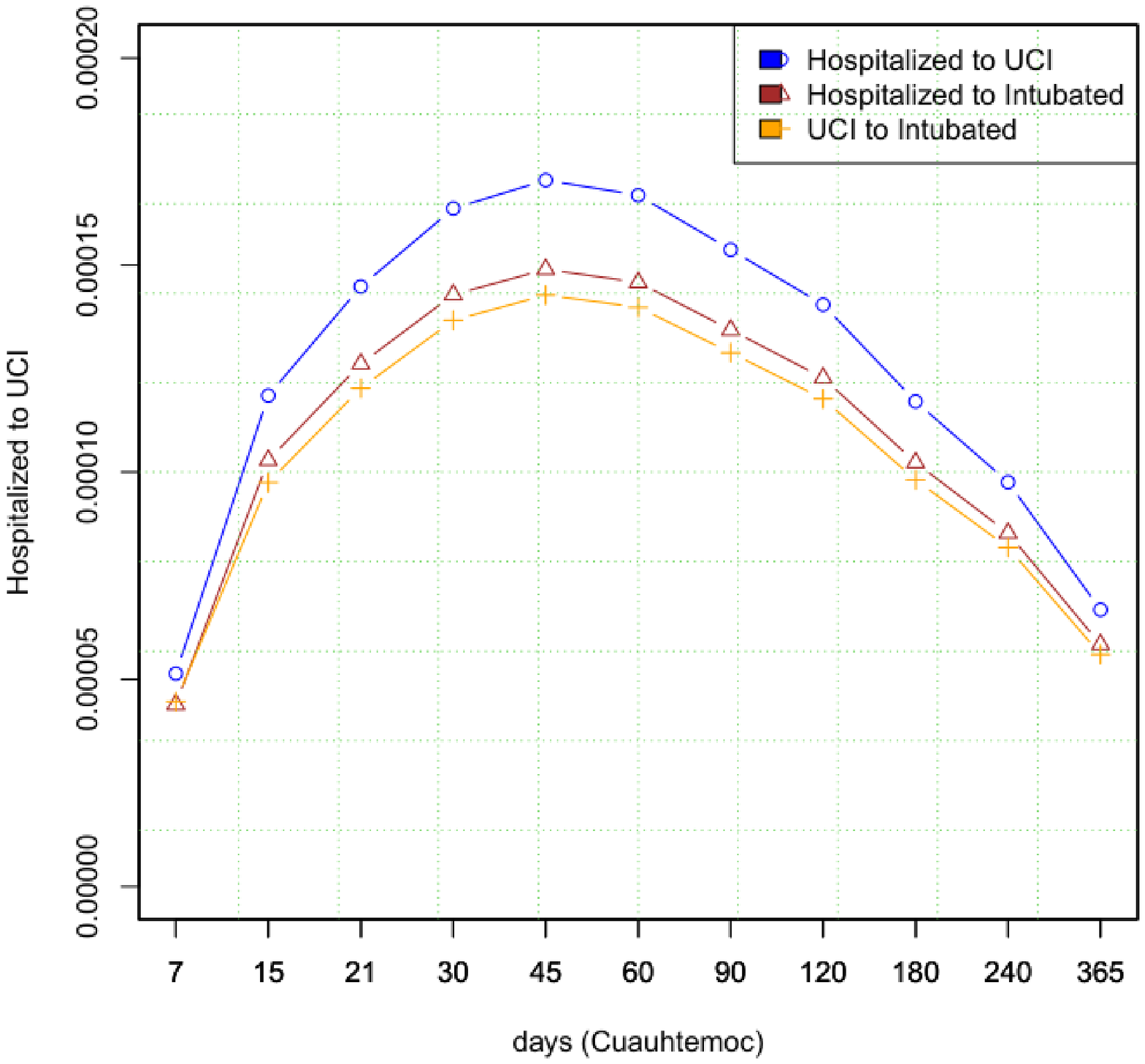}}\hspace{3mm}
\caption{Transition probabilities from hospitalized and UCI to intubated state for delegations: Benito J. ,  Miguel H., Milpa A.,  Tlahuac,  Tlalpan,  Xochimilco, Venustiano C.  and Cuauhtemoc } \label{fig:SextasOcho}
\end{figure}

\end{document}